# Xanthene Dyes for Cancer Imaging and Treatment: A Material Odyssey

**Osman Karaman[1,‡], Gizem Atakan Alkan[1,‡], Caglayan Kizilenis[1], Cevahir Ceren Akgul[1], Gorkem Gunbas[1,2,*]**

[1]Department of Chemistry, Middle East Technical University, Ankara, Turkey
[2]Biochemistry Graduate Program, Middle East Technical University, Ankara, Turkey

*** Correspondence:** ggunbas@metu.edu.tr

## Abstract

Cancer is still among the leading health issues today, considering the cost, effectiveness, complexity of detection/treatment modalities and survival rates. One of the most important criteria for higher survival rates is the early and sensitive diagnosis of the disease that can direct the treatment modalities effectively. Fluorescence imaging agents emerged as an important alternative to the current state of the art due to their spatial and temporal resolution, high sensitivity and selectivity, ease of modification towards generating activatable agents, ease of operation, and low cost. In addition to imaging, light-based treatment modality, photodynamic therapy (PDT), attained remarkable attention, as it is minimally invasive and has fewer side effects compared to the current standard of care treatments. Even though fluorescence imaging and PDT have these significant advantages, light that needs to excite the agent has limited penetration in tissues, hindering widespread utilization. Hybrid xanthene dyes, particularly ones bearing silicon or phosphine oxide as the bridging unit of xanthene moiety, gained significant interest not only due to their excellent photochemical properties in aqueous media, high fluorescence quantum yield, photostability but also their proper absorption and emission maxima that allow for deep tissue imaging and therapy. Here, the design and synthesis strategies, key photophysical properties, their application in fluorescence imaging applications, and surprisingly limited utilization as PDT agents of hybrid xanthene dyes that emerged in the last decade have been reviewed in detail.

Contents

# 1. Introduction

Cancer is still among the leading health issues today, considering the cost, effectiveness, complexity of detection/treatment modalities and survival rates.[1] Even though there is an immense effort in cancer treatment research worldwide, millions of people are diagnosed with cancer every year, and a significant ratio of these patients lose their lives.[2] One of the most important criteria for higher survival rates is the early and sensitive diagnosis of the disease that can direct the treatment modalities effectively, allowing patients to reach the right treatment as early as possible.[3] Although many scientific and technological advances have occurred in health sciences, cancer diagnosis still relies on tissue sampling and histopathological methods.[4] Clinical applications of magnetic resonance imaging (MRI), positron emission tomography (PET), and computer-assisted tomography (CAT) are generally used for the determination of spread amount and metastasis centers more than the precise diagnosis of cancer at early stages. Additionally, the agents used in these high-cost and complicated imaging methods have limited selectivity for cancer cells. Moreover, repetitive scanning is avoided due to the radioactive nature or issues related to the bio-compatibility of these agents. At this point, fluorescence imaging agents emerged as an important alternative due to their spatial and temporal resolution, high sensitivity and selectivity, ease of modification towards generating activatable agents, ease of operation, and low cost.[5] Additionally, its non-invasive nature and real-time measurement capabilities without harming the cells and tissues are among other important advantages of fluorescence imaging.[5]

Photodynamic therapy (PDT) is an effective treatment modality for various cancer types. It has attained remarkable attention during the last decade as it is minimally invasive and has fewer side effects than current state-of-the-art therapies.[6] In PDT, a photosensitizer (PS) is administered to the patient, followed by its activation by light, which triggers singlet oxygen generation and eventually cell death [7] (Figure 1). In addition, PDT also damages the vasculatures around tumor regions and, at the same time, activates the immune system against cancer cells.[8] The technique offers localized treatment since singlet oxygen has a short lifetime in the aqueous medium, and the excitation light can be precisely delivered to the tumor region, leaving most of the healthy cells unaffected.

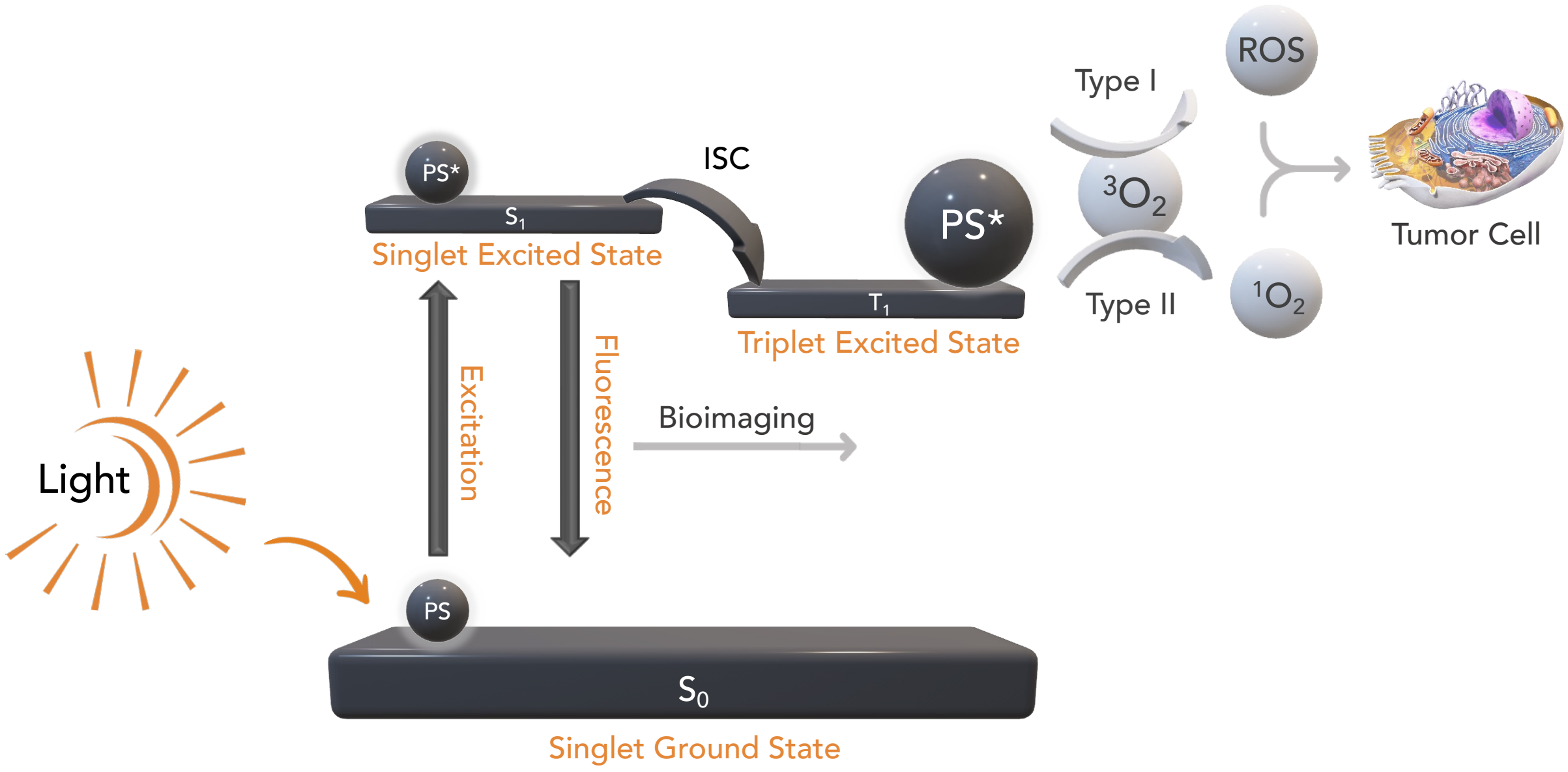

**Figure 1**. Schematic illustration of PDT and bioimaging, including the Jablonski diagram

Even though fluorescence imaging and PDT have these significant advantages, a vital issue hinders widespread use of this method: The light that needs to excite the agent has limited penetration in tissues (*in vivo* applications). To increase the utilization areas of fluorescence imaging agents, fluorophores that absorb and emit in the NIR region (650-900 nm) are needed. The light used in imaging studies using these agents is absorbed minimally by bio-molecules (hemoglobin, water, etc.) and tissues; hence the auto-fluorescence is mainly eliminated. All these advantages of NIR fluorophores result in lower background signal, directly affecting imaging resolution and better tissue penetration.[9] Another important benefit of NIR imaging and PDT is the lower energy of NIR light, resulting in a diminished photo-toxic effect on cells. A majority have used cyanine derivatives as NIR fluorophores and PDT agents for a long time. However, their low quantum fluorescence yields and low photostabilities hinder their widespread use.[10,11] Another important family of NIR fluorophores and PDT agents is based on squaraine derivatives. Although this group of imaging agents gives bright emissions in the NIR region, issues commonly arise due to their highly reactive nature. Some fluorophores that emit green light (ex: BODIPY) can be modified structurally to result in fluorophores that emit in the NIR region. There are several studies on fluorescence imaging on BODIPY dyes [12–14]; however, due to the hydrophobic nature of this core, difficulties arise during *in-vivo* applications.[15,16] Although this problem can be circumvented to an extent by introducing hydrophilic groups on the core, these modifications generally prove cumbersome and limit design flexibility. Classical xanthene-based dyes such as fluoresceins, rhodamines, and rhodols have been successfully utilized in fluorescence imaging[17] and modified with heavy atoms for PDT action, and very promising results were obtained.[18–20] However, due to their comparatively high band-gap, absorption and emission maxima are generally observed in the green region of the spectrum, which limits their use for *in vivo* studies.[21]

In the light of all these developments, it is clear that water-soluble fluorophore cores that are photostable, emit in the NIR region with high quantum yields, and offer simple modification routes to targeted fluorescence imaging and/or PDT agents are still highly sought after. The promising properties of Xanthene-based dyes, such as water solubility, photostability, and ease of modification, motivated researchers to investigate improvements on classical dyes such as fluorescein and rhodamine B towards realizing NIR absorbing/emitting dyes with highly cancer-selective activation for imaging and treatment opportunities.[22] These efforts proved fruitful, and several novel xanthene-based dyes were developed with remarkable properties. This review aims to highlight these exciting advancements towards searching for "the core", focusing on xanthene dyes bearing heteroatoms other than oxygen at the 10 position due to their significantly red-shifted absorption and emission maxima with preserved optimal features of their ancestors, which we believe are paving the way for utilization of fluorescence imaging and PDT among the *standard of care* in cancer imaging and therapy in the coming years.

While coordination chemistry is mainly regarded as the realm of transition metal complexes, its scope now spreads to the binding of a plethora of substrates, including charged and neutral molecular species, blurring the line between coordination chemistry and supramolecular chemistry.[23] Although much success has been attained by utilizing reaction-based activation of fluorophores with over-expressed cancer metabolites such as thiols and $H_2O_2$, high-performance selective bioimaging and/or treatment of cancerous cells heavily depends on the coordination of carefully designed substrate-fluorophore couples to the relatively over-expressed and/or mutated enzymes followed by the relevant reaction for releasing of the "active" agent.[24] Our survey of the last decade clearly demonstrated that the community is betting on the success of this approach towards realizing the next generation of imaging/treatment options.

# 2. Fluorescein Derivatives

## 2.1. Si-Fluoresceins

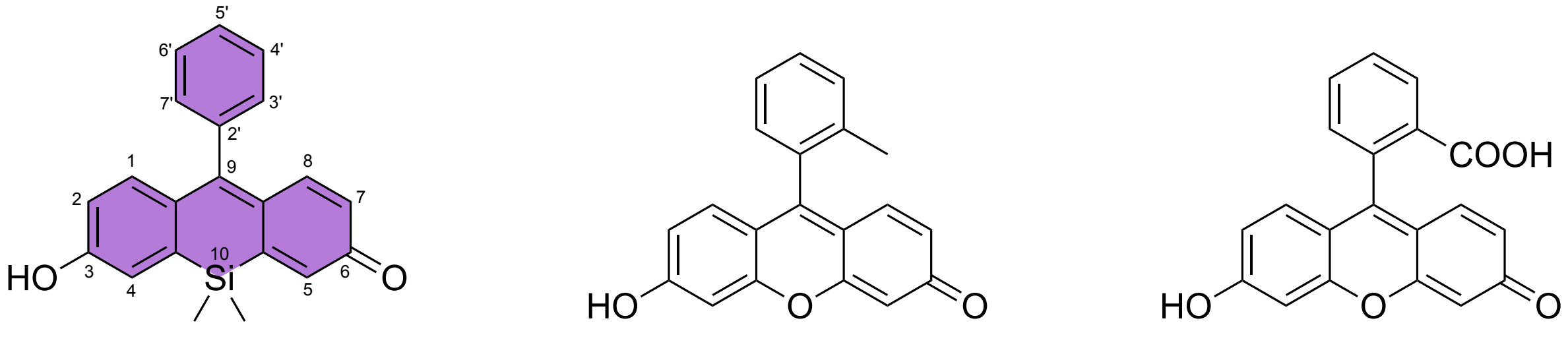

Si-Fluoresceins  Tokyo Green  Fluorescein

In 2011 Nagano's group introduced the Si atom at the 10-position of the widely popular Tokyo Green dye structure.[25] This was the first known example of replacing the bridging oxygen atom with dimethylsilane in the fluorescein framework. The novel dye series was called Tokyo Magentas (TMs), and the synthesis was achieved via the addition of the appropriate aryllithium to the xanthone precursor followed by loss of water yielding **2-Me TM** (Table 1:E2 Scheme 1). The synthetic pathway was equivalent to the pathway for rhodamine analogs with the extension of converting amine groups on the 3 and 6 positions to alcohols with the Sand-Meyer reaction before the nucleophilic addition of aryllithium reagents (Scheme 2, Blue pathway). The absorption and emission maxima of the neutral form of **2-Me TM** were 472 and 593 nm ($\Phi_f$ = 0.10 in phosphate buffer containing 1% DMSO), whereas a 110 nm red-shift of the absorbance and a 4.2-fold increase in the fluorescence quantum yield was observed for the deprotonated form ($\lambda_{max}$=598 nm). To investigate whether the hydroxy group derivatization alone can be used to develop off/on type fluorescence probes, **2-Me TM βgal** (Table 1:E1, Scheme 1) was designed for β-galactosidase activation. **2-Me TM βgal** was almost nonfluorescent with an absorption maximum of 448 nm. Cultured HEK293 cells (lacZ(+) or lacZ(-)) were incubated with **2-Me TM βgal** for 30 min, and a large increase in fluorescence intensity was observed in the intracellular region of lacZ(+); however, not on lacZ(-) cells. Tokyo magenta was proved to be an important milestone for the development of NIR probes for a range of applications in cancer imaging and therapeutics.

Later in 2015, Crovetto *et al.* investigated the aqueous solution of **2-Me TM** (Table 1:E3, Scheme 1) via absorption, steady-state, and time-resolved emission spectroscopy.[26] Three different absorbing and emitting species were identified. $pK_{N-A}$ of **2-Me TM** was measured from the ground-state equilibrium between the neutral and anionic forms using absorption and fluorometric titration data. To reveal the buffer-mediated two-excited-state photophysics system, steady-state and time-resolved emission spectra (TRES) methodologies were employed. Results showed that the main process was the buffer-mediated deprotonation of the neutral form of the dye. In contrast, the buffer-mediated protonation of the anion form of the dye was a much slower process. The reason for the very low value of the $pK^*_{N-A}$ was the higher rate constant for deprotonation than protonation. And lastly, to describe the system's dynamic behavior, the fundamental kinetic rate constants were determined from the pertinent kinetic excited-state photon transfer (ESPT) reaction model, and its dynamics were resolved by collecting the fluorescence decay traces. It was demonstrated that **2-Me TM** undergoes a phosphate-mediated ESPT reaction, and a rapid interconversion between the neutral and anionic species was observed. With all data obtained from this in-depth analysis of the photophysical properties of the dye, the applicability of **2-Me TM** for bioimaging was tested on HeLa 229 cells. Results showed that upon excitation at 530 nm, the dye accumulated visibly inside the cytoplasmic structure, whereas it accumulates in the cytosol and nucleus less effectively but homogenously. This in-depth analysis was critical for fully understanding the photophysics of **2-Me TM** for its use in effective imaging applications.

*Table 1. Selected features of Si-Fluorescein derivatives.*

| Entry (E) | Fluorophore* | Absorption | | Emission | $\Phi_f$ | Intracellular Localization | Properties | Applications | Ref. |
|---|---|---|---|---|---|---|---|---|---|
| | | $\lambda_{max}$ (nm) | ε ($M^{-1}\cdot cm^{-1}$) | $\lambda_{max}$ (nm) | | | | | |
| 1 | 2-Me TM[a] | 582[b]<br>472[c] | 110000[b]<br>29000[c] | 598[b]<br>472[c] | 0.42[b]<br>0.10[c] | - | photostable,<br>water-soluble,<br>pH range = 3-11 | pH responsive | [25] |
| 2 | 2-Me TM βgal | n.d.<br>582[b]*<br>472[c]* | n.d.<br>110000[b]*<br>29000[c]* | n.d.<br>598[b]*<br>472[c]* | n.d.<br>0.42[b]*<br>0.10[c]* | - | cell permeable,<br>activatable | *in vitro* imaging (HeLa, RAW 264.7) | [25] |
| 3 | 2-Me TM | 470[d] | n.d. | 595[e] | n.d. | cytoplasmic structures<br>less accumulation<br>in cytosol and nucleus | water-soluble | *in vitro* imaging (HeLa 229 cells) | [26] |
| 4 | 2-COOH TM[f] | 580 | 110000 | 598 | 0.38 | - | working pH>7.8 | - | [27] |
| 5 | 2-COOH DCTM[f] | 591 | 120000 | 607 | 0.48 | - | working pH>6 | - | [27] |
| 6 | 2-COOH DFTM[f] | 581 | 120000 | 596 | 0.54 | - | working pH>6 | - | [27] |
| 7 | 2-COOH DCTM βgal[g] | 460<br>591* | 1200<br>120000* | 555<br>607* | 0.009<br>0.48* | - | pH-dependent,<br>on/off probe,<br>activatable | red colored strong fluorescence in the presence of $\beta$-galactosidase | [27] |
| 8 | 2-COOH TM | 579 | 93000 | 599 | 0.53 | - | - | - | [28] |
| 9 | 2-COOH 2,7-DFTM | 596 | 60900 | 618 | 0.42 | - | - | - | [28] |
| 10 | Maryland Red | 597 | 65000 | 613 | 0.67 | - | - | - | [28] |
| 11 | 2-COOH 3,4,5,6-TFTM | 598 | 84100 | 614 | 0.50 | - | - | - | [28] |
| 12 | 2-COOH-3,4,5,6-TF-DFTM | 617 | 67900 | 635 | 0.32 | - | - | - | [28] |
| 13 | 2,6-COOH TM | n.d. | n.d. | n.d. | n.d. | - | - | - | [28] |
| 14 | photoFAD-1[h]<br>2-Me TM* | 442<br>582* | 16800<br>43600* | -<br>599* | n.d.<br>0.62* | - | 40-fold fluorescent turn-on, | - | [29] |
| | | | | | | | photobleaching observed | | |
| **15** | **photoFAD-2**[h]<br>2-$CF_3$ TM* | 445<br>590* | 20300<br>52500* | -<br>608* | n.d.<br>0.58* | - | 205-fold fluorescence turn-on,<br>not retained relative to the unactivated donor | *in vitro* imaging (HEK293 cells) | [29] |
| **16** | **photoFAD-3**[h]<br>2-$CF_3$ DCTM* | 451<br>601* | 22300<br>53000* | -<br>616* | n.d.<br>0.60* | cytoplasm | greater cellular retention,<br>139-fold fluorescence turn-on,<br>photostable (thiols and ROS),<br>no cytotoxicity | *in vitro* imaging (HEK293 cells) | [29] |
| **17** | **AM-FAD-3**[h] | 452 | 19300 | 564 | n.d. | cytoplasm | photostable,<br>no cytotoxicity | *in vitro* imaging (HEK293 cells) | [29] |
| **18** | **2-Me-4-OMe-TM** | 456 and 613 | n.d. | 555[i]<br>590[j] | 0.071[i]<br>0.107[j] | cytoplasm / nucleus | solvatofluorochromic | *in vitro* imaging (HEK293 cells) | [30] |
| **19** | **2-Me-4-COH TM** | n.d. | n.d. | n.d. | 0.32 | - | high ALDH1A1, selectivity / medium ALDH1A3 selectivity | - | [31] |
| **20** | **2-F-4-COH TM** | n.d. | n.d. | n.d. | n.d. | - | n.d. | - | [31] |
| **21** | **2-$CF_3$-4-COH TM** | n.d. | n.d. | n.d. | 0.25 | - | medium ALDH1A1 selectivity / low ALDH1A3 selectivity | - | [31] |
| **22** | **2-$NO_2$-4-COH TM** | n.d. | n.d. | n.d. | 0.00 | - | n.d. | - | [31] |
| **23** | **2-Me-4-COH-5-F TM** | n.d. | n.d. | n.d. | 0.21 | - | high ALDH1A1 selectivity / low ALDH1A3 selectivity | - | [31] |
| **24** | **2,5-F-4-COH TM** | n.d. | n.d. | n.d. | 0.13 | - | high ALDH1A1 selectivity / high ALDH1A3 selectivity | - | [31] |
| **25** | **red-AlDeSense** | 594 | - | 614 | 0.21 | - | high ALDH1A1 selectivity / low ALDH1A3 selectivity | *In vitro* imaging<br>(A549 and k562 cells) | [31] |
| **26** | **2-Me-3,6-F-4-COH TM** | n.d. | n.d. | n.d. | 0.20 | - | high ALDH1A1 selectivity / medium ALDH1A3 selectivity | - | [31] |

[a] Photophysical properties were measured in sodium phosphate buffer at pH 9 for the anion form and pH 3 for the neutral form. [b] Anion form of 2-Me TM. [c] Neutral form of 2-Me TM. [d] Estimated from the spectrum (pH=3.04). [e] estimated from the spectrum (pH=6.18-9.00) [f] Measured in 100 mM sodium phosphate buffer at pH 9.0 in the presence of 1% DMSO as cosolvent. [g] Measurements were made in PBS (pH=7.4). [h] Measured in PBS (0.1% DMSO) at pH=7.4. [i] Ex=440nm. [j] Ex= 530 nm. * Values for the activated form. n.d.: not detected. [*]all the fluorophores listed above are emphasized with bold font in the text.

The replacement of the bridging oxygen atom in the classical spiro lactonized fluorescein core with dimethylsilane moiety was achieved by Urano group in 2015.[27] The absorption and emission maxima of the novel Tokyo Magenta analog **2-COOH TM** were 580 and 598 nm, respectively (**$\Phi_f$** = 0.38 in phosphate buffer pH = 9 containing 1% DMSO) (Table 1:E4, Scheme 1). The results were comparable with **2-Me TM**. Interestingly, **2-COOH TM** showed a larger $pK_{a1}$ than $pK_{a2}$, contrary to common fluorescein derivatives. In the fluorescein structure, it was known that the introduction of electron-withdrawing groups such as chlorine and fluorine lower the $pK_a$ values. To **2-COOH TM**, chlorine and fluorine were introduced to the core yielding **2-COOH DCTM** (Table 1:E5, Scheme 1) and **2-COOH DFTM** (Table 1:E6, Scheme 1), which showed similar photophysical properties and successfully reversed the $pK_a$ order. Similar to **2-Me TM βgal**, the fluorescent probe for β-galactosidase was also prepared from **2-COOH DCTM** (Table 2, E7, Scheme 1). In the enzymatic reaction of **2-COOH DCTM βgal** with β-galactosidase, **2-COOH DCTM βgal** was almost colorless and nonfluorescent before the reaction and became red-colored and strongly fluorescent with time (figure 2).

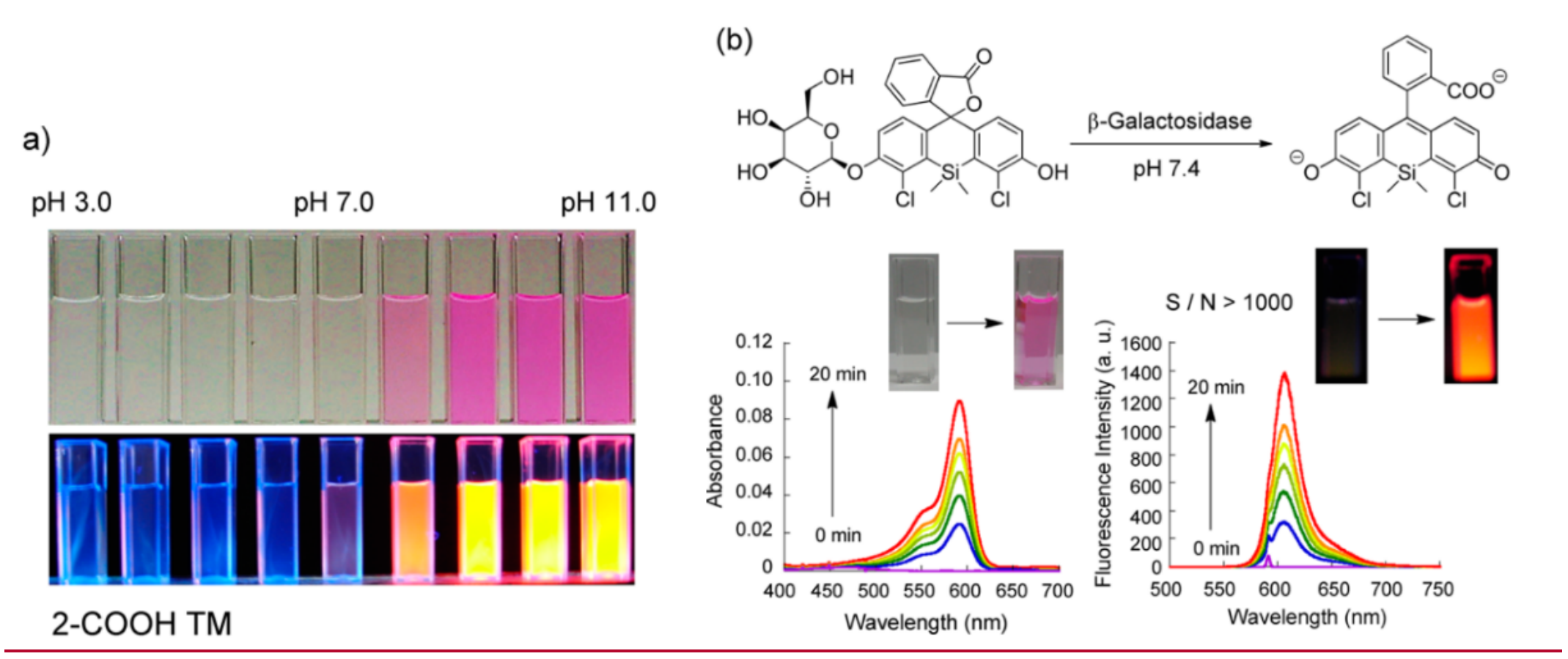

**Figure 2.** (a) Photographs of 2-COOH TM and fluorescein solutions under white light or under irradiation with a handy UV lamp (**λ**em = 365 nm) at various pH values between 3.0 and 11.0. (b) Absorption (left) and emission (right) spectra of 1 μM 2- COOH DCTM **β**gal after addition of **β**-galactosidase (0.3 unit). All experiments were performed at 37 °C in 3.0 mL total volume of 100 mM sodium phosphate buffer at pH 7.4, containing 0.1% DMSO as a cosolvent. The excitation wavelength of 2-COOH DCTM **β**gal was 591 nm. The reaction scheme of 2-COOH DCTM **β**gal and photographs of the sample solutions before and after the enzymatic reaction is also shown. Reproduced with permission [27] Copyright 2017, American Chemical Society

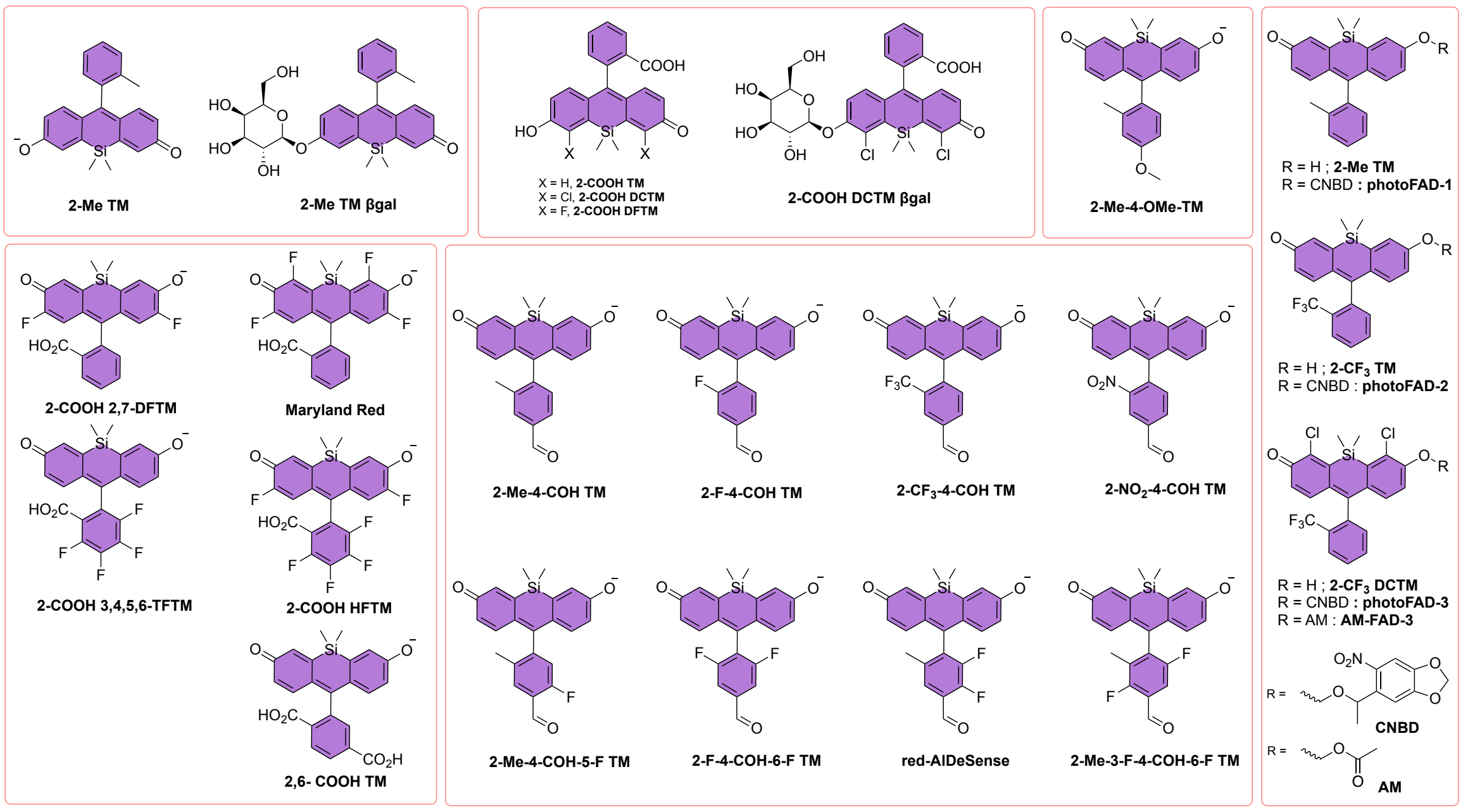

**Scheme 1.** Chemical structures of Si-Fluorescein derivatives.

In 2017 Lavis' group generated an alternative synthetic strategy for a wider variety of Si-fluorescein and Si-rhodamine derivatives.[28] In literature, the pendant aryl ring was introduced by adding metallated aryl species to Si-xanthone in most of the synthesis of Si-rhodamine and Si-fluorescein derivatives (Scheme 2, blue pathway). This approach represented an electronic mismatch between reacting species where the electron-rich ketone acted as the electrophile, and the electron-poor arylmetal served as the nucleophile. Hence, the synthetic pathways were limited by using stronger arylmetal reagents that are incompatible with many functional groups and required protection strategies. In the new strategy, a bismetalled biaryl derivative was added to an ester or anhydride twice to introduce the pendant aryl ring (Scheme 2, green pathway). Hence the electronic match was achieved via an electron-rich bis-nucleophile and an electron-poor electrophile. The silicon analog of fluorescein, **2-COOH TM** was synthesized from this new synthetic strategy with a higher yield (Table 1:E8, Scheme 1). Other synthesized Si-fluorescein analogs were the difluorinated derivative of **2-COOH TM** from positions 2 and 7 (**2-COOH 2,7-DFTM**, Table 1:E9, Scheme 1), the tetrafluorinated derivative of **2-COOH TM** from the positions 2,4,5 and 7 (**Maryland Red**, Table 1:E10, Scheme 1), the tetrafluorinated derivative of **2-COOH TM** from the positions 4′,5′,6' and 7' (**2-COOH 3,4,5,6-TFTM**, Table 1:E11, Scheme 1), the hexafluorinated derivative of **2-COOH TM** from the positions 2,7,3′,4′,5′ and 6′ (**2-COOH-3,4,5,6-TF-DFTM**, Table 1:E12, Scheme 1) and the carboxylic acid bearing derivative of 2-COOH TM at the 5′ position (**2,6-COOH TM**, Table 1:E13, Scheme 1). The introduction of fluorine atoms reduced the dyes' $pK_a$ values, which provided biologically useful fluorogenic probes and pH sensors. This new synthetic strategy by an electronically matched protocol starting from a simple dibromide intermediate, serves as a milestone in the synthesis of Si-fluorescein based dyes.

Smaga *et al.* developed a series of photoactivable formaldehyde (FA) donors (**photoFADs**) in 2019 using the silicon substituted fluorescein cores.[29] Each derivative contained a photolabile nitrobenzyl group connected to the Si-xanthene dye through an acetal linkage. The dye units of the photoFAD series included Si-xanthenes bearing a 2-methylphenyl group at the 9-position (**photoFAD-1**), a 2-trifluoromethylpheny group at the 9-position (**photoFAD-2**), and a 4,5-dichloro derivative of **photoFAD-2** (**photoFAD-3**) with dimethylsilane group as the bridging moiety. The photoFADs were nonfluorescent, and upon irradiation with light, the photocleavage was achieved to yield the unstable hemiacetal intermediate, which was hypothesized to generate the corresponding emissive dye and formaldehyde spontaneously. A 40-fold enhancement in the fluorescence signal was observed for **photoFAD-1** (Table 1:E14, Scheme 1)), however photobleaching of the dye was also in effect. The trifluoromethyl derivative **photoFAD-2** showed improved photostability with an enhanced turn-on fluorescence (Table 1:E15, Scheme 1). However, when it was further evaluated in HEK293 cells, the dye did not retain in the cells relative to the inactivated donor. By introducing electron-withdrawing chloro substituents to **photoFAD-3**, the $pK_a$ value was decreased, and greater cellular retention was achieved. In this way, the dye remained in an anionic state under physiological conditions, reducing the dye's passive diffusion through the cell membrane. *In vitro* studies revealed that **photoFAD-3** was nonfluorescent in its inactivated form with a 139-fold fluorescence turn-on response (Table 1:E16, Scheme 1). A 10.4-fold fluorescence enhancement was observed when **photoFAD-3** was exposed to irradiation in HEK293 cells. In addition to developing the first photoactivable FA donor, this work highlighted the importance of synthetic modification of fluorescent probes for desired properties and/or circumventing issues arising in live cell studies.

Later in 2019, another study on bioimaging was reported. Espinar-Barranco *et al.* synthesized a Si-xanthene derivative bearing 2-methyl-4-methoxyphenyl group at the 9-position (**2-Me-4-OMe-TM**, Table 1:E18, Scheme 1) with dimethylsilane group as bridging moiety.[30] To fully investigate the solvatochromic properties, the prepared dye was dissolved in 16 different solvents, and the absorption spectra, steady-state fluorescence spectra, quantum yield, and fluorescence lifetime were measured. According to the spectroscopic data, two different groups of solvents were distinguished. A dual absorption band, one centered between 434 and 474 nm and the other between 592 and 613 nm, was observed in the first group of solvents, including alcohols which are polar protic except for DMSO and acetonitrile. The main feature of these solvents was their acidic character. On the contrary, a single absorption band, centered between 420 and 446 nm, was observed for the other solvents, including aprotic ones. The highest quantum yield values corresponded to the dye dissolved in protic solvents where the dual absorption was observed. Using the Catalan approach,[32] it was concluded that both solvent groups and related absorption bands were mainly affected by the acidity. However, emission characteristics were affected by different physicochemical parameters. For the first band, the solvent polarity was the primary physicochemical parameter that created the red shift in emission. The result revealed that the second absorption and emission bands have arisen from the hydrogen bonding interactions with the solvent. Results from the ($E_T30$) solvent scale, polarity analysis of the absorption and emission maxima demonstrated that the first band corresponded to a $\pi \rightarrow \pi^*$ transition, whereas the red-shifted band was due to an $n \rightarrow \pi^*$ transition. The acid-base equilibrium around physiological pH was investigated using absorption, steady-state, and time-resolved fluorescence spectroscopy. The ions present in the phosphate buffer at near-neutral pH effectively promote the excited state photon transfer (ESTP) reaction, which creates a phosphate-dependent decrease in the fluorescence lifetime of the dye. The results indicated the usage of **2-Me-4-OMe-TM** as a fluorescence lifetime imaging microscopy (FLIM) sensor for phosphate in biological samples was not fruitful due to the low sensitivity of the dye when utilized intracellularly. Since the cytoplasm resembled a complex matrix consisting of multiple compartments and microenvironments, studies were conducted to evaluate the effects of membranous systems on the photophysical properties of **2-Me-4-OMe-TM** in SDS micelles at different phosphate concentrations. A decrease in the 595 nm absorption maximum and an increase at 460 nm were observed. Lastly, the experiments in HEK293 cells showed a clear accumulation pattern. The FLIM images revealed that the anionic form was predominantly excited and showed a longer lifetime in a nonpolar environment. The dye accumulated in nuclei showed longer fluorescence than those dissolved in the cytoplasm but lower than those incorporated in the plasma membrane structure. The reason for this observation was attributed to the microenvironment being highly polar in the cytoplasm and less polar in the nucleus.

In 2020, a Si-xanthene core was utilized for an aldehyde dehydrogenase 1A1 (ALDH1A1) activity monitoring fluorescent sensor by Bearrood *et al*.[31] A cell-permeable, red-shifted, activity-based sensor named **red-AlDeSense** was developed based on the TokyoMagenta core. The core unit consisted of Si-xanthene bearing 4-formylphenyl group bearing at the 9-position with dimethylsilane group as the bridging moiety. Eight derivatives were synthesized by functionalizing the aryl group at positions 3', 4', 6', and 7' (Table 1:E19-26 Scheme 1). These derivatives were analyzed from data based on comparing their fluorescence quantum yields, enzymatic turnover numbers, and activities based on ALDH1A1 and ALDH1A3. As a result, the derivative bearing a 2-methyl-4-formyl-5,6-difluorophenyl group at the 9-position, **red-AlDeSense**, was shown to be selective for ALDH1A1 over ALDH1A3 and displayed a 3.2-fold turn-on response when treated with ALDH1A1. The constitutional isomer of **red-AlDeSense** was also synthesized as the last probe to test the impact of the different fluoro substitution patterns; however, positioning one of the fluoro groups at position 6' instead of 4' resulted in the loss of selectivity between ALDH1A1 and ALDH1A3. The absorbance and emission maxima of **red-AlDeSense** were centered at 594 and 614 nm, respectively ($\Phi_f$ = 0.21), and the dye employed high cell permeability associated with the $pK_a$ value of the phenolic alcohol (6.68). The sensor required only the action of ALDH1A1 for both accumulation and fluorescence turn on. Compared to its fluorescein analog ALDH1A1 sensor, **AlDeSense**, **red-AlDeSense** showed faster response times. Results of flow cytometry and confocal microscopy revealed that **red-AlDeSense** stained A549 lung adenocarcinoma cells with high ALDH1A1 activity. Also, the studies showed independent staining for ALDH1A1 and CD44 levels in 549 lung adenocarcinoma cells. This work was one of the first examples of a design approach employed in the development of a red fluorescent Si-xanthene type sensor based on the donor-photoinduced electron transfer (d-PeT) mechanism.

## 2.2. Phospa-Fluoresceins

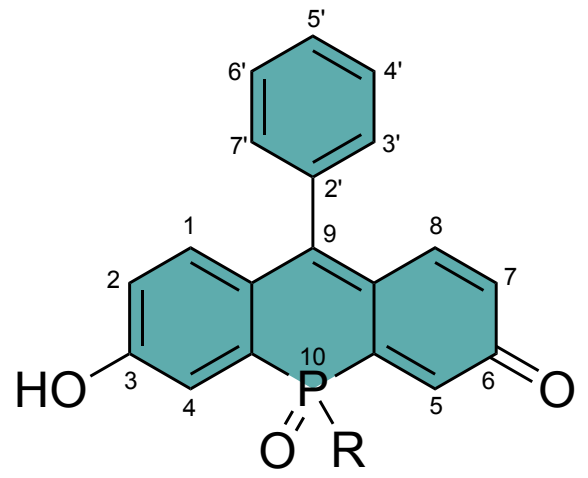

The first introduction of the phosphine oxide moiety into the fluorescein core by substitution of the endocyclic oxygen atom was envisioned by Yamaguchi's group in 2016.[33] Phospha-fluorescein (**POF**) with phenylphosphine oxide moiety bearing 2-methylphenyl group at the 9-position revealed a pH-depended absorption spectrum due to the equilibrium between the protonated and deprotonated forms, similar to many reported fluorescein derivatives.[25,34] The absorption and emission maxima of **POF** were found as 627 nm and 656 nm respectively (**Φf** = 0.33 $Na_2CO_3$ / $NaHCO_3$ buffer containing 1% DMSO) (Table 2:E1, Scheme 5). This result revealed that the introduction of the phosphine oxide moiety yielded a 136 nm shift with respect to its oxygen analog TokyoGreen[34] and a 45 nm shift with respect to its silicon analog TokyoMagenta.[25] However, the fluorescence quantum yields were decreased by 2.6 and 1.3 fold, respectively. Quantum chemical calculations of **POF** affirmed the results by predicting a significant decrease of the LUMO level due to the **σ***- **π*** conjugation, which results in a narrower HOMO-LUMO gap compared to the oxygen and silicon analogs. Live cell applications of **POF** were performed on HeLa cells using its acetyl-protected derivative **AcPOF** (Table 2:E2, Scheme 5) since **POF** was impermeable to the cell membrane due to its anionic character. The studies showed that **POF** revealed no cytotoxicity. RAW 264.7 cells were stained with **AcPOF,** and the obtained fluorescence intensities were comparable between pH 4.5 and 6.5 upon excitation at 532 nm. Upon excitation at 627 nm, a significant increase in fluorescence intensity was observed for pH 6.5, which shows **POF** should be a promising pH-responsive fluorescent probe. These characteristics showed that the electron-withdrawing character of the phosphine oxide moiety provided a significantly red-shifted absorption and fluorescence compared to the corresponding oxygen and silicon analogs and rendered **POF** a promising fluorescent probe.

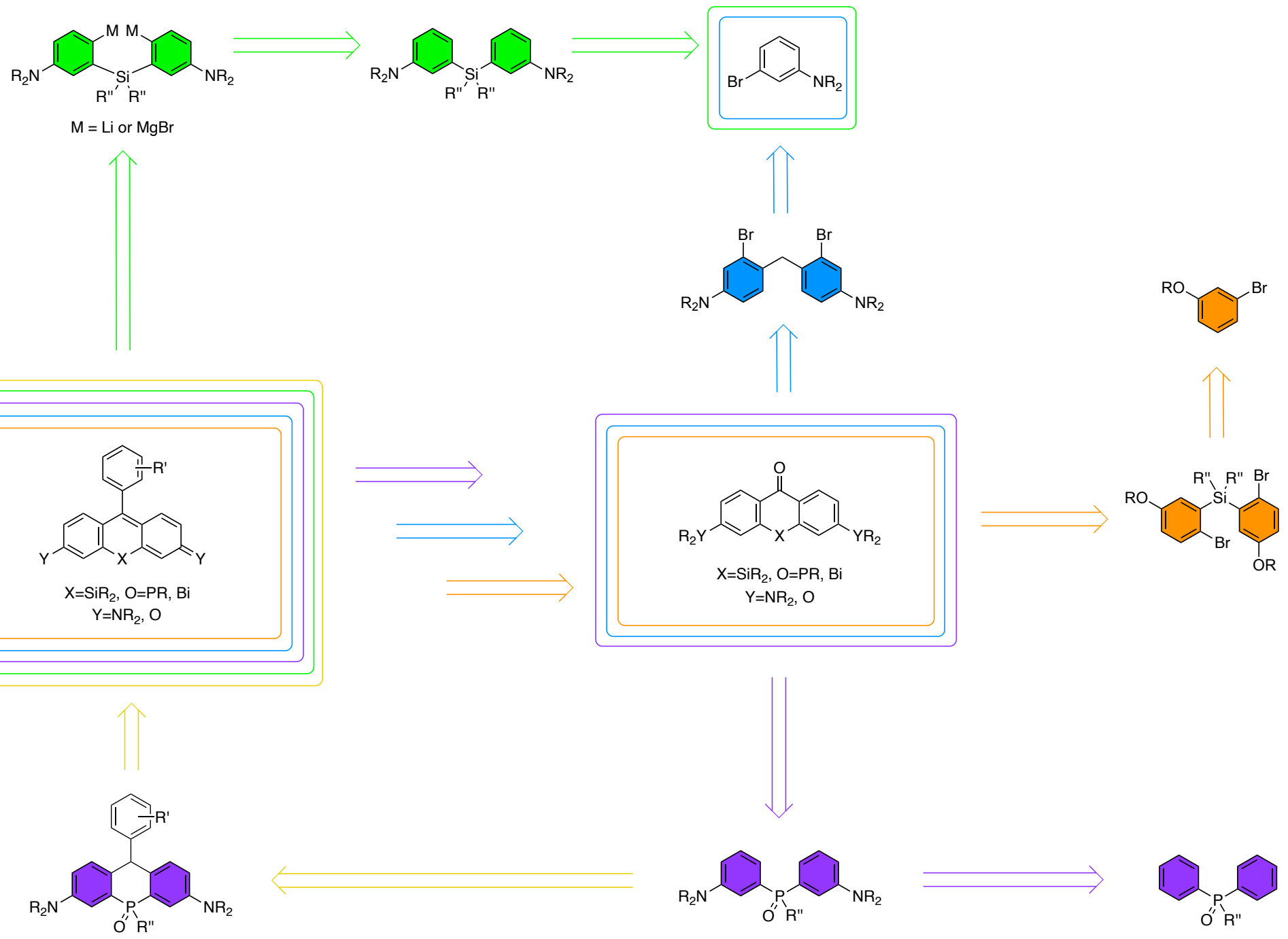

**Scheme 2.** General synthetic route for Si-Fluoresceins, Si-Rhodamines, P-fluoresceins, and P-Rhodamines,

**POF** and other phospha-fluorescein derivatives reported by Wang and coworkers were synthesized from xanthone precursors in which the introduction of bulky aryl groups in the xanthones was achieved by nucleophilic addition of arylmetal reagents, albeit in generally low yields.[35] The second reported synthetic method in literature was based on the condensation of m-bromoanilines with benzaldehydes which is limited to synthesizing phospha-rhodamines.[36] Synthesis of unsymmetrical phospha-fluorescein analogs from either way was nontrivial (Scheme 2, blue pathway). Later in 2017, Yamaguchi's group reported a robust synthetic method to yield phospha-fluorescein dyes.[37] In the first part of the synthesis, the reaction of the two aryl groups onto a phenylphosphine moiety starting from p-dichlorophenylphosphine was achieved via four steps in one pot with a 47% overall yield (Scheme 3).

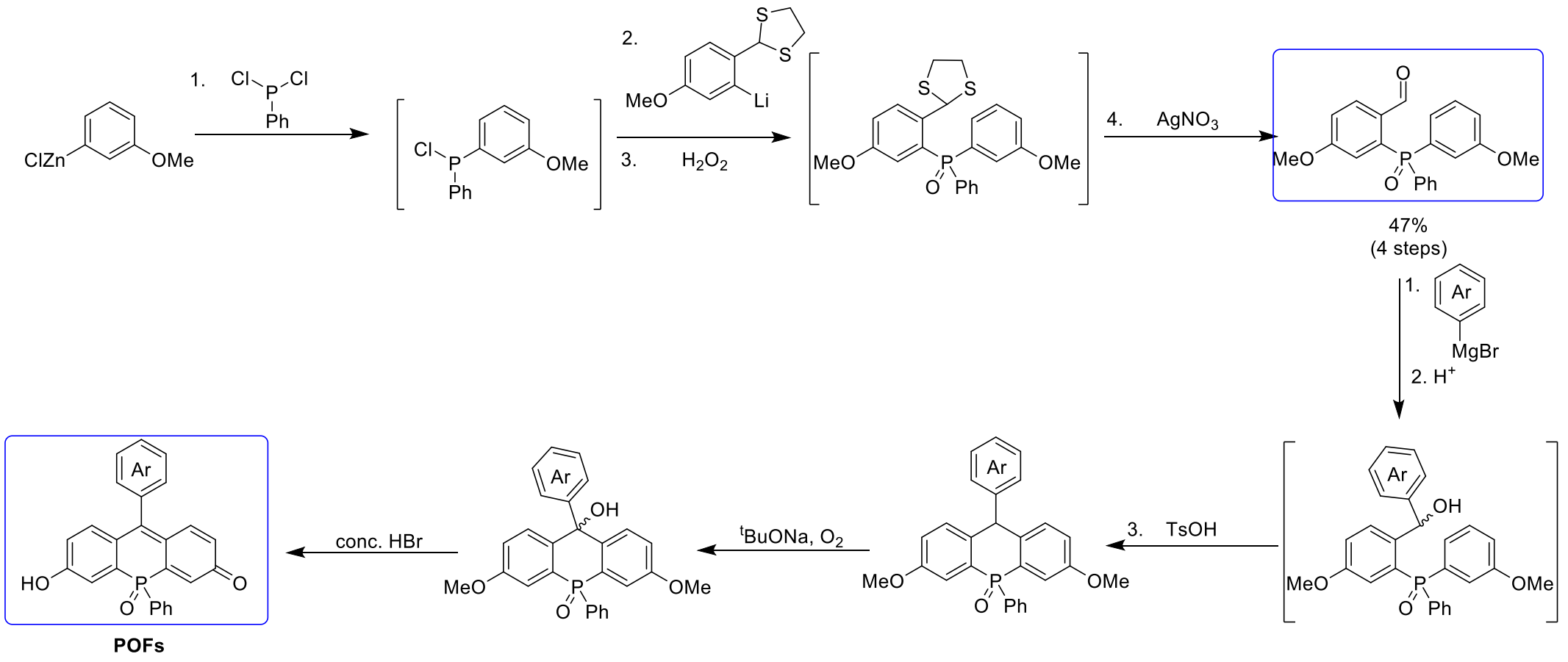

**Scheme 3**. An alternative synthetic route to POF

*Table 2. Selected features of P-Fluorescein derivatives.*

| Entry (E) | Fluorophore* | Absorption | | Emission | $\Phi_f$ | Intracellular Localization | Properties | Applications | Ref. |
|---|---|---|---|---|---|---|---|---|---|
| | | $\lambda_{max}$ (nm) | ε ($M^{-1}\cdot cm^{-1}$) | $\lambda_{max}$ (nm) | | | | | |
| 1 | POF [a] | 627 | 53000 | 656 | 0.33 | - | Photostable, water-soluble, pH range = 3-11 | pH responsive probe | [33] |
| 2 | AcPOF | n.d. 627* | n.d. 53000* | n.d. 656* | n.d. 0.33* | - | cell permeable, activatable | *in vitro* imaging (HeLa, RAW 264.7) | [33] |
| 3 | POF-Me[b] (POF) | 627 | 51300 | 656 | 0.32 | - | water-soluble | - | [37] |
| 4 | POF-$Me_2$[b] | 628 | 54100 | 656 | 0.32 | - | water-soluble | - | [37] |
| 5 | SNAPF-$Me_2$[b] | 645 | 28700 | 744 | 0.03 | - | water-soluble high $pK_a$ value | - | [37] |
| 6 | POF-$Me_2$[c] | 628 | 53900 | 656 | 0.32 | - | water-soluble | - | [38] |
| 7 | POF-$OMe_2$[c] | 632 | 54500 | 661 | 0.24 | - | water-soluble, | - | [38] |
| | | | | | | | low photobleaching resistance | | |
| 8 | **POF-OMe$_2$COOH** | n.d. | n.d. | n.d. | n.d. | - | - | - | [38] |
| 9 | **CAPF-1**$^d$ | 636 | n.d. | 663 | 0.02 $^e$<br>0.22 $^f$ | - | $Ca^{2+}$ ion selective,<br>fluorescence turn on ratio = 13 | - | [39] |
| 10 | **CAPF-1-AM** | n.d.<br>636* | n.d. | n.d.<br>663* | n.d.<br>0.02 $^{e,*}$<br>0.22$^{f,*}$ | - | $Ca^{2+}$ ion-selective,<br>cell permeable | *in vitro* cytosolic calcium imaging (HeLa) | [39] |
| 11 | **NR$_{600}$**$^g$ | 600 | 67570 | 619 | 0.68 | - | photostable | - | [40] |
| 12 | **NR$_{604}$**$^g$ | 604 | 45740 | 627 | 0.48 | - | photostable | - | [40] |
| 13 | **diAM-NR$_{600}$**$^g$ | n.d.<br>600* | n.d.<br>67570* | n.d.<br>619* | n.d.<br>0.68* | - | photostable,<br>cell permeable,<br>activatable | *in vitro* imaging (HeLa, NIH-3T3, RAW 264.7) | [40] |
| 14 | **POF-SO$_3$H**$^h$ | 632 | 33700 | 662 | 0.24 | MT | photostable | - | [41] |
| 15 | **POF-TPP** | n.d. | n.d. | n.d. | n.d. | - | Photostable,<br>cell permeable | *in vitro* imaging (A431) | [41] |
| 16 | **Et$_2$NPOF** | n.d. | n.d. | n.d. | n.d. | - | pH range=4.5-6.5 | - | [41] |
| 17 | **Et$_2$NPOF-CCH** | n.d. | n.d. | n.d. | n.d. | - | off-on-off fluorescence response | - | [41] |
| 18 | **Et$_2$NPOF-dex** | n.d. | n.d. | n.d. | n.d. | lysosome<br>endosome | Photostable | *in vitro* imaging (HeLa)<br>off-on-off type pH probe | [41] |

Photophysical properties were measured in an aqueous pH-buffer solution containing $Na_2HPO_4/NaH_2PO_4$ at pH=9.0. $^b$ Photophysical properties were measured in an aqueous pH-buffer solution containing 0.1 M $Na_2CO_3/NaHCO_3$ at pH=9. $^c$ Photophysical properties were measured in an aqueous solution containing PBS buffer at pH=7.4 containing 1% (v/v) DMSO as co-solvent. $^d$ Photophysical properties were measured in HEPES buffer from a stock solution of fluorophore in DMSO. $^e$ $Ca^{2+}$-free. $^f$ $Ca^{2+}$-bound. $^g$ All experiments were conducted in PBS buffer (10 mM, pH = 7.4 containing 1% DMSO). $^h$ All measured in 50 mM HEPES buffer containing 0.1 DMSO as a co-solvent. $^*$ Values for the activated form. MT: Mitochondria. n.d. : Not detected. *$^*$all the fluorophores listed above were emphasized with bold font in the text.

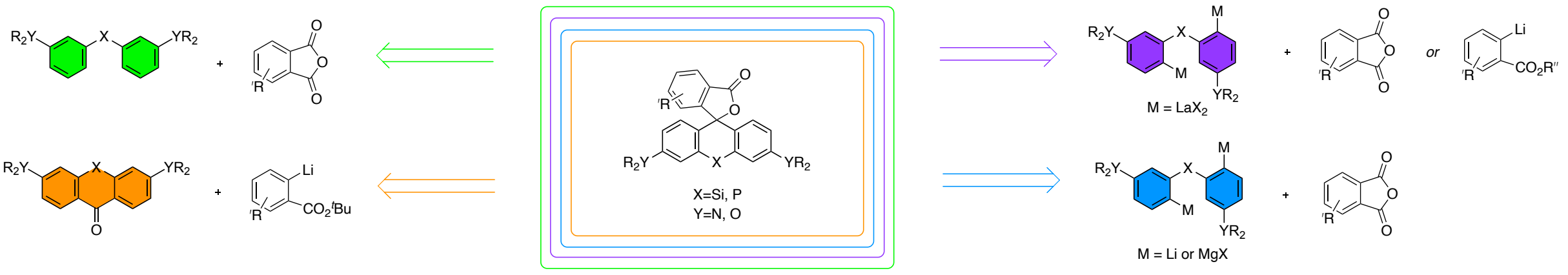

**Scheme 4.** Alternative synthetic route for P-fluoresceins, P-Rhodamines, Si-Fluoresceins, and Si-Rhodamines.

In the second part, the third aryl group was introduced by nucleophilic addition of Grignard reagents, which provided the functionalization at the 9-position of the dye. Without isolation, the diastereomeric mixture was submitted to intramolecular Friedel-Crafts cyclization using p-toluenesulfonic acid as the catalyst. The cyclization process produced both regioisomers and yielded a mixture of cis/trans isomers. Trans products of the targeted regioisomers were isolated with yields between 23% and 38%. The final step included the oxidation of trans products with t-BuONa under an oxygen atmosphere to yield the corresponding benzylic alcohols and treatment with HBr to afford the corresponding phospha fluorescein derivatives. Since the synthetic pathway started with p-dichlorophenylphosphine, all the dye derivatives had phenylphosphine oxide moiety as the bridging group. Phospha fluorescein bearing 2-methylphenyl group at the 9-position (**POF**) was synthesized from the new synthetic approach again and named **POF-Me** (Table 2:E3, Scheme 5). In addition to **POF-Me**, a 2,6-dimethyl group bearing derivative (**POF-Me$_2$**) was synthesized (Table 2:E4, Scheme 5). A seminaphto-phosphafluorescein derivative bearing 2,4-dimethyphenyl group **SNAPF-Me$_2$** (Table 2:E5, Scheme 5) was also synthesized. Compared to the spirolactonized seminaphtofluorescein derivative SNAF-COOH from the literature,[42] **SNAPF-Me** showed a maximum 118 nm shift in absorption and a 12-fold increase in fluorescence quantum yield. This result confirmed the contribution of the facile synthesis of unsymmetrical phospha fluoresceins. Also, in 2017, another synthetic method was realized by Yamaguchi's group.[38] The method was designed to generate phospha rhodols (PORLs) from phospha rhodamines (PORs) by treatment with an aqueous NaOH solution (0.1 M) under ambient temperatures. Conversion of PORLs to their corresponding phospha-fluoresceins (POFs) was slow and required harsher conditions. The bridging moiety for the phospha-rhodamine derivatives in this study was composed of phenylphosphine oxide. Only the phospha-rhodamine derivatives bearing 2,6-dimethylphenyl and 2,6-dimethoxyphenyl groups at the 9-position resulted in direct conversion to their POF analogs, **POF-Me$_2$** (Table 2:E6, Scheme 5) and **POF-OMe$_2$** (Table 2:E7, Scheme 5) respectively. Also, the 2,6-dimethoxy-4-carboxyphenyl bearing phospha-rhodamine derivative yielded its POF analog **POF-OMe$_2$COOH** (Table 2:E8, Scheme 5) as water-soluble sodium salt, albeit in low yield.

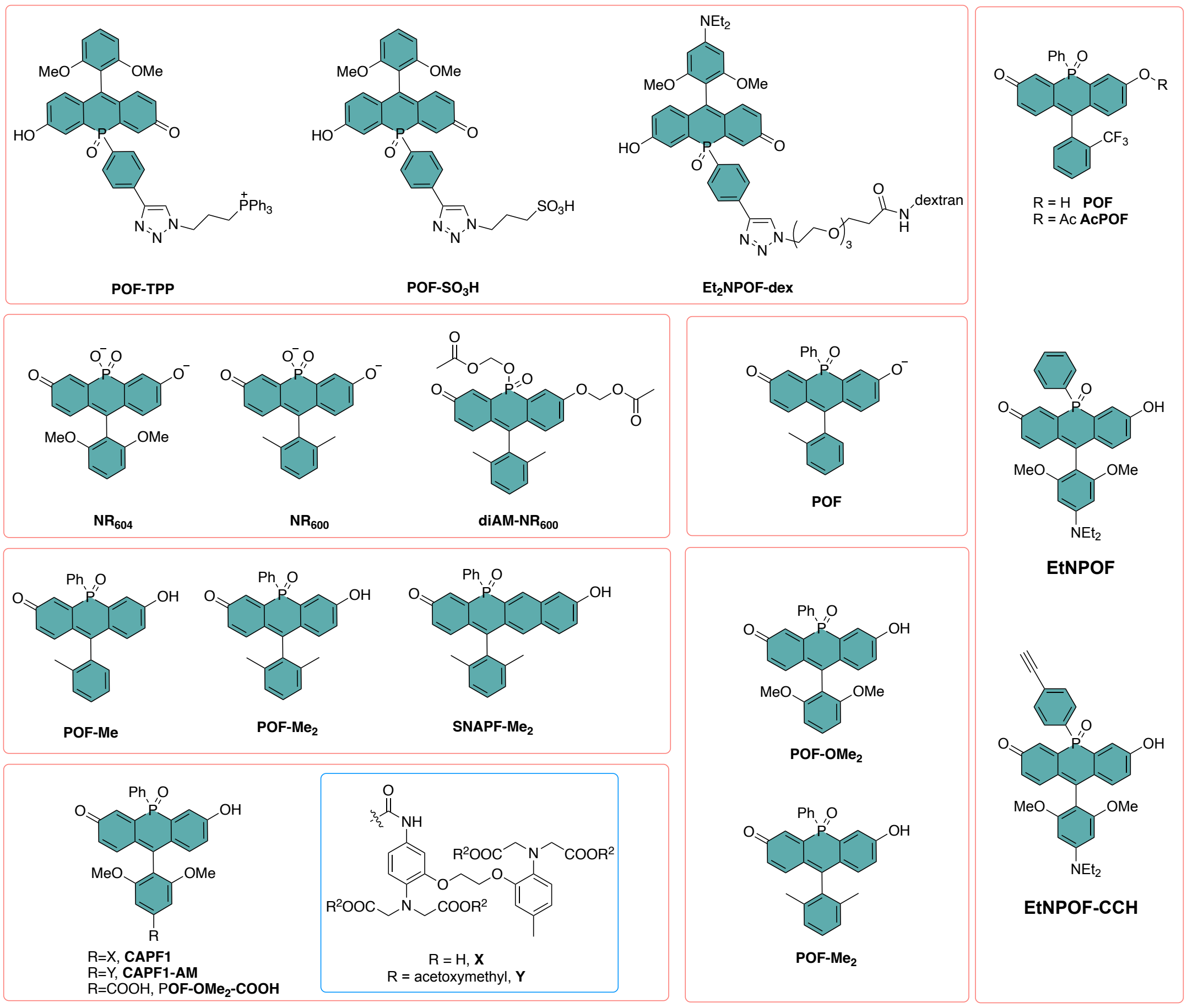

**Scheme 5.** Chemical structures of P-Fluorescein derivatives.

Despite the synthetic efforts on phospha-fluorescein derivatives, their applications as fluorescence probes were yet to be investigated. In 2018 Yamaguchi's group reported the first study on a phospha-fluorescein derivative as a fluorescent probe for calcium imaging.[39] The phospha-fluorescein derivative having phenylphosphine oxide as bridging moiety and bearing 2,6-dimethoxyphenyl at the 9-position was functionalized from the 5′-position with 1,2-bis(*o*-aminophenoxy)ethane *N,N,N',N'*-tetraacetic acid (BAPTA) moiety, which acts as a binding site specific to $Ca^{2+}$ and as an electron donor in the photo-induced electron transfer (PET) system. The probe was called **CaPF-1** (Table 2:E9, Scheme 5). A 13-fold increase of the fluorescence intensity at 663 nm was observed upon complexation with $Ca^{2+}$ ($Ca^{2+}$ free $\Phi_f$ = 0.02, $Ca^{2+}$-bound $\Phi_f$ = 0.22 in HEPES buffer solution). For the applications in mammalian cells, all carboxylic acid moieties of **CaPF-1** were converted into acetoxymethyl (AM) esters yielding **CaPF-1-AM** (Table 2:E10, Scheme 5). The intracellular localization of **CaPF-1** was examined by staining with calcein-AM, which diffuses into the cytosol after hydrolysis by esterase in living cells. HeLa cells were incubated with **CaPF-1-AM**, then fluorescence images were recorded with a confocal microscope which revealed good co-localization with calcein showing that the new fluorescent probe enables the detection of the concentration changes related to cytosolic $Ca^{2+}$ signaling.

In 2019 Stains' group reported the phospha-fluorescein analog of the Nebraska Red dye series and developed a bio-imaging probe.[40] Tetramethyl phospha-rhodamine with phosphinate group as the bridging moiety bearing 2,6-dimethylpheny group and 2,6-dimethoxyphenyl group at the 9-position was treated with 1M NaOH at room temperature for three days yielding their fluorescein analogs **$NR_{600}$** and **$NR_{604}$** (Table 2:E11-12, Scheme 5) respectively. To test the utility of these new NR dyes, a bio probe for imaging enzymatic activity was prepared by AM-protection of **$NR_{600}$** yielding **diAM-$NR_{600}$** probe (Table 2:E13, Scheme 5). *In vitro* analysis in the presence of pig liver esterase (PLE) revealed a 13.8-fold increase in fluorescence. On the contrary, no increase in fluorescence was observed in the presence of PBS alone. Imaging studies after incubation with an esterase probe and HeLa cells using confocal microscopy showed the formation of **$NR_{600}$** in the cytosol with a 10-fold increase in fluorescence signal, which confirmed the success of **diAM-$NR_{600}$** as a no-wash probe for cellular esterase activity.

Lastly, in 2021, Yamaguchi's group reported a series of off-on-off type pH probes, including phospha fluoresceins, rhodamines, and rhodols.[41] These probes were functionalized with 2,6-dimethoxyphenyl at the 9-position and all bearing phenylphosphine oxide as the bridging moiety. For this series of NIR phospha-xanthene dyes, 4-ethynylphenyl group was installed on the phosphorous atom. This newly inserted terminal alkyne could be modified with various azide-containing molecules of interest via copper-catalyzed azide-alkyne cycloaddition (CuAAC). CuAAC reaction of the alkynylated phospha fluorescein derivative with 3-azidopropane-1-sulfonic acid yielded **POF-$SO_3H$** (Table 2:E14, Scheme 5). The absorption, emission characteristics, and fluorescence quantum yield of **POF-$SO_3H$** were almost identical to its unfunctionalized form (**POF**). These showed that the substituents on the phenyl group attached to the phosphorous atom did not perturb the electronic structure. POF bearing the mitochondria targeting group, triphenylphosphonium (TPP), was synthesized and named **POF-TPP** (Table 2:E15, Scheme 5). The staining patterns revealed selective labeling. To achieve an off-on-off type probe, the aryl group at the 9-position was further functionalized with a diethylamine ($Et_2N$) moiety prom the para position (**$Et_2$NPOF-CCH**) (Table 2:E17, Scheme 5). The $Et_2N$ group acted as an electron donor in the PET mechanism, and fluorescence was quenched over a large pH range where the $Et_2N$ group was not protonated. The core OH moiety acted as the other switching site responsive to a pH change. Firstly, a model molecule without the alkyne group, **$Et_2$NPOF** (Table 2:E16, Scheme 5), was synthesized. Both acidic (pH 3.7) and neutral (pH 7.4) solutions showed weak fluorescence upon excitation at 633 nm, confirming the off state of the probe as expected. However, a 14-fold increase in emission intensity was observed at pH 5.5, verifying the on-state. Then, **$Et_2$NPOF-CCH** was synthesized, followed by CuAAC reaction to yield **$Et_2$NPOF-dex** (Table 2:E18, Scheme 5). In the images of HeLa cells, the dots are associated with acidic compartments such as lysosomes and endosomal vesicles, whereas the intense spots represent endosomes with a pH of around 5.5 (figure 3). Overall, this novel design served as a new example for modifying phospha-xanthene dyes and yielded a promising off-on-off type dextran-conjugated probe.

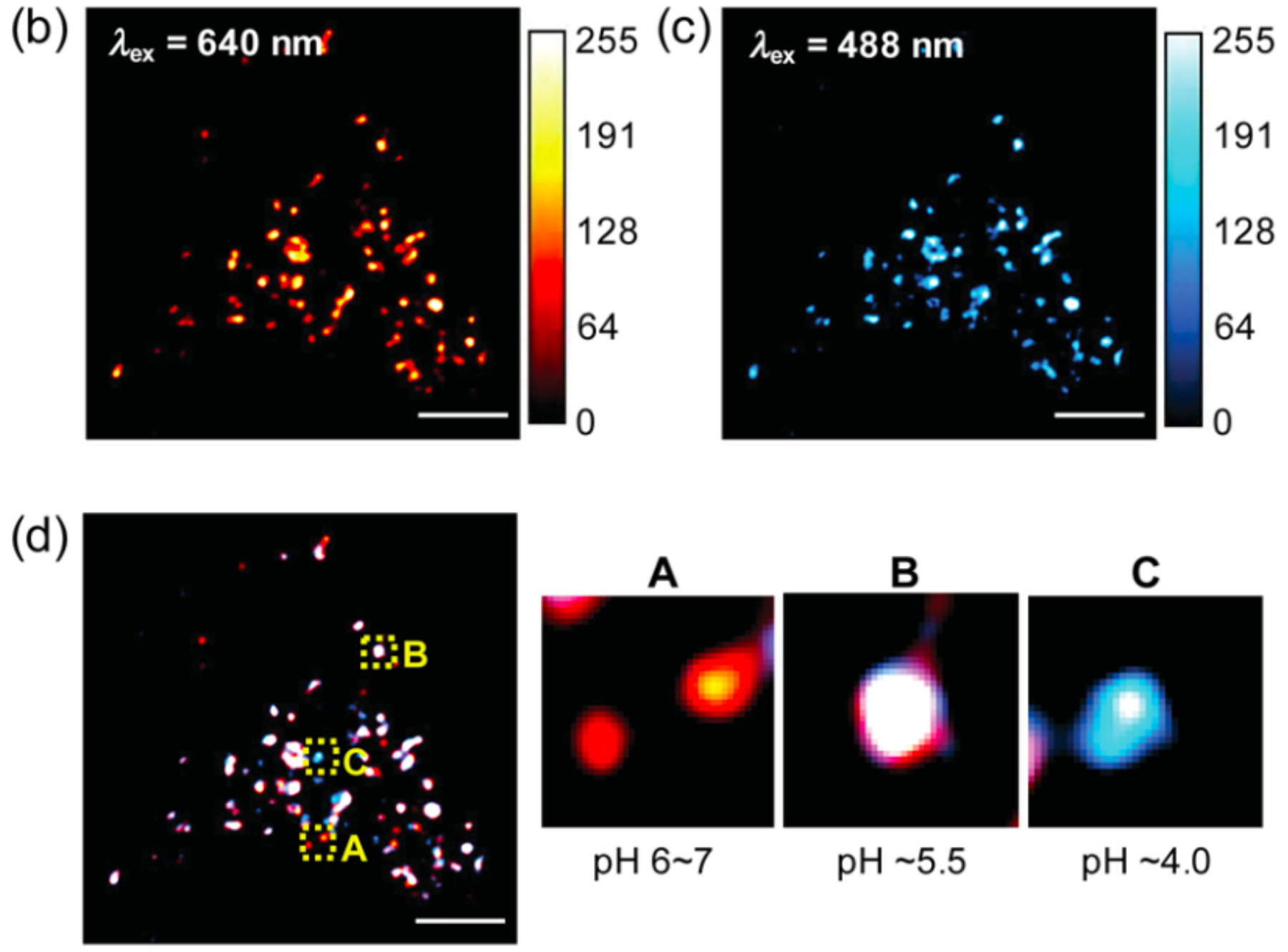

**Figure 3.** (b and c) Endosomal images in HeLa cells incubated with Et2NPOF-dex. Images were acquired at 640 nm ((b), quasi-colored in red) and 488 nm ((c), quasi-colored in cyan) without washing the cells; scale bar = 5 μm. (d) Merged image of the two channels. The areas shown in yellow boxes are magnified on the right. Reproduced with permission [41] copyright 2021, The Royal Society of Chemistry

# 3. Rhodamine Derivatives

## 3.1. Silicon-Rhodamines

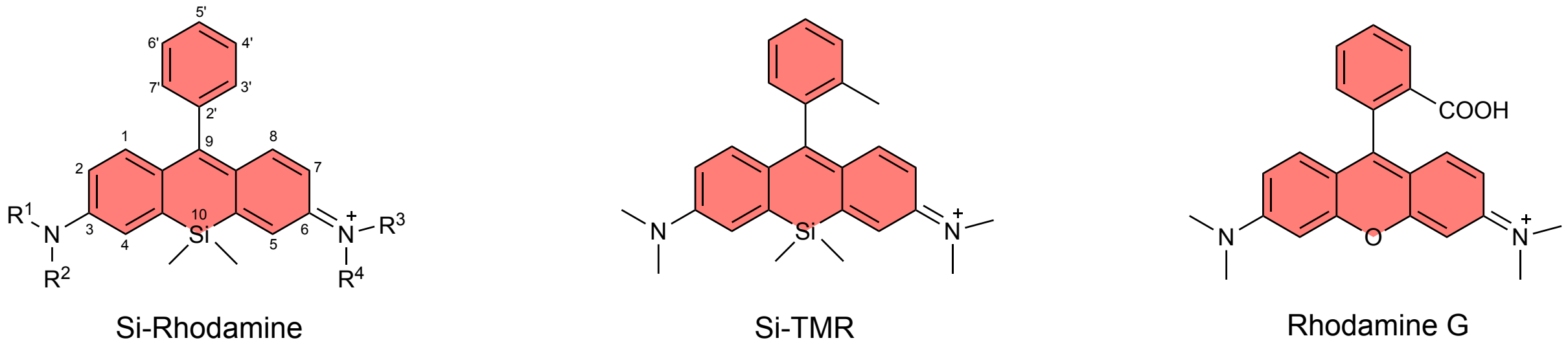

In 2011 Nagano's group introduced the dimethylsilane group to classical rhodamine derivative (Si-rhodamine), which exhibited a large bathochromic shift while preserving advantages of classical rhodamine dyes like high quantum yield in an aqueous medium, stability against photobleaching and high hydrophilicity.[43] Initially, they synthesized a tetramethyl Si-rhodamine bearing 2-methylphenyl group at the 9-position with dimethylsilane as the bridging moiety and studied its photophysical properties. The cyclic voltammetry and Stern-Volmer quenching experiments demonstrated that far-red to NIR fluorescence of SiR could be controlled through PeT. To examine the utility of group 14 rhodamine as a fluorophore for PeT-based probes in detail, ten more SiR derivatives were synthesized with various HOMO energy levels of the aryl moiety at the 9-position by introducing methyl, methoxy, and dimethylamine groups. The experiments showed that the fluorescence quantum yield of SiR could be nearly equal to zero. The threshold level for on/off switching of SiRs was obtained at around 1.3-1.5 V. Overall, the results concluded that the fluorescence of SiR could be activated using a PeT strategy. 2-Me SiR derivative was applied to living HeLa cells yielding a bright fluorescence in the cells with no indication of toxicity. The probe was easily loaded into cells localizing mostly in mitochondria. By functionalization of the 2-Me SiR dye from the 5′-position, a $Zn^{2+}$ sensor (**SiR-Zn**, Table 3:E1, Scheme 6) was generated. As the concentration of $Zn^{+2}$ was increased, **SiR-Zn** emitted greater fluorescence and showed high selectivity towards $Zn^{2+}$. The fluorescence imaging was also demonstrated in HeLa cells yielding fluorescence increase upon addition of $Zn^{2+}$ and fluorescence decrease upon addition of $Zn^{2+}$ chelator *N,N,N′,N′*-tetrakis(2-pyridylmethyl)ethylenediamine (TPEN).

In 2012, Myochin *et al.* introduced four NIR absorbing probes for matrix metalloproteinases (MMPs), and one of the NIR fluorophores was based on the SiR core (**SiR-MMP,** Table 3:E2, Scheme 6).[44] However, after the enzymatic reaction, **SiR-MMP** remained in the extracellular environment, which hindered the visualization of intracellular MMP activity. In 2015, a new series of SiR-based dark FRET quenchers for MMP probes were developed.[45] Various aryl amine groups were introduced to the iodinated SiR core via the Buchwald-Hartwig Coupling method. The fluorescence of both **SiNQ660** (Table 3:E3, Scheme 6) and **SiNQ780** (Table 3:E4, Scheme 6) probes was quenched due to free rotation around amino groups. Next, the water stability of sulfonated **SiNQ780** (**wsSiNQ780,** Table 3:E5, Scheme 6) derivatives was investigated. 2′,6′-dimethoxy analog showed more decent stability in PBS buffer in all pH values. To visualize MMP activity, **wsSiNQ780** coupled with DYD730 analog fluorescent dye, which has a strong fluorescence at Cy7 region (above 780 nm), via enzymatically cleavable peptide sequence. Furthermore, MMP activity was monitored in HT-1080 cells, and fluorescence increase was reported in both the inner and outer surface of the cell due to cleavage of the probe at the outside of the cell and partial transfer of DYD730 into the cell. Also, *in vivo* fluorescence imaging of MMP activity was demonstrated in HT-1080 tumor-bearing mice and resulted in a 14-fold enhancement of fluorescence in the tumor region after 180 min, which confirms the practical usefulness of **SiNQ780** as a NIR dark quencher in the Cy7 region

*Table 3. Selected features of Si-Rhodamine derivatives.*

| Entry (E) | Fluorophore* | Absorption | | Emission | $\Phi_f$ | Intracellular Localization | Properties | Applications | Ref |
|---|---|---|---|---|---|---|---|---|---|
| | | $\lambda_{max}$ (nm) | ε ($M^{-1}cm^{-1}$) | $\lambda_{max}$ (nm) | | | | | |
| 1 | SiR-Zn[a] | 650/ 651** | 98000/ 110000** | 665 | 0.009/ 0.12** | n.d. | Photostable, 15 fold $\Phi_f$ enhancement** | $Zn^{2+}$ probe | [43] |
| 2 | SiR-MMP[b] | 650 | n.d. | 680 | 0.001/ 0.169** | n.d. | Cell-impermeable | MMP probe | [44] |
| 3 | SiNQ660[c] | 660 | n.d. | n.d. | <0.001[d] | n.d. | n.d. | Dark quencher | [45] |
| 4 | SiNQ780[c] | 679 | n.d. | n.d. | <0.001[d] | tumor cell | 14 fold $\Phi_f$ enhancement** | Dark quencher | [45] |
| 5 | wsSiNQ780[d] | 680/ 780** | n.d. | 720/ 780* | n.d. | n.d. | Water soluble, Photostable, 20 fold $\Phi_f$ enhancement | MMP probe | [45] |
| 6 | SiR650 (SiTMR)[e] | 646 | 110000 | 660 | 0.31 | n.d. | - | - | [46] |
| 7 | SiR680[e] | 674 | 130000 | 689 | 0.35 | n.d. | photostable | - | [46] |
| 8 | **SiR700**[e] | 691 | 100000 | 712 | 0.12 | n.d. | photostable | - | [46] |
| 9 | **SiR720**[e] | 721 | 160000 | 740 | 0.05 | n.d. | - | - | [46] |
| 10 | **SiR700-RCB-1**[e] | 691 | 100000 | 712 | 0.12 | Malignant gliomas | Photostable | Tanascin-C imaging | [46] |
| 11 | **SiR-carboxyl**[e] | 645 | 100000 | 660 | 0.39 | n.d. | Solvent polarity sensitive | - | [47] |
| 12 | **SiR-SNAP**[e] | 650* | n.d. | 668* | 0.30** | Microtubules and centrosomes | 7.5 fold $\Phi_f$ enhancement** | Cellular protein imaging | [47] |
| 13 | **SiR-CLIP**[e] | 652* | n.d. | 668* | 0.46** | n.d. | 6 fold $\Phi_f$ enhancement** | Cellular protein imaging | [47] |
| 14 | **SiR-Halo**[e] | 648* | n.d. | 668* | 0.39** | n.d. | 6 fold $\Phi_f$ enhancement** | Cellular protein imaging | [47] |
| 15 | **HMSiR**[d] | 650 | 100000 | 671 | 0.39 | n.d. | pH-dependent off/on mechanism | SLM imaging of proteins | [48] |
| 16 | **AMSiR**[d] | 656 | n.d. | ~675 | n.d. | n.d. | pH-dependent off/on mechanism | - | [48] |
| 17 | **MMSiR**[d] | 653 | n.d. | ~675 | n.d. | n.d. | pH independent | - | [48] |
| 18 | **SiTMR-HaloTag**[f] | 643** | 140000[h] | 662* | 0.41** | n.d. | 6.8 fold $\Phi_f$ enhancement** | Imaging Halo Tag protein interaction | [49] |
| 19 | **$JF_{646}$-HaloTag**[f] | 646** | 152000[h] | 664* | 0.54** | n.d. | 21 fold $\Phi_f$ enhancement** | Imaging Halo Tag protein interaction | [49] |
| 20 | **Si-DMA**[c] | ~640 | n.d. | ~660 | 0.01/ 0.17** | MT | 18 fold $\Phi_f$ enhancement** | $^1O_2$ imaging in PDT | [50] |
| 21 | **$SiRh_Q$**[g] | 637 | 77000 | 654 | 0.38 | n.d. | | | [51] |
| 22 | **SiR-palloidin**[g] | 640** | n.d. | 655* | n.d. | MT | High photon count, 1st photoactivatable SiR | Imaging F-actin | [51] |
| 23 | **SiRB**[h] | 652 | 160000 | 670 | 0.17 | n.d. | - | - | [52] |
| 24 | **SiRB-3**[h] | 670 | 120000 | 689 | 0.08 | n.d. | - | - | [52] |
| 25 | **SiRB-2**[h] | 693 | 110000 | 713 | 0.023 | n.d. | - | - | [52] |
| 26 | **Br-SiRB**[h] | 658 | 140000 | 674 | 0.14 | n.d. | - | - | [52] |
| 27 | **$NO_2$-SiRB**[h] | 664 | 160000 | 680 | 0.09 | n.d. | - | - | [52] |
| 28 | **CN-SiRB**[h] | 661 | 140000 | 677 | 0.12 | n.d. | - | - | [52] |
| 29 | **COOH-SiRB**[h] | 656 | 92000 | 671 | 0.16 | n.d. | - | - | [52] |
| 30 | **HC≡C-SiRB**[h] | 656 | 140000 | 672 | 0.16 | n.d. | - | - | [52] |
| 31 | **Mito-SiRB*** ** | n.d. | n.d. | n.d. | n.d. | MT | - | New synthetic approach | [52] |
| 32 | **SiRB-NO**[i] | 667** | n.d. | 680** | n.d. | n.d. | 800 fold $\Phi_f$ enhancement** | NO detection | [53] |
| 33 | **Lyso-SiRB-NO**[i] | n.d. | n.d. | 680** | n.d. | lysosome | 1000 fold $\Phi_f$ enhancement** | Lysosomal NO detection | [53] |
| **34** | **SiRB-Cu**[j] | 664** | n.d. | 680** | n.d. | lysosome | 150 fold $\Phi_f$ enhancement** | Lysosomal $Cu^{2+}$ detection | [54] |
| **35** | **Vinblastine-SiR-S**[d] | 640 | n.d. | ~670 | 0.40 | n.d. | Cell permeable | - | [55] |
| **36** | **Vinblastine-SiR-L**[d] | 640 | n.d. | ~670 | 0.40 | n.d. | Cell permeable | Imaging microtubule reorganization | [55] |
| **37** | **FAP-1**[k] | 645** | 190000 | 662** | 0.36** | Non-nuclear | 45 fold $\Phi_f$ enhancement** | Formaldehyde probe | [56] |
| **38** | **$SiOH_2R$**[f] | 663 | 105000 | 681 | 0.43 | MT | 5000 fold exo, 4.26 endo $\Phi_f$ enhancement* | $H_2O_2$ probe | [57] |
| **39** | **$JF_{635}$-HaloTag** | 635 | 167000 | 652 | 0.56 | n.d. | 113 fold $\Phi_f$ enhancement | Fine tuning for live cell and *in vivo* imaging | [58] |
| **40** | **SiRD**[l] | 690 | n.d. | 710 | 0.31 | MT | Both -endo and exogenous mitochondrial NO detection | NO probe | [59] |
| **41** | **PSiR**[m] | 620** | n.d. | 680** | n.d. | lysosome | photostable | ROS-induced LCD in tumor imaging | [60] |
| **42** | **Ph-SiR650**[n] | 645 | n.d. | 668 | 0.14 | n.d. | - | - | [61] |
| **43** | **2′Me SiR600**[n] | 593 | n.d. | 613 | 0.38 | n.d. | - | - | [61] |
| **44** | **2′Me SiR610**[n] | 604 | n.d. | 627 | 0.31 | n.d. | - | - | [61] |
| **45** | **QG0.6**[o] | 595 | n.d. | 617 | 0.28 | n.d. | 30 fold $\Phi_f$ enhancement** | - | [61] |
| **46** | **QG3.0**[o] | 609 | n.d. | 632 | 0.27 | n.d. | 30 fold $\Phi_f$ enhancement** | GSH imaging | [61] |
| **47** | **SiNH**[f] | 470** | n.d. | 595** | 0.32** | n.d. | Water soluble,<br>Introduced with micelle | endogenous peroxynitrite detection | [62] |
| **48** | **EP-2OMe-SiR600**[k] | 500/ 595** | n.d. | n.d./ 612** | 0.002/ 0.32** | n.d. | 138 fold $\Phi_f$ enhancement** | probe for DPP-IV | [63] |
| **49** | **SiR-Mito-8**[f] | 651 | n.d. | 668 | 0.33 | MT | 5.9 fold higher in cancer cells | Selective cancer cell imaging | [64] |
| **50** | **SiR-Mito-11**[f]** | 654 | n.d. | 665 | 0.40 | MT | ***Cell, Type, $IC_{50}$ (mM):**<br>HT22, Neuron, 9.1<br>H4, Neuroglioma, 2.6<br>A172, Glioblastoma, 0.4<br>N2a, Neuroblastoma, 5.8<br>C6, Astrocytoma, 3.2<br>SK-N-MC, Neuroblastoma, 0.3<br>U87-MG, Glioblastoma, 1.1 | Cancer diagnosis and therapeutic | [65] |
| **51** | **Py-SiRh**[l] | 655 | n.d. | 680 | 0.12 | lysosome | 70 fold $\Phi_f$ enhancement** | Lysosomal cell death imaging | [66] |
| **52** | **Folate-SiR-1**[p] | 652 | n.d. | 674 | 0.076 | n.d. | 83 fold $\Phi_f$ enhancement** | FR-expressing tumor imaging | [67] |
| **53** | **Folate-SiR-2**[p] | 656 | n.d. | 676 | 0.051 | n.d. | - | - | [67] |
| **54** | **$JF_{669}$**[r] | 669 | 112000 | 682 | 0.37 | n.d. | $K_{L-Z}$:0.262 | - | [68] |
| **55** | **CaSiR-1 AM**[o] | 650** | n.d. | 664** | 0.20 | lysosome | Cell permeable | $Ca^{2+}$ detection | [69] |
| **56** | **CaSiR-2 AM**[s] | 637** | n.d. | 664** | 0.01 to 0.26 | cytosol | Cell permeable | $Ca^{2+}$ detection | [69] |
| **57** | **KMG-501**[t] | 663** | 83900 | 684** | 0.004/ 0.05** | cytoplasm | - | $Mg^{2+}$ detection | [70] |
| **58** | **KMG-502**[t] | 670** | 105000 | 690** | 0.003/ 0.13** | cytoplasm | - | $Mg^{2+}$ detection | [70] |
| **59** | **Py-SiRh-HP**[f] | 655** | n.d. | 680** | n.d. | lysosome | Cancer cell and tumor-selective | ROS detection | [71] |
| **60** | **ATP-pH**[u] | 700** | n.d. | 740** | n.d. | tumor cell | Dual-stimulus responsive toward ATP and pH | fluorescent and photoacoustic tumor imaging | [72] |

Photophysical properties were measured in [a]100 mM HEPES buffer containing 100 mM $NaNO_3$ and 10 mM NTA at pH 7.4. [b]in TCN buffer (pH 7.4), referring to that of cresyl violet (Φf = 0.54) in methanol as a standard. [c]in MeOH. [d]in PBS buffer containing 0.1% DMSO [e]in TBS buffer containing 0.1% SDS [f]in aqueous solution [g]in 10 mM HEPES at pH 7.3 [h]in EtOH solution containing 0.1 mM HCl. [i]in 0.1 M PBS containing 20% $CH_3CN$ at pH 7.4 [j]in 20 nM HEPES containing 20% $CH_3CN$ [k]Cy5.5 in PBS at pH 7.4 (Φf = 0.23) or ICG in DMSO (Φf = 0.13) as a fluorescence standard. [l]in 50 mM HEPES containing 30% $CH_3CN$ at pH 7.4 [m]in 20 mM PBS containing 1% $CH_3CN$ at pH 7.4 [n]in 0.2 M sodium phosphate buffer, pH 7.4, containing 1% DMSO [o]in 0.2 M sodium phosphate buffer, pH 7.4, containing 5% DMSO [p]in 100 mm sodium phosphate buffer at pH 7.4 [r]in 10 mM HEPES, pH 7.3 [o]in 30 mM MOPS buffer containing 100 mM KCl and 10 mM ethylene glycol tetraacetic acid (EGTA) (pH 7.2) [s]in HBSS (Hank's balanced salt solution) containing 0.03% Pluronic and 0.45% DMSO [t]in 100 mM HEPES buffer (pH 7.2) with 130 mM KCl and 10 mM NaCl [u] in HEPES buffer (25 mM HEPES/EtOH = 1:1, v/v.) at pH:6.0

*all the fluorophores listed above are emphasized with bold font in the text.

** Determined from the activated form

***Table of $IC_{50}$ for SiR-Mito11 against various cell types

**** derived from entry 29

$\Phi_f$ = fluorescence quantum yield, ε = molar extinction coefficient, n.d. = not detected, MT = mitochondria

Koide *et al*. introduced a series of novel silicon-containing rhodamine derivatives covering a wide range of the NIR-I region.[46] The main strategy in this work was to induce red-shift in both absorption and emission maxima by manipulating the groups on the amine moieties. The first dye of the series was the **SiR650** (Table 3:E6, Scheme 3) which bears dimethyl amine groups and has an absorption maximum at 646 nm and an emission maximum at 660 nm in PBS buffer at pH 7.4 ($\varphi_F$=0.31). The second dye, **SiR680** (Table 3:E7, Scheme 6), utilized fused 6-membered methylcycloamines, resulting in bathochromic shift and the absorption and emission maxima at 674 nm and 689 nm, respectively ($\varphi_F$ = 0.34). **SiR700** (with fused 5-membered methylcycloamine) showed decreased $\varphi_F$ of 0.12, whereas **SiR720** gave a significantly lower $\varphi_F$ of 0.05 (for photophysical properties, see Table 3:E6-9). **SiR680** and **SiR700** (Table 3:E8, Scheme 6) has been chosen for further studies according to their high emission maximum and reasonable quantum yield. An amine-reactive unit containing **SiR700** (**2-Me-4-COOSu SiR700**) was prepared and modified with a corresponding antibody for targeting the extracellular matrix glycoprotein Tanascin-C (TN-C). Both free **SiR700** and labeled **SiR700** (**SiR700-RCB1,** Table 3:E10, Scheme 6) showed extraordinary photostability. In vivo imaging of xenograft tumor model mice (prepared with human malignant meningioma HKBMM cells) was successfully achieved using **SiR700-RCB1** (figure 4).

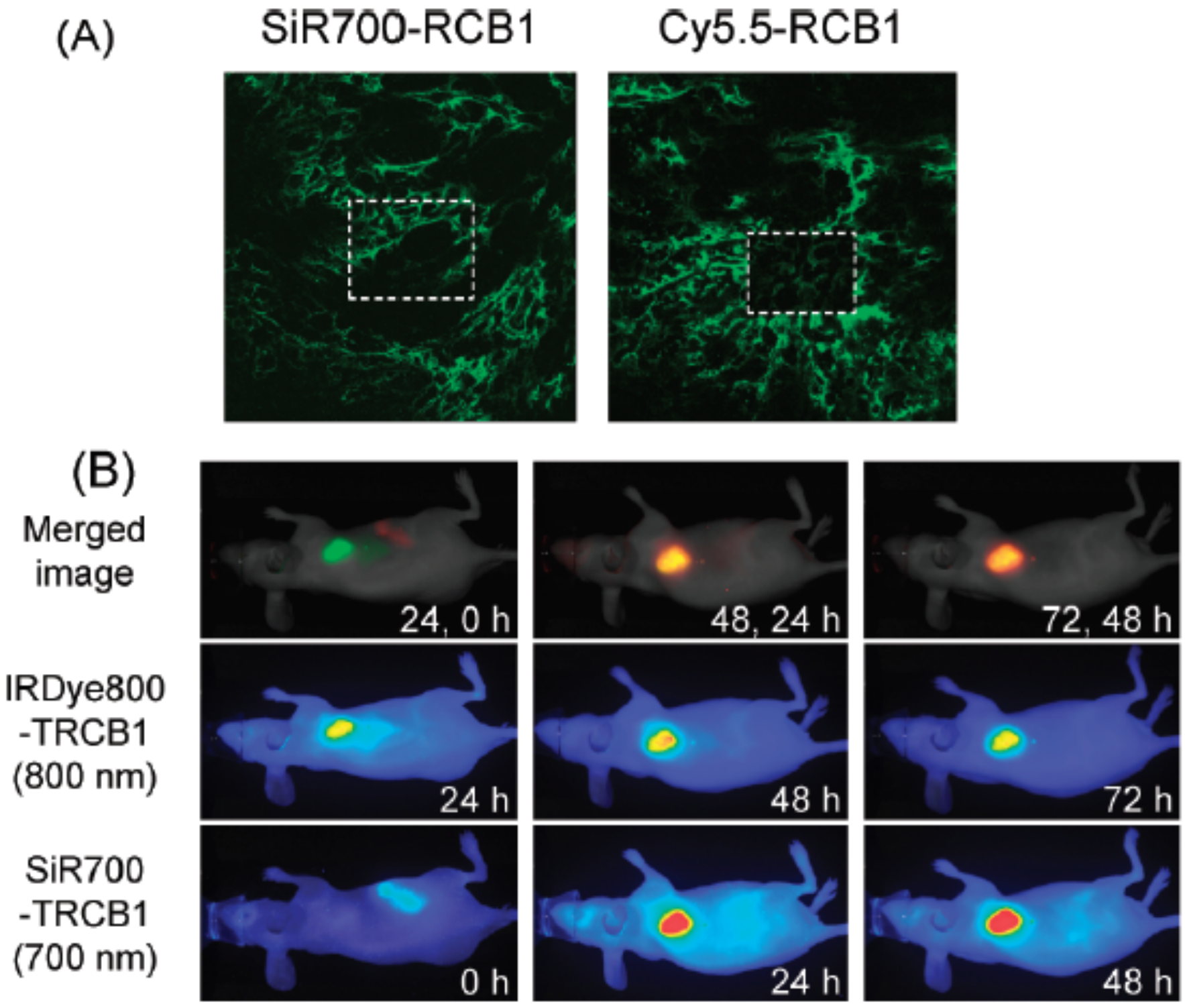

**Figure 4.** (A) Fluorescence images of frozen sections of tumors harvested three days after intravenous injection of SiR700- or Cy5.5-labeled anti-tenascin-C antibody (SiR700- or Cy5.5-RCB1). Dotted squares were irradiated at 670 nm for 5 min. Fluoromount/Plus (COSMO BIO) was used as a mounting medium for preparing the slide samples. (B) In vivo imaging using a Pearl Impulse Imager with a 700 nm channel and an 800 nm channel. SiR700-RCB1 was iv injected into a mouse xenograft tumor model prepared with human malignant meningioma HKBMM cells at 24 h after the iv injection of IRDye800- RCB1. The tumor was at the base of the right forefoot. SiR700-RCB1 (bottom row) and IRDye800-RCB1 (middle row) were separately monitored in two wavelength windows, and merged images (800 nm: green, 700 nm: red) are also shown (top row). Wavelength maxima for excitation/emission filters were 685/720 nm (700 nm channel) for SiR700 and 785/820 nm (800 nm channel) for IRDye800. Fluorescence images were acquired at 24 h (left column), 48 h (middle column), and 72 h (right column) after injection of IRDye800- RCB1 and at 0 h (left column), 24 h (middle column,) and 48 h (right column) after injection of SiR700-RCB1 into the mouse xenograft tumor model. Reproduced with permission [46] Copyright 2012, American Chemical Society

Following Wang's work in 2012,[73] Lukinavičius *et al*. introduced the dicarboxylic acid derivative of Si-rhodamine (**SiR-carboxyl**, Table 3:E11, Scheme 7), which can be coupled specifically to proteins by utilizing different techniques.[47] Spiro-lactonization ability facilitates cellular uptake, and the second carboxylic acid provides a handle for modifying the probe for specific proteins. **SiR-carboxyl** was converted to protein-targeting dyes with well-known protein tags to yield **SiR-SNAP**, **SiR-CLIP,** and **SiR-HALO** (Table 3:E12-14, Scheme 7). These derivatives had comparable $\varphi_F$ with **SiR-carboxyl** with high photostability and membrane permeability. The delicate, environmentally sensitive zwitterion-lactone equilibrium was utilized for protein labeling with negligible background staining for different live cells.

Uno *et al*. presented spirocyclized Si rhodamine derivatives for live cell super-resolution imaging.[48] Three derivatives with pH-dependent cyclization utilizing pendant group ($-CH_2OH$, $-CH_2NH_2$, $-CH_2SH$) on the ortho position of the aromatic ring at the 9 position named as **HMSiR**, **AMSiR**, **MMSiR** (Table 3:E15-17, Scheme 7) respectively. **HMSiR** and **AMSiR** have cyclization pKs ($pK_{cycl}$) at 5.8 and 4.7, respectively, whereas **MMSiR** has $pK_{cycl}$ lower than 2 due to the strong nucleophilicity of the sulfur. The predominantly longer lifetime of **HMSiR**, high $\varphi_F$ (0.39 in aqueous solution), and red-shifted absorption and emission maxima at 650 nm and 671 nm, respectively, made it the most suitable candidate for single molecule localization microscopy (SLM).

Grimm *et al.* introduced several improved fluorophores for live cell and single molecule microscopy, where two derivatives belong to the Si-Rhodamine family.[49] One was the formerly synthesized tetramethyl Si-rhodamine (**SiTMR**, Table 3:E6, Scheme 7), while the other was a novel azetidine containing Si-Rhodamine ( $JF_{646}$ ).  $JF_{646}$ showed similar absorption and emission maxima with its tetramethyl analog (**$JF_{646}$-HaloTag**, **SiTMR-HaloTag,** Table 3:E18-19, Scheme 7, respectively) exhibited higher $\varphi_F$ (0.54 v.s 0.41) in aqueous solution. For cellular imaging with **$JF_{646}$**, Halo Tag ligand was incorporated, and upon treatment with the target protein, a 21-fold increase in absorption was observed as opposed to a 6.8-fold increase in **SiTMR**. Moreover, **$JF_{64}$** showed better localization in super-resolution microscopy and lower background in conventional microscopy compared to its tetramethyl analog.

Kim *et al*. designed and synthesized a 9,10-dimethlyanthracene modified Si rhodamine derivative (**Si-DMA,** Table 3:E20, Scheme 6) to detect singlet oxygen during photodynamic therapy.[50] In methanol, **Si-DMA** showed an 18-fold increase in $\varphi_F$ upon reaction with singlet oxygen. In PBS buffer, the probe forms H-aggregates evident from the blue-shifted absorption peak at 625 nm, which is common for rhodamine chromophores. Upon treatment with $^1O_2$, an even higher increment in fluorescence was observed, which was attributed to additional fluorescence quenching upon H-aggregate before photoirradiation. The probe was also shown to be selective to $^1O_2$ and localized at mitochondria. Although **Si-DMA** was the first probe that was obtained a clear image of singlet oxygen generation during PDT at the subcellular level, it was limited due to dim fluorescence before activation and low hydrophilicity.

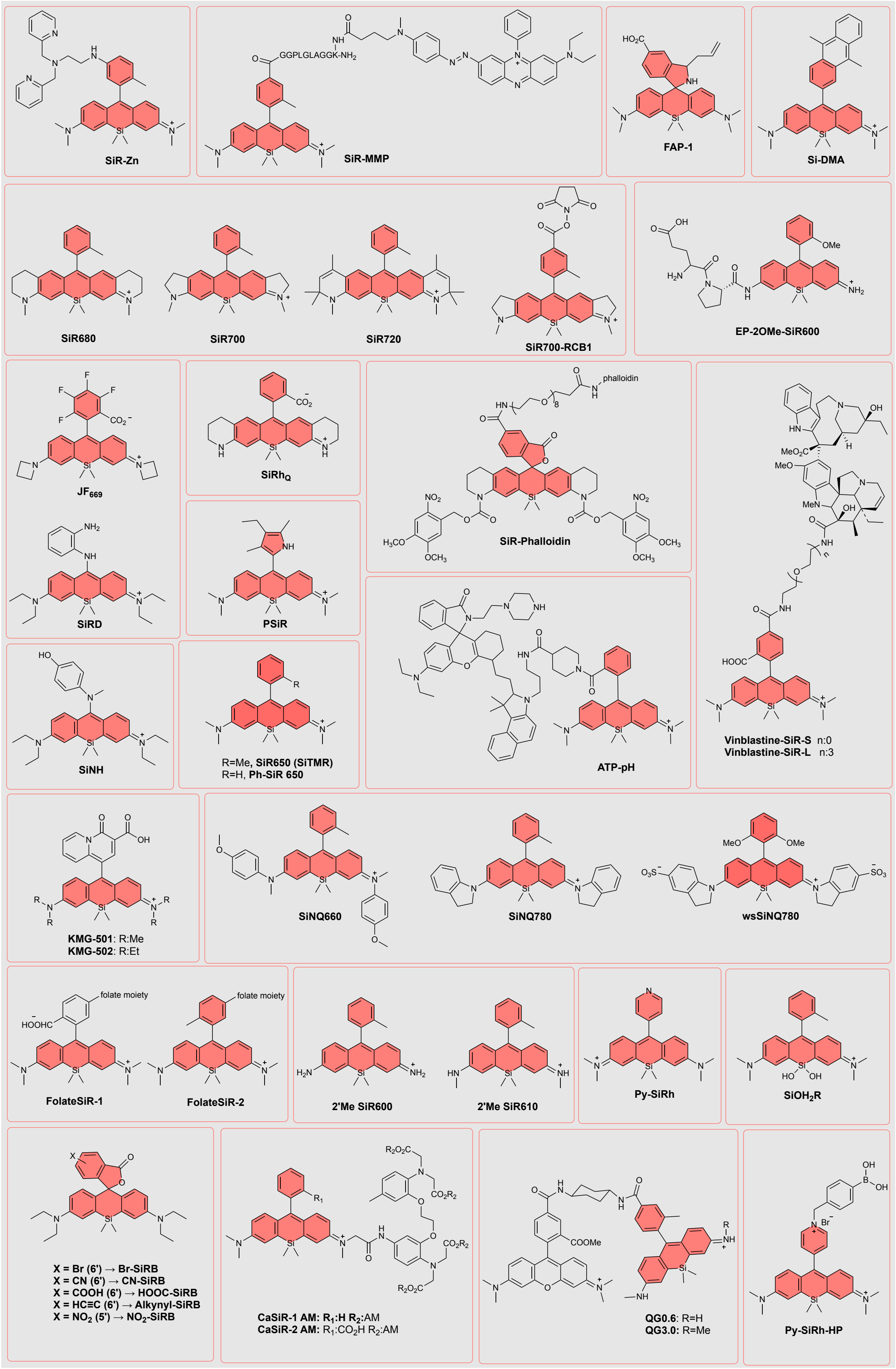

**Scheme 6.** Chemical structures of Si-Rhodamine derivatives.

Meimetis *et al*. generated a fluorescent vinblastine probe utilized from the SiR core in 2013. The tetramethyl Si-rhodamine bearing 2-carboxyphenyl group at the 9-position was functionalized with desacetyl vinblastine hydrazide using linkers from the 4'-position.[55] The length of the linkers between the drug and fluorophore was varied to evaluate the binding affinity and cellular distribution characteristics. Hence, two SiR vinblastine probes were prepared using a short (**Vinblastine-SiR-S,** Table 3:E35, Scheme 6) and a long (**Vinblastine-SiR-L,** Table 3:E36, Scheme 6) linker. The half maximal inhibitory concentrations of the probes were measured in the ovarian cancer cell line OVCA429 yielding 140.7 and 180.7 nM for **Vinblastine-SiR-S** and **Vinblastine-SiR-L,** respectively. Next, the inhibitory effect of the probes in polymerization was determined using a tubulin polymerization assay. The formation of microtubule paracrystals was observed at high concentrations. This observation was realized to act as a pathway for imaging phenotype specific to vinblastine. After washing the incubated probes in OVCA429 cells transfected with CellLight® tubulin-RFP, imaging showed that the long linker derivative provided good density and paracrystal length. However, no paracrystal formation was observed for the **Vinblastine-SiR-S**. The vinblastine SiR probes provided the only live cell imaging and cellular distribution of the mitotic inhibitor vinblastine.

In 2014, Wang *et al.* developed an alternative pathway for spirolactonized Si-rhodamines (**SiRB**).[52] Unlike the general synthetic route, the first silicon bridge was settled between either identical monomers or different monomers to yield symmetric and asymmetric diaryl silly ether (DASE). The condensation reaction of DASE with excess 2-formyl benzoic acid derivatives at high temperatures resulted in **SiRB** fluorophores. With this approach, several derivatives of **SiRB** were synthesized, and their photophysical properties were investigated (Table 3:E23-30 Scheme 6). Also, **Mito-SiRB** was synthesized from **HC≡C-SiRB** and colocalization assay with commercially available mitotracker rhodamine 123 in bEnd3 and HepG2 cells proven the mitochondrial localization of dye in both tumor and healthy cells (Table 3:E31, Scheme 7). This approach provided a more straightforward synthesis of **SiRB** dyes and enabled the synthesis of asymmetrical **SiRB** dyes.

In 2016, Grimm *et al.* developed another Si-rhodamine derivative that bears fused tetrahydroquinolines on 3 and 6 positions and dicarboxylic acid containing aromatic moiety on the 9 position of the xanthene core.[51] In this study, the first caged probe application of a Si rhodamine derivative was realized utilizing the nitroveratryl oxycarbonyl (NVOC) group to increase photon count and nuclear localization in super-resolution microscopy applications. Also, phalloidin conjugate of the caged probe (**SiR-palloidin,** Table 3:E22, Scheme 6) was prepared for imaging cellular F actin on COS7 cells and its activity compared with commercially available **SiR-palloidin** which is widely used in localization microscopy. In PBS buffer, **SiR-palloidin** had shown better photon count per localization event but with lower theoretical localization precision due to higher fluorescence background. The introduction of an anti-bleaching agent, Trolox, resulted in a significant performance increase in oxygen-deprived buffer and the cellular actin network was visualized with considerably enhanced localization precision due to markedly improved photon counts (figure 5).

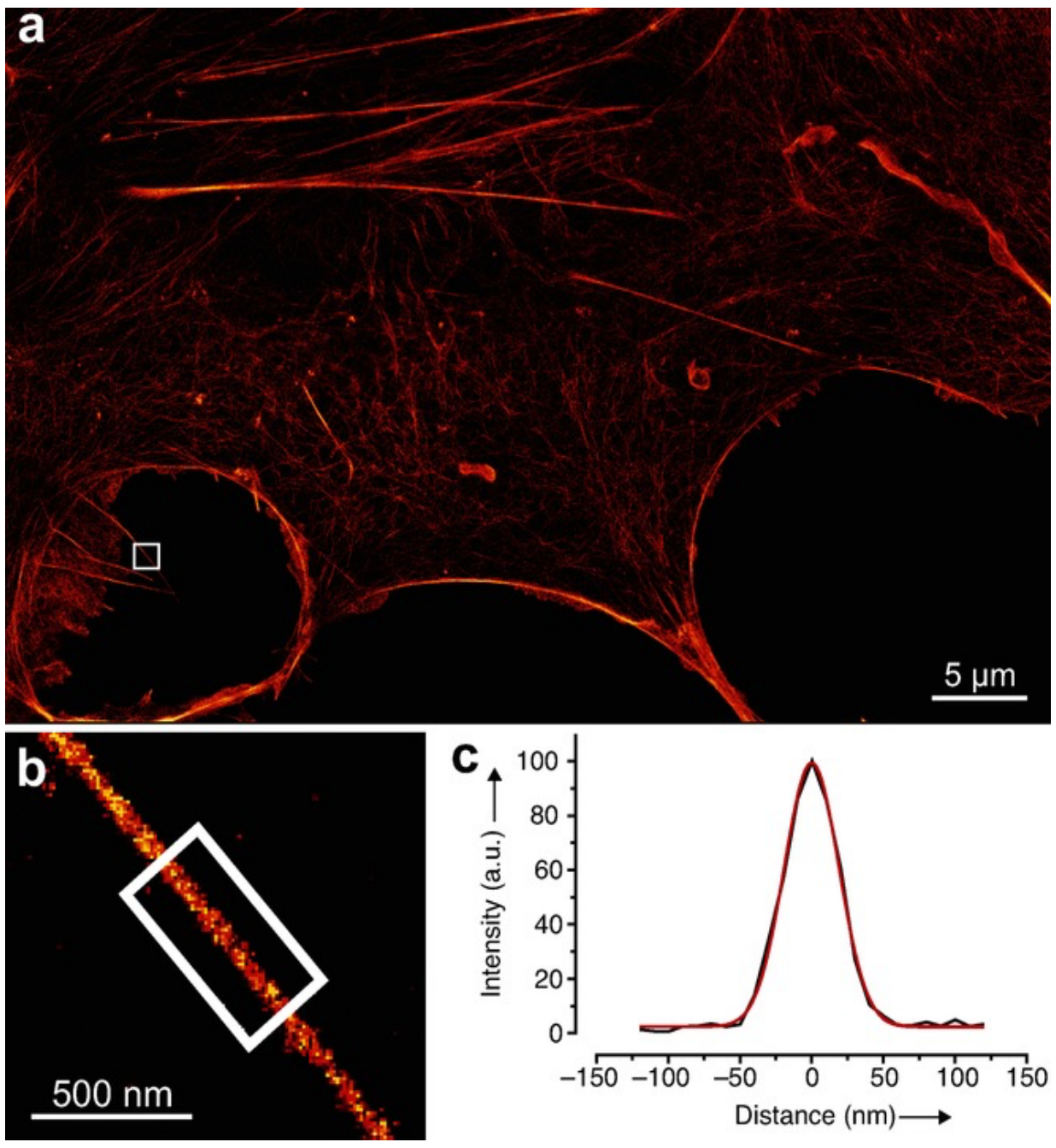

**Figure 5**. a) Localization microscopy image (HILO) of COS-7 cell stained with phalloidin conjugate 17; scale bar: 5 µm. b) Expanded image of boxed region in (a) showing a protruding filopodial structure. c) Line-scan intensity across the filopodial structure in (b; black) and Gaussian fit (red). Full width at half maximum (FWHM)=45 nm. Reproduced with permission [51] copyright 2016, John Wiley & Sons, Inc.

In 2016, Wang *et al.* reported fluorescent nitric oxide (NO) sensing probe utilized from **SiRB.**[53] Introducing the 1,2-diaminobenzene moiety as a reactive site to **SiRB** yielded the **SiRB-NO** probe (Table 3:E32, Scheme 7). Conversion of the aryldiamine into aryltriazole in the presence of NO upon the PeT process resulted in fluorescence enhancement. Furthermore, morpholine moiety was introduced to improve the resolution and selectivity of the probe for monitoring lysosomal NO. Although the lysosome targeted probe, **Lyso-SiRB-NO** (Table 3:E33, Scheme 7) exhibited a similar response toward NO like **SiRB-NO** at physiological pH, **Lyso-SiRB-NO** showed superior tolerance to $H^+$ than **SiRB-NO**. Finally, **Lyso-SiRB-NO** was tested in HepG2 and LO2 cells, and successive lysosomal NO detection was confirmed in both diseased and healthy cells. In the same year, Wang's group also introduced the lysosome-targeted and $Cu^{2+}$ selective probe, **SiRB-Cu** (Table 3:E34, Scheme 7), which incorporated thiosemicarbazide as the cage unit at the 9 position.[54] Upon reaction of the probe with $Cu^{2+}$ in an acidic environment, conversion of thiosemicarbazide to isothiocyanate was observed, resulting in 150-fold fluorescence enhancement at 680 nm. Basicity of the probe, which was associated with the amino group in the thiosemicarbazide resulted in a perfect match with the physiological pH of the lysosome. Due to its chelating ability with $Cu^{2+}$ and lysosome targeting group, lysosome localization and monitoring of the intracellular accumulation of $Cu^{2+}$ in lysosome induced by pyrrolidine dithiocarbamate (PDTC) was successfully demonstrated with **SiRB-Cu**.

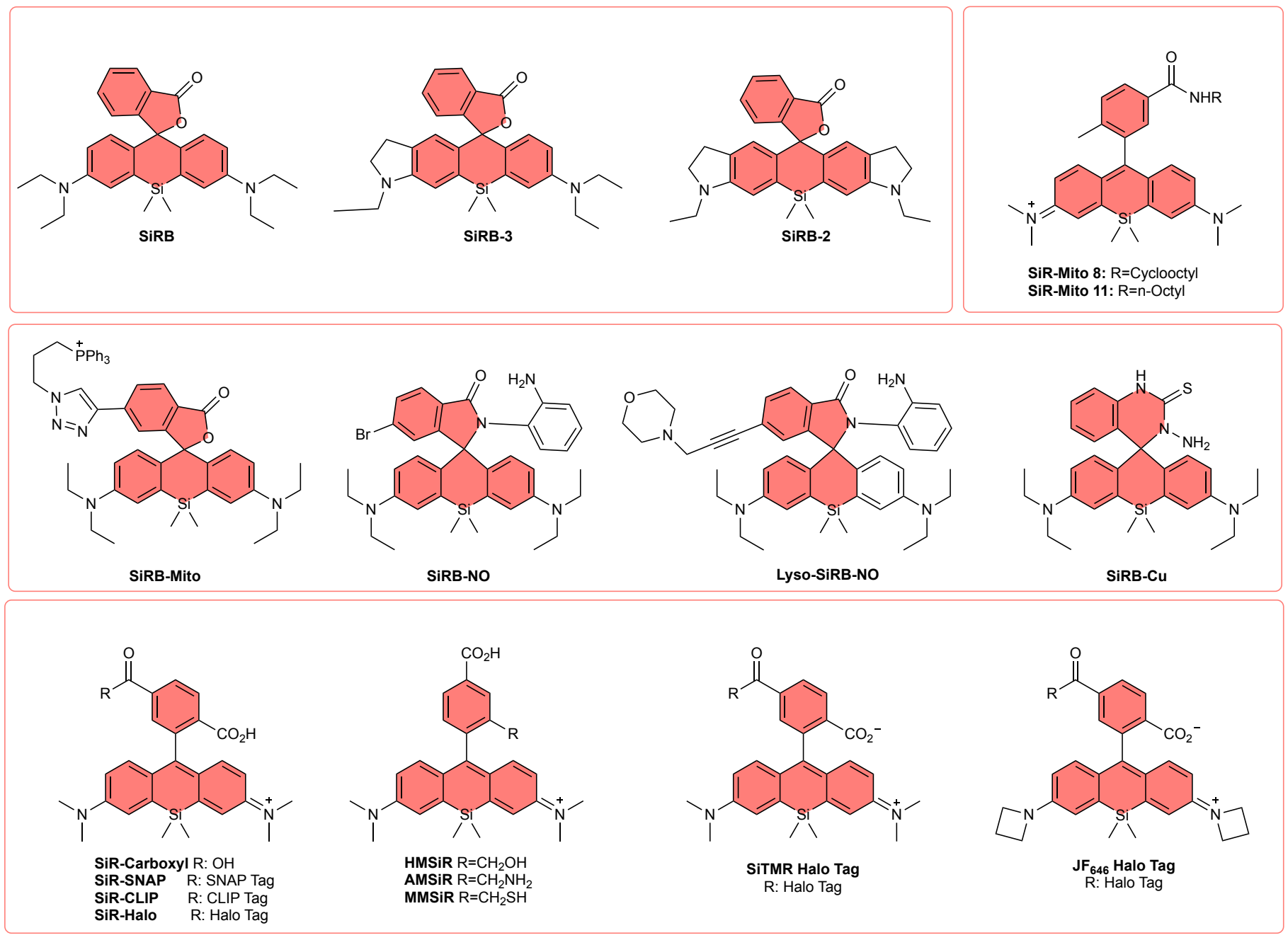

**Scheme 7.** Chemical structures of Si Rhodamine derivatives.

Brewer and Chang developed a spiro SiR based formaldehyde (FA) probe, **FAP-1**, in 2015.[56] (Table 3:E37, Scheme 6). Upon reaction with formaldehyde, it was envisioned that imine formation and subsequent aza-Cope rearrangement occurred and yielded the corresponding aldehyde. Fluorescence intensity at 662 nm was increased 8 to 45-fold depending on the incubation time. Detection of formaldehyde in live cells and photostability of the probe were demonstrated with the treatment of HEK293T cells with **FAP-1**. Confocal images confirmed the formaldehyde detection and fluorescence enhancement with excellent photostability at relatively low laser power. Furthermore, pharmacological inhibition of LSD1, which was responsible for the elevated FA levels in certain cancers, was successfully monitored in the MCF7 human breast cancer cell line, which was known to overexpress LSD1.

Zhou *et al.* designed and synthesized borinate and silanediol-containing rhodamine derivatives for ratiometric $H_2O_2$ detection in 2017.[57] Using chemoselective alteration fluorophore scaffolds (CAFS), conversion of corresponding probes to tetramethylrhodamine (**TMR**) was aimed. The idea behind this work was to avoid spectral overlap between absorption or emission of the corresponding product and reactants by providing a large blue shift upon reaction with $H_2O_2$. For this purpose, the Si-Rhodamine derivative was converted to the corresponding silanediol derivative (**SiOH$_2$R**, Table 3:E38, Scheme 6) by tetrabutylammonium hydroxide. **SiOH$_2$R** has absorption, and emission maxima at 663 nm and 681 nm, respectively, which provided 113 nm and 111 nm blue shift upon conversion to **TMR,** and these results were formerly supported with theoretical calculations. Tamao oxidation was proposed for **SiOH$_2$R** oxidation to **TMR,** and kinetic studies were in line with this mechanism. After photophysical and kinetic studies, the **SiOH$_2$R** probe was utilized in *in vitro* studies. **SiOH$_2$R** was incubated in exogenous H2O2-exposed HeLa cells, and a 5000-fold increase in the **TMR**/**SiOH$_2$R** emission ratio was reported. Additionally, mitochondrial localization of resulting TMR was confirmed with colocalization experiments.

In 2017, Grimm *et al.* reported a series of fluorescent xanthene dyes by incorporating four-membered azetidine rings into classic fluorophore structures, resulting in the Janelia Fluor (JF) dye series.[58] $JF_{549}$, fluorescein analog, absorbed green light and was a good match for light sources near 550 nm. Replacement of the xanthene oxygen with the $CMe_2$ group yielded the carborhodamine derivative **JF$_{618}$** with a 58 nm shift in absorption and a 56 nm shift in emission maxima. In addition to shifts in absorption and emission maxima, these modifications affected the equilibrium between the nonfluorescent closed lactone form and fluorescent open zwitterionic form. Lastly, the O atom at the 10 position of the xanthene dye $JF_{549}$ moiety was replaced by the $SiMe_2$ group in two of the derivatives yielding novel rhodamine analogs **$JF_{646}$** and the fluorinated derivative **$JF_{635}$** with a novel synthetic approach as shown in Scheme 2 (green pathway). A red shift of 97 nm was achieved by **$JF_{646}$** compared to that of the oxygen analog. A single fluorine atom on each azetidine ring on **$JF_{635}$** yielded a dye with an absorption maximum of 635 nm, a higher degree of fluorogenicity, and lower absorbance in water. **$JF_{635}$-HaloTag** (Table 3:E39, Scheme 7) ligand yielded a small, cell-permeable ligand that showed a 21-fold increase in absorbance following binding to the HaloTag protein compared to the other analogs. The ligand was further investigated in a Drosophila GAL4 line expressing myristoylated HaloTag protein in 'Basin' neurons, and it showed consistent labeling throughout the living tissue and low nonspecific background. In the same year, Tang *et al.* introduced an NO sensing probe, **SiRD** (Table 3:E40, Scheme 6), based on **SiR,** which revealed red-shifted absorption and emission maxima, higher specificity, and faster response time toward NO compared to its classical rhodamine analog.[59] As a proof of concept, the probe was incubated with HeLa cells, and no significant cytotoxicity was observed. Also, colocalization studies revealed mitochondrial endogenous NO detection. Authors showed that substituting oxygen atom with dimethyl silane enabled the NO monitoring with a controllable photoinduced electron transfer mechanism. These enhanced properties made the probe suitable for *in vivo* experiments.

Zhang *et al.* introduced a pyrrole incorporated SiR, **PSiR** (Table 3:E41, Scheme 6), which was designed as a lysosome-targeted fluorescent probe for detecting highly reactive oxygen species (hROS).[60] Unlike most SiRs, alkylated pyrrole moiety was introduced instead of a phenyl ring. With this modification, selective hROS interaction with the nitrogen of the pyrrole ring was observed, resulting in the PET mechanism's turn-off. The high selectivity towards hROS was demonstrated against ROS, metal ions, NADH, and biothiols. Subcellular localization and fluorescence imaging of hROS were successfully demonstrated in HeLa and HepG2 cells. Additionally, in-vivo studies were performed, and successful discrimination of tumor tissue from healthy tissue was confirmed by monitoring the tumor-bearing mouse xenograft model.

Umezawa *et al.* had investigated the dynamics of glutathione (GSH) sensing by a series of novel fluorophores to overcome the limitations in real-time live cell GSH imaging, namely irreversibility and slow reaction rates.[61] Due to the electrophilicity of rhodamine with spirocyclization equilibrium that can be shifted in the presence of nucleophiles such as thiols, the rhodamine core was chosen as the candidate for GSH detection. With the aid of Mayr's reactivity scale, it was envisioned that a reversible GSH attack on the 9 position of rhodamine would occur rapidly and cause a drastic change in optical properties. To validate this approach, kinetic studies of classical tetramethyl rhodamine (**Ph TMR**) and its silicon substituted analog (**Ph SiR650,** Table 3:E42, Scheme 6) with thiols were investigated. **Ph SiR650** exhibited a much lower dissociation constant ($K_d$) for reaction with BnSH compared to **Ph TMR**. Additionally, **Ph SiR650** showed rapid and full recovery of absorbance in the presence of thiol scavenger, indicating the reaction's reversibility. After these promising results, optimization on the SiR platform was performed. Upon analyses of nine different derivatives, **2'Me SiR600** and **2'Me SiR610** (Table 3:E43-44, Scheme 6) were nominated as potential GSH probes for real-time GSH imaging due to their suitable $K_{d,GSH}$, high $\boldsymbol{\varphi}_{F,}$ and high rate constants. FRET-based ratiometric probes for quantitative GSH imaging were realized using **2'Me SiR600** and **2'Me SiR610** as the acceptor and **TMR** unit as the donor, connected with an appropriate cyclohexyl-based linker. The resulting dyads **QG0.6** and **QG3.0** showed a significant decrease in absorbance from SiR fluorophores at around 600 nm, together with a marked shift of the fluorescence spectra with an isosbestic point of 600 nm for **QG0.6** and 613 nm for **QG3.0** (Table 3:E45-46, Scheme 6). Additionally, both dyads showed high $\boldsymbol{\varphi}_F$ compared to commonly utilized TQ Green fluorescent probes (30-fold brighter).

Abnormal $ONOO^-$ expression in vivo is usually associated with cellular dysfunctions due to damage on various biomacromolecules, resulting in diverse pathogenic effects such as cancer. Hence,

Tang et al. designed and synthesized a novel SiR-based probe for endogenous peroxynitrite detection in live cells.[62] Unlike most SiR derivatives, Si-Xanthone core was treated with oxalyl chloride to yield chlorine substitution on position 9 and methyl(4-hydroxyphenyl) amino group introduced via nucleophilic aromatic substitution instead of the commonly utilized tolyl moiety. The synthesized probe (**SiNH,** Table 3:E47, Scheme 6) demonstrated a fast and specific response toward peroxynitrite via PeT based on/off mechanism. The oxidation product had an absorption maximum at 595 nm with $\varphi_F$:0.32, where the non-oxidized probe was almost nonfluorescent. To test the SiNH probe's capability for detecting endogenous peroxynitrite in live cells, **SiNH** was encapsulated with mPEG-DSPE to overcome the poor water solubility of the probe. Fluorescent imaging of HeLa cells incubated with nanoprobe revealed high sensitivity for peroxynitrite with minimal cytotoxicity in live cells.

Formerly synthesized SiR600 derivative (Z-DEVD-SiR600) exhibited high background fluorescence, hindering the *in vivo* and *ex vivo* imaging efforts.56 To suppress background signal by PeT mechanism, Urano and coworkers synthesized several SiR600 and Ac-SiR600 derivatives by modifying phenyl moiety on position 9, which caused alteration in the electronic density of the resulting fluorophores.[63] **2OMe-SiR600** was selected due to its high $\varphi_F$ (0.32), while its acetylated derivative exhibited very low fluorescence efficiency (0.001). To test the utility of **2OMe-SiR600** in ex-vivo experiments, glutamic acid-proline dipeptide modified derivative **EP-2OMe-SiR600** (Table 3:E48, Scheme 6) was prepared to target dipeptidypeptidase IV (DPP-IV) for esophageal cancer detection. Then, *ex vivo* fluorescence imaging of esophageal cancer in clinical specimen with **EP-2OMe-SiR600** demonstrated high fluorescence enhancement of 4-fold upon binding to DPP-IV in tumor regions compared to healthy ones.

Lee and Kim groups developed several **SiR-based** probes with various hydrophobicity for mitochondrial staining.[64] This was achieved by modifying the R group on the amide unit that was incorporated into the tolyl ring at position 9. Among these derivatives, **SiR-Mito 8,** which bears a cyclooctyl amine side chain, exhibited the highest intracellular fluorescence intensity and excellent mitochondrial localization (Table 3:E49, Scheme 7). **SiR-Mito 8** was shown to be an ideal candidate for cancer-specific NIR imaging by *in vitro* studies. Following their work on the **SiR-Mito** series, Kim and Lee *et al.* focused their attention on utilizing these dyes for brain cancer imaging.[65] First, they evaluated the **SiR-Mito** series for cell viability using H4 neuroblastoma and HT22 hippocampal neuron cells. Results showed that probes with cycloalkane substituents exhibit high cellular toxicity. Long, branched alkyl chain containing probe was also cytotoxic, while probes with short alkyl chains showed only marginal selectivity between H4 and HT22 cells. Concerning this preliminary screening, a new derivative, **SiR-Mito 11**, was developed, which bears octylamine substituent to achieve lower cytotoxicity and to sustain mitochondrial targeting (Table 3:E50, Scheme 7). The approach was fruitful, and **SiR-Mito 11** was shown to be a mitochondria-targeted NIR fluorescent probe. More interestingly, **SiR-Mito 11** showed highly selective toxicity (90% cell viability vs. negligible) between cancerous H4 cells and healthy HT22 cell lines. The selectivity was also demonstrated among other brain cancer cells, including glioblastoma, neuroblastoma, and astrocytoma.

Kun Li *et al.* incorporated pyridine at position 9 of SiR, which provided ease of modification for photophysical properties compared to standard **SiR**s.[66] Lysosomes targeting **Py-SiRh** probe were realized using the new platform, which revealed absorption and emission maxima at 655 nm and 680 nm, respectively (Table 3:E51, Scheme 6). For targeted imaging studies, pyridine moiety was modified with 4-ethylphenyl acrylate, a cystine-specific cage. Localization and photostability studies were carried out in HeLa cells. Comparable results with commercially available Lyso-Tracker Green were obtained at much higher wavelengths. Additionally, *in vivo* studies revealed selective cysteine detection in B16 tumor-bearing Balb/c mouse. The study showed that **Py-SiRh** is a promising platform for studying lysosomal death and related pathological events.

Imaging of Folate receptors has been pursued since their alpha isomer is overexpressed in ovarian and endometrial cancer cells. Existing NIR probes showed non-specific tissue adsorption and required prolonged washout procedures to visualize tumors. To overcome these difficulties, Urano and coworkers developed **SiR-based** probes where a folate ligand moiety was conjugated to the **SiR** core via a negatively charged tripeptide to visualize folate-receptor-expressing tumors (**Folate-SiR-1** and **Folate-SiR-2,** Table 3:E52-53 Scheme 6).[67] Imaging studies revealed that **Folate-SiR-1** was superior due to its lower background fluorescence which could be attributed to the non-fluorescent spirocyclized nature of **Folate-SiR-1** in the absence of alpha isoform of folate receptor (FR-**α**). **Folate-SiR-1** exhibited highly specific tissue binding, resulting in high contrast tumor images in mouse models *in vivo.* Low background fluorescence facilitated a high tumor to background ratio up to 83 in FR-expressing tumor-bearing mice within 30 minutes.

Lavis group recently introduced a new **SiR** derivative to Janelia Fluor (JF) family.[68] Here all remaining positions of the carboxylic acid containing pendant phenyl ring were modified with fluorines to yield Janelia Fluor derivative (**$JF_{669}$,** Table 3:E54, Scheme 6) towards obtaining both red-shifted $\lambda_{max}$ and enhanced $K_{L-Z}$ ratio, which indicates lactone-zwitterion equilibrium constant. In addition to achieving these goals, it was shown that fluorinated phenyl ring could also serve as an electrophile in nucleophilic aromatic substitution reactions ($S_NAr$). Results revealed that **$JF_{669}$** could undergo $S_NAr$ type reactions with $N_3^-$, $CN^-$, $NH_3$, and $NH_2OH$ regioselectively from the 3′ position and yielded useful intermediates for further bioconjugation.

In 2020, Urano's group developed another $Ca^{2+}$ sensing probe based on SiR (**CaSiR-2 AM,** Table 3:E56, Scheme 6) to overcome the poor reproducibility of the formerly introduced probe **CaSiR-1 AM** (Table 3:E55, Scheme 6).[69] It was speculated that organelle targeting, particularly lysosome localization due to its cationic nature, was the main reason for poor reproducibility. Introducing -COOH on the 2′ position of the pendant phenyl ring was envisioned to result in a neutral probe at physiological pH (**CaSiR-2 AM**) compared to positively charged probe **CaSiR-1 AM** and therefore limit lysosome and mitochondria targeting. The approach was successful and resulted in cytosolic localization with a high S/N ratio in detecting cytosolic $Ca^{2+}$ ions. Better performance of **CaSiR-2 AM** was also confirmed by a fluorescence measurement in HeLa cells where **CaSiR-1 AM** exhibited large background fluorescence. Furthermore, neuronal activity in rat hippocampal cultured slice was also monitored. **CaSiR-2 AM** was uniformly distributed into the cytosol, consistent with live cell fluorescence imaging. **CaSiR-2 AM** showed sharp fluorescence peaks with a strong and high S/N ratio, indicating spontaneous neural firing visualization.

Murata *et al.* introduced novel SiR-based NIR fluorescence probes with appropriate ligands on position 9 to visualize the dynamics and functions of $Mg^{2+}$.[70] Both tetramethyl SiR derivative (**KMG-501,** Table 3:E57, Scheme 6) and tetraethyl analog (**KMG-502,** Table 3:E58, Scheme 6) were modified with a charged β-diketone binding site for selective $Mg^{2+}$ detection. **KMG-501** was found to be cell-impermeable, hence the carboxylic acid group on the binding site was substituted with acetoxymethylester (KMG-501 AM), which was cell-permeable and showed high intracellular retention due to AM cleavage. KMG-501 AM exhibited partially lysosomal and mitochondrial localization and was mainly localized in the cytoplasm in cultured rat hippocampal neurons. On the other hand, **KMG-502** showed intracellular permeability even without AM modification, however, the reverse pathway was also quite favorable. Hence, the dye did not retain in the cells during imaging. KMG-501 AM detects intracellular $Mg^{2+}$ concentration and is suitable for imaging $Mg^{2+}$ dynamics in the NIR fluorescence region. Additionally, the interaction of intracellular signals by simultaneous multi-color imaging using four different chromophores was successfully demonstrated.

Following Zheng's work in 2017[57] and Zhang's work in 2019[66], Zhou *et al.* introduced a novel pyridine-incorporated **SiR** derivative (**Py-SiRh-HP**, Table 3:E59, Scheme 6) toward realizing selective ROS sensing probes.[71] ROS selective oxidation of phenylboronic acid moiety of **Py-SiRh-**

**HP** resulted in highly fluorescent **Py-SiRh**. Lysosomal localization was confirmed with the co-staining assay in HeLa cells, and the ROS detection capability of **Py-SiRh-HP** was demonstrated both in HeLa cells and in the tumor-bearing mouse xenograft model.

Liu *et al.* developed a new ATP/$H^+$ dual stimulus responsive NIR probe, **ATP-pH** (Table 3:E60, Scheme 6), for fluorescent and photoacoustic ratiometric imaging of tumors.[72] SiR had chosen as a donor fluorophore and conjugated to CS dye with a piperidine-4-carboxylic acid linker to avoid interference from high GSH levels in the tumor region. Accurate fluorescent and photoacoustic tumor imaging were accomplished with reversible **ATP-pH** *in vitro* and in mouse tumor models due to activation of the probe in the tumor region with both $H^+$ and ATP.

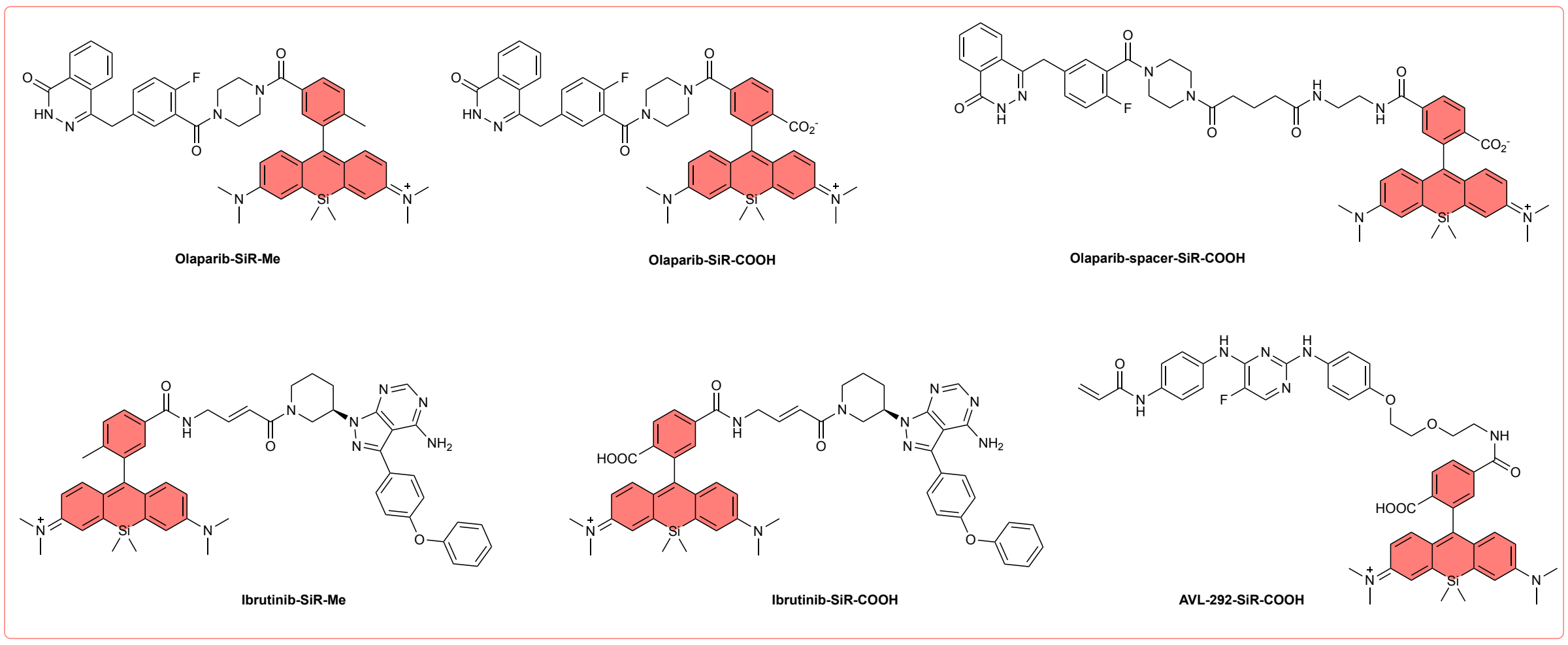

**Scheme 8.** Chemical structures of SiR-based CIDs

Kim *et al.* designed and synthesized a series of fluorophore-drug conjugates based on Si rhodamine and PARP inhibitor Olaparib to visualize their activity in live cells.[75] For this purpose, in addition to Olaparib conjugated **SiR-Me** and **SiR-COOH** derivatives, **Olaparib-spacer-SiR-COOH** (Table 4:E1-3, Scheme 8) was synthesized. These three derivatives had absorption maxima at 653 nm, while their emission maxima were observed between 670-678 nm. To test these companion imaging drugs (CID), three GFP expressing cell lines (HT1080 / GFP-H2B – Nucleus staining, OVCA429 / GFP-Mito – Mitochondria staining, MDA-MB-231 /PARP1-GFP – Colocalization) were utilized. Cationic **SiR-Me** was localized to mitochondria as expected. On the other hand, **SiR-COOH** was shown to be mostly localized in the cytoplasm. **Olaparib-spacer-SiR-COOH** was synthesized to reduce steric hindrance with the drug and sensitizer via insertion of a linear spacer. However, a negligible difference in $IC_{50}$ values between **SiR-COOH** and **Olaparib-spacer-SiR-COOH** was observed. However, in vitro studies revealed that **Olaparib-spacer-SiR-COOH** has greater nucleus localization. This pharmacokinetic difference could be resulted from increased hydrophobicity due to the chemical spacer, which provided nuclear permeability. All in all, zwitterionic **SiR-COOH** with an appropriate spacer has shown to be a promising CID for pharmacokinetic studies in live cells.

*Table 4. Properties of SiR-based CIDs*

| Entry (E) | Fluorophore* | Absorption λmax (nm) | Emission λmax (nm) | Tissue Target | $IC_{50}$ (nM) | Target Protein | Ref |
|---|---|---|---|---|---|---|---|
| 1 | Olaparib-SiR-Me[a] | 653 | 675 | nucleus | 99.2 | - | [75] |
| 2 | Olaparib-SiR-COOH[a] | 653 | 670 | nucleus | 100.3 | - | [75] |
| 3 | Olaparib-spacer-SiR-COOH[a] | 653 | 678 | nucleus | 112.9 | PARP | [75] |
| 4 | Ibrutinib-Si-Me | 653 | 670 | MT | n.d. | Btk | [76] |
| 5 | Ibrutinib-Si-COOH | 651 | 671 | Btk-mCherry | 122.8 | Btk | [76] |
| 6 | AVL-292-Si-COOH | 646 | 668 | Btk-mCherry | 283.5 | Btk | [76] |

[a]in PBS buffer containing 0.1% DMSO. MT: mitochondria. * all the fluorophores listed above are emphasized with bold font in the text.

After developing drug conjugates based on Si rhodamine for PARP inhibitor visualization, Kim *et al*. developed a series of fluorophores for Bruton's Tyrosine Kinase (Btk) imaging.[76] Three of these fluorophores were based on Si rhodamine. Ibrutinib and AVL-292 were chosen as the two highly potent and selective inhibitors with sub-nanomolar inhibitor activities for conjugating with the fluorophores. Ibrutinib conjugated SiR-Me, ibrutinib conjugated SiR-COOH and AVL-292 conjugated Si-COOH were synthesized (**Ibrutinib-SiR-Me**, **ibrutinib-SiR-COOH**, **AVL-292-SiR-COOH**, Table 4:E4-6, Scheme 8). Although all fluorophores revealed satisfactory imaging characteristics and colocalization with Btk-mCherry in cells, only ibrutinib conjugated Si-COOH had shown pharmacokinetics and target localization that enabled Btk-specific single-cell imaging (figure 6). Compared to previously described green Ibrutinib-BFL,[77] enhanced binding and imaging characteristics were observed, which was attributed to improved pharmacokinetics, lower background accumulation, and lower autofluorescence due to red-shifted emission.

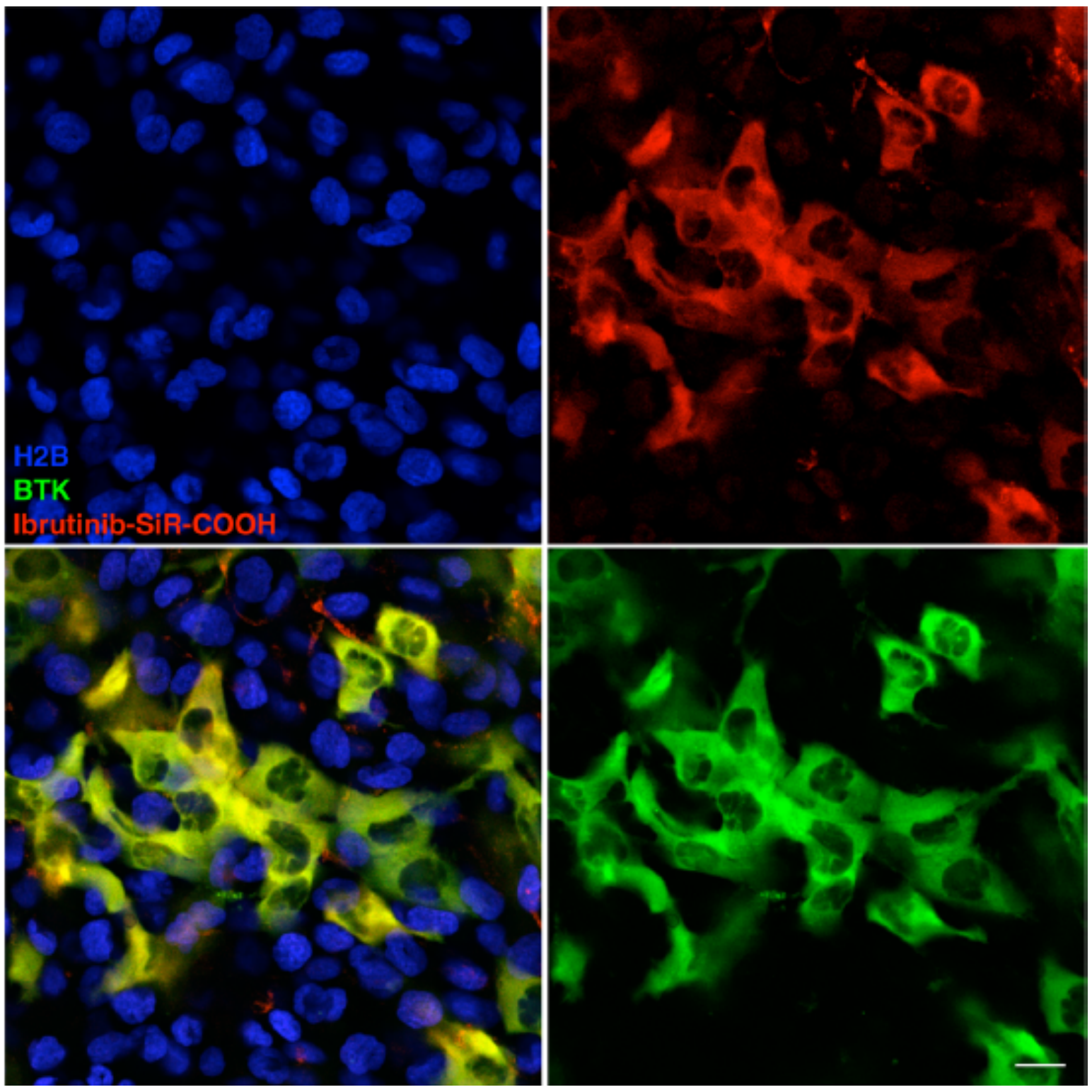

**Figure 6.** A single cell in vivo imaging of the Ibrutinib-SiR-COOH CID. Mice with a mixed HT1080 tumor containing both Btk negative and positive cells (green: Btk-mCherry cells, blue: H2B-GFP cells) were imaged 24 h after i.v. administration of Ibrutinib-SiR-COOH (75 nmol). Note the excellent colocalization between the CID and Btk-mCherry, persisting even 24 h post-injection. Scale bar: 20 μm. Reproduced with permission [76] Copyright 2015, American Chemical Society

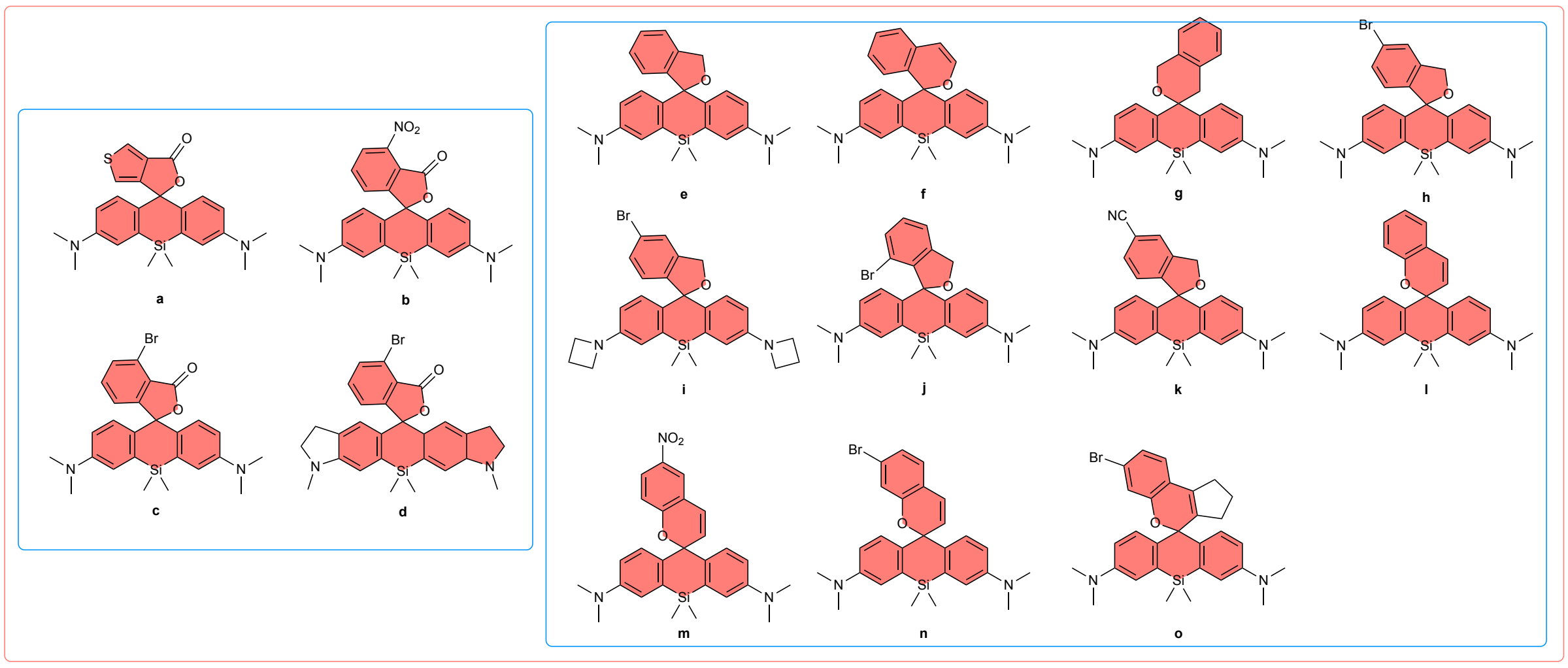

**Scheme 9.** Chemical structures of SiR derivatives.

In 2021, Butkevich [71] developed a new synthetic approach toward synthesizing SiR analogs based on regioselective double nucleophilic addition of aryllanthanum to several carbonyl-containing dye precursors. [78] The proposed reaction strategy was similar to the literature methods utilizing transmetalation (RLi, RMgX), [28,79] however it was stated that these reagents are restricted with respect to their high basicity and possible single electron transfer pathways, and undesired reactions are commonly observed. It was envisioned transmetalation with more Lewis acidic aryllanthanide reagents of type $RMX_2$ (M=La, Ce) could overcome these difficulties with their higher nucleophilicity toward carbonyl compounds. To test the applicability of the new approach, a diverse analog of cell permeable, spontaneously blinking, and photoactivable derivatives of **SiR** were synthesized with superior yields, and their behavior in an aqueous solution at various pH values was investigated. (Table 5, Scheme 9).

*Table 5. Properties of Selected SiR Fluorophores* [78]

| Entry (E) | Dye* | Absorption $\lambda_{max}$ (nm) | Emission $\lambda_{max}$ (nm) | pKa[a] | $D_{0.5}$[b], $K_{L-Z}$ |
|---|---|---|---|---|---|
| 1 | a | 646 | 674 | open | $D_{0.5}$:8, $K_{L-Z}$:5, |
| 2 | b | 658 | 682 | $pK_{s1}$:1.2 | $D_{0.5}$:38, $K_{L-Z}$:0.36 |
| 3 | c | 656 | 676 | $pK_{s1}$:1.3<br>$pK_{s2}$:2.3 | $D_{0.5}$:48, $K_{L-Z}$:0.0041 |
| 6 | d | 708 | 738 | [c]$pK_{s1}$:<1 | $D_{0.5}$:39, $K_{L-Z}$:0.016 |
| 7 | e | 650 | 674 | $pK_{s1}$:2.0<br>$pK_{s2}$:4.2 | $D_{0.5}$:41, $K_{L-Z}$:0 |
| 8 | f | 652 | 676 | $pK_{s1}$:1.4<br>$pK_{s2}$:5.1 | $D_{0.5}$:36, $K_{L-Z}$:0 |
| 9 | g | 650 | 647 | $pK_{s1}$:2.6 | $D_{0.5}$:-, $K_{L-Z}$:- |
| | | | | $pK_{s2}$:4.0 | |
| 10 | h | 658 | 678 | $pK_{s1}$:1.9<br>$pK_{s2}$:4.4 | $D_{0.5}$:42, $K_{L\text{-}Z}$:0 |
| 11 | i | 658 | 678 | $pK_{s1}$:1.2<br>$pK_{s2}$:2.9 | $D_{0.5}$:41, $K_{L\text{-}Z}$:0 |
| 12 | j | 664 | 688 | $pK_{s1}$:1.7<br>$pK_{s2}$:4.7 | $D_{0.5}$:41, $K_{L\text{-}Z}$:0 |
| 13 | k | 658 | 680 | $pK_{s1}$:1.8<br>$pK_{s2}$:4.5 | $D_{0.5}$:47, $K_{L\text{-}Z}$:0 |
| 14 | l | 642 | - | $pK_{s1}$:2.2<br>$pK_{s2}$:4.0 | $D_{0.5}$:-, $K_{L\text{-}Z}$:- |
| 15 | m | 658 | - | $pK_{s1}$:1.7<br>$pK_{s2}$:3.5 | $D_{0.5}$:-, $K_{L\text{-}Z}$:- |
| 16 | n | 658 | - | $pK_{s1}$:1.7<br>$pK_{s2}$:3.6 | $D_{0.5}$:-, $K_{L\text{-}Z}$:- |
| 17 | o | 650 | 674 | $pK_{s1}$:1.3<br>$pK_{s2}$:2.2 | $D_{0.5}$:-, $K_{L\text{-}Z}$:- |

[a]Apparent pKa values in 20% (v/v) DMSO– 100 mM Na phosphate buffer. [b]With 1% (v/v) DMSO. [c]3m forms a colorless water adduct at pH > 1.5. [*]all the fluorophores listed above are emphasized with bold font in the text.

## 3.1. Phospa-Rhodamines

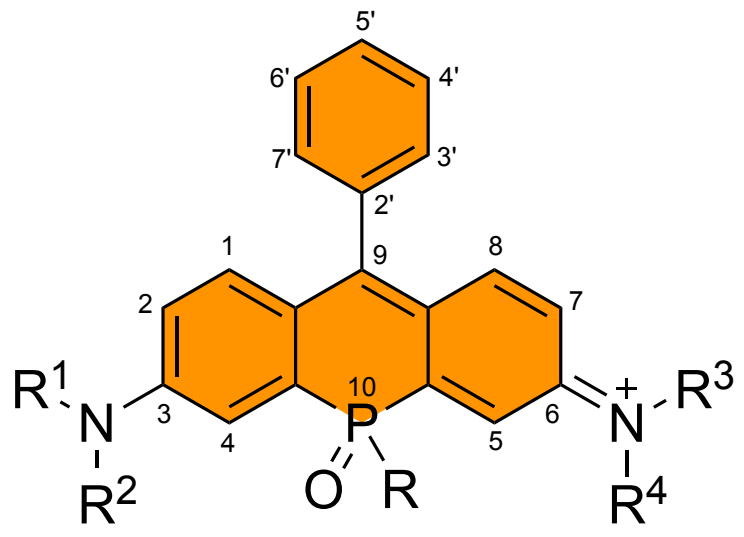

In addition to group 4 elements Si and Ge and group 6 elements S, Se, and Te, Phosphorus (group 5) was used for the electronic modifications of π-conjugated systems due to the orbital interaction between π-skeleton and significant perturbation of the electronic structure. The oxidation of the phosphorus atom enhanced the electron-accepting ability and chemical stability of the π-skeleton.

Wang's group demonstrated the first introduction of phosphorus into xanthene dyes in 2015.[35] Replacement of bridging oxygen atom in the classic rhodamine framework with phosphine oxide yielded a series of phosphorous substituted tetraethyl rhodamines (PRs). Replacement of the bridging oxygen atom with silicon yielded near-infrared (NIR) Si-rhodamine with an emission maximum of around 650 nm. When phosphorus was integrated into conjugated ring systems, **σ*- π*** orbital interactions increased, and the energy level of the LUMO decreased, resulting in greater shifts towards the NIR region. The classical phosphorus substituted tetraethyl rhodamine (**PR**) showed absorption and emission maxima at 694 and 712 nm in PBS buffer (**$\Phi_f$** = 0.06 in PBS, 0.36 in DCM) (Table 6:E1, Scheme 10). The 2′-Methyl derivative **Me-PR** showed the same absorption and emission values with a nearly two-fold increase in fluorescence quantum yield (**$\Phi_f$** = 0.11 in PBS, 0.42 in DCM) (Table 6:E2, Scheme 10). The 2′,4′,6′-triMethyl derivative **tMe-PR** yielded 696 and 713 nm absorption and emission maxima (**$\Phi_f$** = 0.15 in PBS) (Table 6:E3, Scheme 10). Two synthetic methods were established in the synthesis of PRs. For the first method, $NEt_2$-MDPP was coupled with formaldehyde and oxidized to yield the phosphorous-containing xanthone core (Scheme 2, purple pathway). The insertion of the phenyl group at the 9-position to yield the desired PR was achieved via reaction of the core with the corresponding phenyllithium derivative followed by treatment with HCl. The second method provided the insertion of the phenyl group at the 9-position by the reaction of $NEt_2$-MDPP with the corresponding aldehyde derivative catalyzed by *p*-toluenesulfonic acid, followed by oxidation with chloranil (Scheme 2, yellow pathway). In this method, $NEt_2$-MDPP acts as the electron-rich bis-nucleophile, and the aldehyde acts as the electron-poor electrophile creating a better electronic match. Both methods led to desired PRs; the first method through a simpler procedure, and the second with higher yields. PRs showed fluorescence in the pH 2 to 12 range conforming to pH-independent emission. To evaluate the NIR bioimaging capacity of PRs, **Me-PR** and **tMe-PR** were applied to stain living HepG2 cells and normal mice, confirming the similar membrane permeability of the PRs to parent rhodamine. This new series of rhodamines opened a new chapter in NIR-fluorescence probe design.

*Table 6. Selected features of P-Rhodamine derivatives.*

| Entry (E) | Fluorophore*** | Absorption | | Emission | $\Phi_f$ | Intracellular Localization | Properties | Applications | Ref. |
|---|---|---|---|---|---|---|---|---|---|
| | | $\lambda_{max}$ (nm) | ε ($M^{-1}\cdot cm^{-1}$) | $\lambda_{max}$ (nm) | | | | | |
| 1 | PR[a] | 694 | 65000 | 712 | 0.06 | - | low cytotoxicity | - | [35] |
| 2 | Me-PR[a,b] | 694 | 92000 | 712 | 0.11 | - | photostable, low cytotoxicity | *in vitro* (HepG2 cells) and *in vivo* imaging | [35] |
| 3 | tMe-PR[a] | 696 | 68000 | 713 | 0.15 | - | photostable, low cytotoxicity | *in vitro* (HepG2 cells) and *in vivo* imaging | [35] |
| 4 | $NR_{666}$[c] | 666 | 165000 | 685 | 0.38 | - | photostable, non-cell permeable | *in vitro* imaging (HeLa cells) | [36] |
| 5 | $NR_{700}$[c] | 700 | 71000 | 722 | 0.11 | - | photostable, cell permeable | *in vitro* imaging (HeLa cells) | [36] |
| 6 | $NR_{698}$[c] | 698 | 25900 | 712 | 0.32 | - | photostable | - | [36] |
| 7 | $NR_{744}$[c] | 744 | 80000 | 764 | 0.16 | - | photostable | - | [36] |
| 8 | NR-HOCl[c] | n.d.<br>705**<br>673* | n.d.<br>n.d.<br>n.d. | n.d.<br>730**<br>690* | n.d.<br>n.d.<br>n.d. | - | HOCl selective, activatable, low cytotoxicity | *in vitro* imaging (HeLa cells and RAW 264.7) | [36] |
| 9 | $NR_{675}$[c] | 675 | 73200 | 695 | 0.24 | - | - | - | [40] |
| 10 | $NR_{669}$[c] | 669 | 77200 | 689 | 0.30 | - | - | - | [40] |
| 11 | **EtPR-Me**[d] | 707 | 94500 | 731 | 0.12 | - | GSH selective,<br>low chemical stability at pH=4-10 | - | [80] |
| 12 | **EtPR-Me$_2$**[d] | 708 | 99300 | 734 | 0.13 | - | high chemical stability at pH=4-10 | - | [80] |
| 13 | **PREX 710**[d] | 712 | 94200 | 740 | 0.13 | MT<br>lysosome | high chemical stability at pH=4-10,<br>photostable,<br>water-soluble,<br>high resistance to photobleaching,<br>membrane-permeable | *in vitro* (HeLa cells) and *in vivo* imaging | [80] |
| 14 | **PREX 710 NHS**[d] | 718 | - | 739 | 0.058[e]<br>0.047[f] | MT<br>lysosome | photostable,<br>resistance to photobleaching | *in vitro* (HeLa cells) and *in vivo* imaging | [80] |
| 15 | **EtPR-Me$_2$**[g] | 708 | 99000 | 734 | 0.22 | - | - | - | [81] |
| 16 | **ThioMorPR-Me$_2$**[g] | 683 | 98000 | 711 | 0.82 | - | - | - | [81] |
| 17 | **MePOPhR**[g] | 698 | 22000 | 726 | 0.25 | - | - | - | [82] |
| 18 | **MePOPhR-Me**[g] | 699 | 68000 | 725 | 0.33 | | | | [82] |
| 19 | **MePOPhR-OMe**[g] | 701 | 26000 | 731 | 0.16 | - | - | - | [82] |
| 20 | **MePOPhR-OMe$_2$**[g] | 703 | 66000 | 735 | 0.34 | - | - | - | [82] |
| 21 | **MePOPhR-Et**[g] | 699 | 73000 | 724 | 0.31 | - | - | - | [82] |
| 22 | **MePOPhR-Me$_2$**[g] | 700 | 55000 | 735 | 0.35 | - | - | - | [82] |
| 23 | **MePOPhR-Me$_3$**[g] | 700 | 42000 | 733 | 0.32 | - | - | - | [82] |
| 24 | **MePOPhR-CF$_3$**[g] | 709 | 76000 | 738 | 0.13 | | | | [82] |
| 25 | **MePO$^t$BuR-Me**[g] | 693 | 45000 | 719 | 0.32 | - | - | - | [82] |
| 26 | **MePO$^t$BuR-OMe$_2$**[g] | 697 | 42000 | 720 | 0.23 | - | - | - | [82] |
| 27 | **AzPOPhR-Me**[g] | 699 | 52000 | 726 | 0.43 | - | - | - | [82] |
| 28 | **AzPOPhR-CF$_3$**[g] | 710 | 58000 | 735 | 0.31 | | | | [82] |
| 29 | **PyPOPhR-Me**[g] | 708 | 63000 | 729 | 0.25 | - | - | - | [82] |
| 30 | **PyPOPhR-CF$_3$**[g] | 718 | 63000 | 744 | 0.18 | - | - | - | [82] |
| 31 | **PyMePOPhR-Me**[g] | 703 | 54000 | 730 | 0.27 | - | - | - | [82] |
| 32 | **PyMePOPhR-Et**[g] | 703 | 72000 | 729 | 0.28 | - | - | - | [82] |
| **33** | **PyMePOPhR-CF$_3$**[g] | 713 | 66000 | 741 | 0.21 | - | - | - | [82] |
| **34** | **PyMePOPhR-OMe**[g] | 706 | 54000 | 733 | 0.25 | - | - | - | [82] |
| **35** | **PyMePOPhR-OMe$_2$**[g] | 706 | 56000 | 737 | 0.16 | - | - | - | [82] |
| **36** | **HT-NR$_{666}$** | n.d.<br>666* | n.d.<br>165000* | n.d.<br>685* | n.d.<br>0.38* | cell membrane | cell permeable,<br>HaloTag protein labeling,<br>selective labeling of the membrane population of hOX2R | *in vitro* imaging (Chinese Hamster Ovary cells) | [83] |
| **37** | **mHT-spiroNR$_{666}$** | n.d. | n.d. | n.d. | n.d. | cell membrane | HaloTag protein labeling,<br>selective labeling of the membrane population of hOX2R | increase in fluorescence intensity in the presence of fetal bovine serum<br>*in vitro* imaging (Chinese Hamster Ovary cells) | [83] |
| **38** | **MePR 705**[c] | 705 | 93800 | 733 | 0.13 | lysosome | high chemical stability at pH=4-10,<br>photostable | *in vitro* imaging (HeLa cells) | [84] |
| **39** | **AzPR 710**[c] | 708 | 89300 | 734 | 0.14 | lysosome | high chemical stability at pH=4-10,<br>photostable | *in vitro* imaging (HeLa cells) | [84] |
| **40** | **PyPR 710**[c] | 712 | 97500 | 739 | 0.13 | MT<br>lysosome | high chemical stability at pH=4-10,<br>low photostability | *in vitro* imaging (HeLa cells) | [84] |
| **41** | **BcPR 705**[c] | 704 | 53000 | 733 | 0.21 | lysosome<br>(brighter and more selective) | pH independent NIR probe for staining lysosomes in living cells,<br>high photostability | *in vitro* imaging (HeLa, COS-7, A431, and 3T3-L1 cells) | [84] |
| **42** | **InPR 770**[c] | 770 | 56200 | 816 | 0.017 | - | low chemical stability at pH=4-10,<br>photostable | - | [84] |
| **43** | **InPR 775**[c] | 773 | 51600 | 821 | 0.011 | - | high chemical stability at pH=4-10,<br>photostable | - | [84] |
| **44** | **TMPOR-methyl**[h] | 688 | 98000 | 709 | 0.16 | - | - | - | [85] |
| **45** | **JF$_{668}$**[i] | 669 | 26700 | 687 | 0.34 | - | - | - | [68] |
| **46** | **JF$_{704}$**[i] | 704 | <200 | 723 | n.d. | - | - | - | [68] |
| **47** | **JF$_{690}$**[i] | 690 | 150000 | 707 | 0.24 | nucleus | HaloTag protein labeling | *in vitro* imaging (U2OS cell nuclei expressing HaloTag-histone H2B) | [68] |
| **48** | **JF$_{722}$**[i] | 722 | 87200 | 743 | 0.11 | nucleus | HaloTag protein labeling,<br>good loading kinetics in live cells | *in vitro* imaging (U2OS cell nuclei expressing HaloTag-histone H2B) | [68] |
| **49** | **JF$_{711}$**[i] | 711 | 12400 | 732 | 0.17 | nucleus | HaloTag protein labeling,<br>high brightness | *in vitro* imaging (U2OS cell nuclei expressing HaloTag-histone H2B) | [68] |
| **50** | **EtPOMeR-*meta*BrSO$_3^-$**[j] | 704 | n.d. | 728 | 0.22 | cellular membrane | voltage-sensitive cellular fluorescence | *in vitro* imaging (HEK cells) | [86] |
| **51** | **EtPOMeR-*para*BrSO$_3^-$**[j] | 704 | n.d. | 728 | 0.19 | cellular membrane | voltage-sensitive cellular fluorescence | *in vitro* imaging (HEK cells) | [86] |
| **52** | **EtPOMeR-*meta*X$_1$SO$_3^-$**[j] | 705 | n.d. | 731 | 0.01 | cellular membrane | voltage-sensitive cellular fluorescence | *in vitro* imaging (HEK cells) | [86] |
| **53** | **EtPOMeR-*meta*X$_2$SO$_3^-$**[j] | 703 | n.d. | 728 | 0.01 | cellular membrane | voltage-sensitive cellular fluorescence | *in vitro* imaging (HEK cells and cultured hippocampal neurons isolated from rat embryos) | [86] |
| **54** | **EtPOMeR-*para*X$_1$SO$_3^-$**[j] | 704 | n.d. | 723 | 0.07 | cellular membrane<br>MT | voltage-sensitive cellular fluorescence,<br>high brightness | *in vitro* imaging (HEK cells and cultured hippocampal neurons isolated from rat embryos)<br>voltage and $Ca^{2+}$ imaging and electrode recording in *ex vivo* retinas from *rd1* mice | [86] |
| **55** | **EtPOMeR-*para*X$_2$SO$_3^-$**[j] | 704 | n.d. | 726 | 0.09 | cellular membrane | voltage-sensitive cellular fluorescence | *in vitro* imaging (HEK cells) | [86] |
| **56** | **POR-Hoechst**[k] | n.d. | n.d. | n.d. | n.d. | nucleus | photostable | *in vitro* imaging (A431 cells) | [41] |
| **57** | **POR-SO$_3$H**[k] | 713 | 95300 | 739 | 0.12 | - | photostable | - | [41] |

[a] Photophysical properties were measured in PBS buffer containing 1% DMSO [b] Determined with isomer mixtures without isolation. [c] All experiments were performed in PBS (10 mM, pH=7.4 with 1% DMSO). [d] Photophysical properties were measured in PBS (pH=7.4). [e] DOL=2.2. [f] DOL=3.1. [g] All photophysical properties were measured in PBS buffer pH=7.4 + 1%DMSO as co-solvent. [h] Photophysical properties measured in PBS (20 mM, pH=7.4). [i] All properties was measured in 10 mM HEPES, pH=7.3. [j] Measured in dPBS with 0.1% DMSO. [k] Measured in 50 mM hEPES buffer (pH=7.4) containing 0.1% DMSO as a co-solvent. * Values for the activated form. ** Values for the phosphinate ester form. MT: mitochondria.*** all the fluorophores listed above are emphasized with bold font in the text.

In 2016 Stains' group reported a new series of phosphorous substituted rhodamine analogs called the Nebraska Red (NR) series.[36] Inspired by replacing the bridging oxygen of tetramethylrhodamine (**TMR**) with varying elements, they envisioned replacing the bridging oxygen with phosphorus, similar to **Me-PR** from the Wang group's work.[35] In **Me-PR**, bridging oxygen was replaced with methyl phosphine oxide. However, the replacement was performed with the phosphinate group in the Nebraska Red series. Initially, they synthesized two dyes, phospha TMR analog **$NR_{666}$** (Table 6:E4, Scheme 10) and its ethyl ester counterpart **$NR_{700}$** (Table 6:E5, Scheme 10). Absorption and emission in the NIR region were observed in both dyes. Under the light of their photophysical properties, the dimethylaniline functionality was replaced with julolidine substituent yielding **$NR_{698}$** (Table 6:E6, Scheme 10) and **$NR_{744}$** (Table 6:E7, Scheme 10). Replacement with julolidine substituent showed an increase in absorption maxima by 32 nm for phosphinate analog and 44 nm for ethyl ester analog. However, a 3-fold decrease in the fluorescence quantum yields was observed. Additionally, a reduction in **$NR_{700}$** fluorescence was observed over time. Studies showed that this decrease was due to the autohydrolysis of **$NR_{700}$** to **$NR_{666}$**. This phenomenon was not observed for **$NR_{744}$**, which was considered to arise by the increased steric bulk. To build upon this observation, **$NR_{666}$** and **$NR_{700}$** were incubated in HeLa cells, and cell uptake was only observed for **$NR_{700}$**, indicating the phosphinate ester substituents can be used to modulate cell permeability of NR dyes. Also, by using **$NR_{666}$**, a self-reporting small molecule delivery reagent was prepared. Combining the selective reaction of spirocyclic thioethers with HOCl, non-fluorescent **NR-HOCl** (Table 6:E8, Scheme 10) was designed, and it was triggered by HOCl yielding green-colored fluorescent **$NR_{700}$**, whereafter then hydrolyzed to **$NR_{666}$**. A distinct off-on signal was observed in the presence of HOCl in HeLa cells proving the Nebraska Red template provided a robust scaffold for further developments. Later, in 2019 Stains' group synthesized two more derivatives of NR dyes.[40] Incorporating methyl groups to 2′ and 6′ positions to phosphinate rhodamine core yielded **$NR_{675}$** (Table 6:E9, Scheme 10), and its dimethoxy analog yielded **$NR_{669}$** (Table 6:E10, Scheme 7). Compared to its monomethyl analog **$NR_{666}$** a similar absorption was observed for **$NR_{669}$** with a relatively lower extinction coefficient and fluorescence quantum yield (165000 $M^{-1}$ $cm^{-1}$ vs. 77200 $M^{-1}$ $cm^{-1}$ and 0.38 vs. 0.30). Interestingly, the dimethoxy analog **$NR_{675}$** showed slight red-shifted absorption, however, the brightness was significantly lower than **$NR_{666}$** (62700 $M^{-1}$ $cm^{-1}$ vs. 17568 $M^{-1}$ $cm^{-1}$).

In 2018, Grzybowski *et al.*[80] reported three new phosphatetraethyl rhodamine derivatives (PORs) containing phenyl phosphine oxide moiety as the bridging group with different aryl groups at the 9-position: 2-Methylphenyl group bearing derivative (**EtPR-Me**, Table 6:E11, Scheme 10), 2,6-dimethylphenyl group bearing derivative (**$EtPR\text{-}Me_2$**, Table 6:E12, Scheme 7) and 2,6-dimethoxyphenyl bearing derivative **(PREX 710,** Table 6:E13, Scheme 10). The chemical stability of the dyes against nucleophiles was tested to assess the potential utility as labeling reagents. The dyes' absorption changes in aqueous solution were monitored at pH = 4.2, 7.4, and 10.3. **EtPR-Me** showed a gradual decrease at pH = 10.3 when tested with NaOH and with glutathione (GSH), the most abundant low-molecular-weight thiol in the living cells. However, the **$EtPR\text{-}Me_2$** and **PREX 710** showed high resistance towards nucleophiles and high stability towards fetal bovine serum (FBS), confirming the suitability of sterically protected PORs in various biological systems, including in different cells and blood. Among the synthesized PORs, only **PREX 710** possessed high photostability and was named PREX (Photo-Resistant Xanthene dye). Next, an NHS ester bearing **PREX 710** derivative (**PREX 710 NHS**, Table 6:E14, Scheme 10) was prepared for labeling amino groups of biomolecules. The photostability was retained, and strong signals were observed from imaging of HeLa cells. **PREX 710** is also membrane-permeable and predominantly localized in the mitochondria cells. Finally, *in vivo* imaging of blood vessels in the deep region of the mouse brain was performed using **PREX 710 NHS,** which showed intense fluorescence signals (figure 7).

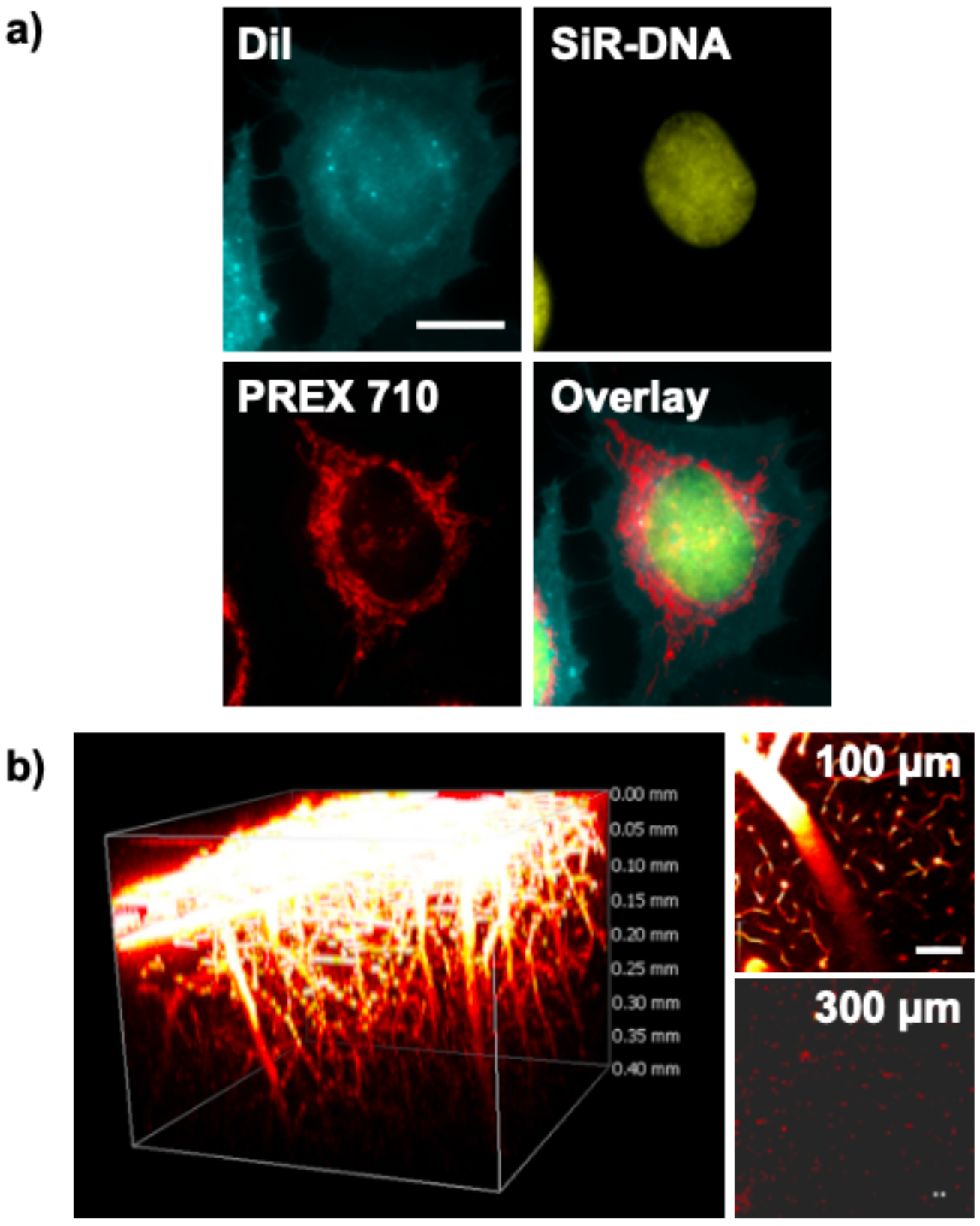

Figure 7. a) Three-color imaging of cell membrane (cyan), nucleus (yellow), and mitochondria (red) in living HeLa cells stained with DiI, SiR-DNA, and PREX710; scale bar: 20 µm. b) 3D image of blood vessels in mouse brain stained with PREX710-dextran conjugate. Maximum intensity projections of the 3D stacks were obtained using a CW laser at 638 nm. Single *xy* frames from the *z*-stack at 100 µm and 300 µm depth are shown on the right. Scale bar: 100 µm. Reproduced with permission [80] copyright 2018, John Wiley & Sons, Inc.

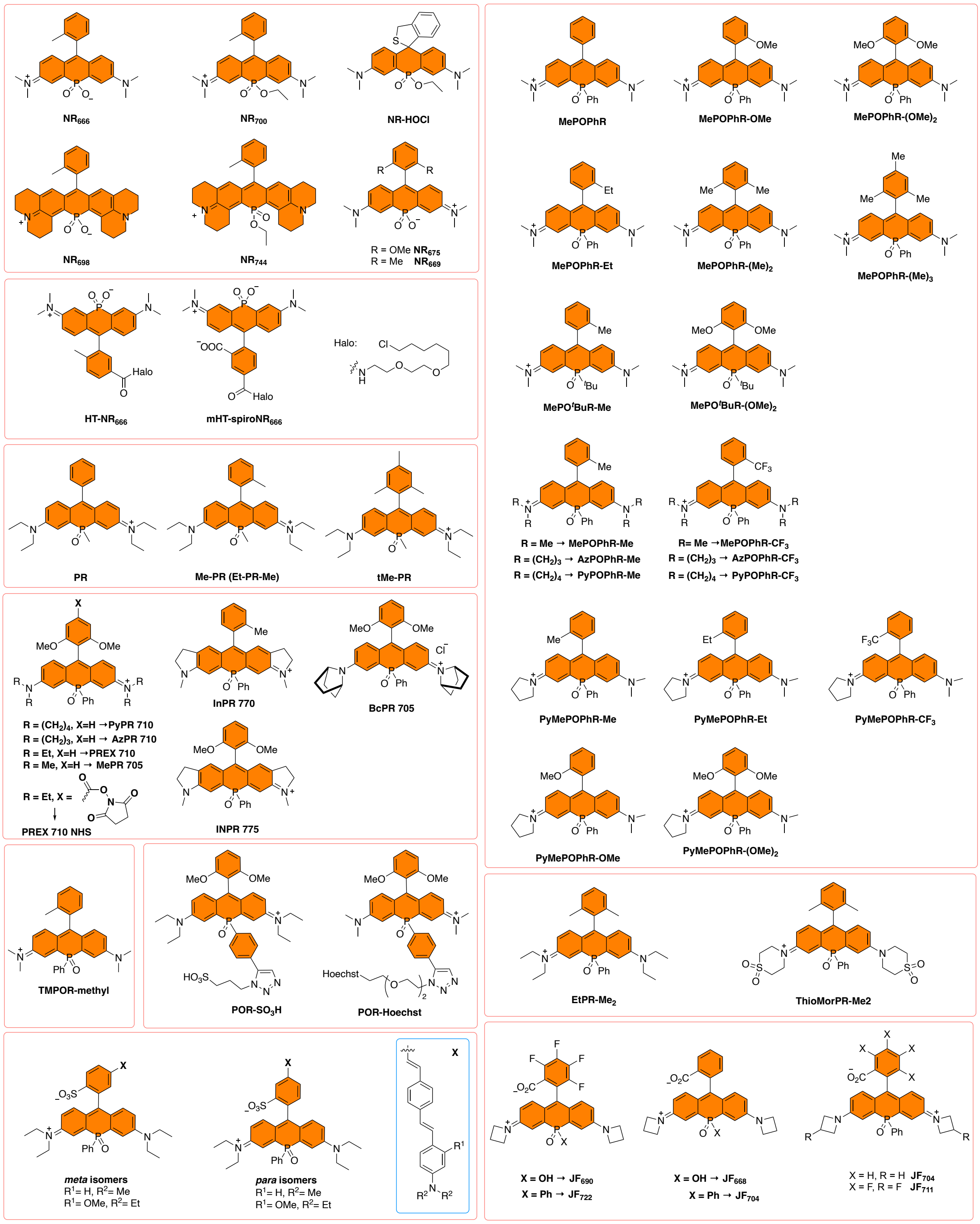

**Scheme 10.** Chemical structures of P-Rhodamine derivatives.

2020 was the start of the golden era for phosphine oxide containing rhodamine dyes. Firstly Lv *et al.* reported a new phospha-rhodamine derivative with increased quantum yields.[81] In 2008, Foley *et al.* reported that upon photoexcitation, the formation of a twisted intramolecular charge transfer (TICT) excited state reduces the quantum yield. The rapid nonradiative decay of the charge separated TICT state to the ground state results in an emission of a photon. Rhodamines tend to form a TICT state in aqueous solutions, resulting in a moderate decrease in the quantum yield since strongly polar water molecules stabilize the charge separated TICT state. In addition, the highly reactive diradical nature of the TICT state may lead the dyes to undergo irreversible bleaching reaction leading to low photostability. Therefore, rhodamine dyes with a low tendency to form a TICT state should exhibit high quantum yields, longer fluorescence lifetimes, and higher photostability.[87] Studies show that the formation of the TICT state can be prevented by improving the ionization potentials (IP) of the amino auxochromes by utilizing geometrically constrained azacyclic substituents such as 7-azabicyclo[2.2.1]heptane, azetidine, and aziridine. The replacement of dimethylamino auxochrome with the azetidine ring showed enhancement in the quantum yields of rhodamine and silicon-rhodamine.[49][58] However, the quantum yield enhancement is less obvious for the red to NIR silicon-rhodamine compared to classical rhodamine. Lv *et al.* proposed that the negative inductive effect (-I effect) exerted by the electron-withdrawing groups can improve the IP values of amino auxochromes. Firstly, piperidine, diethylamino, azetidine, 4-methyl-1,4-azaphosphinane 4-oxide, and thiomorpholine 1,1-dioxide containing derivatives of classical rhodamine core were synthesized. The quantum efficiencies were calculated as 0.06, 0.31, 0.86, 0.92 and 0.99 (in PBS) respectively. These results proved that the phosphine oxide (P=O) and sulfone ($SO_2$) functionalized piperidine moieties improved the IP values of the amino groups and increased the quantum yields. To validate this promising result in NIR phospha-rhodamine cores, diethylamino (**EtPR-Me$_2$**) and thiomorpholine 1,1-dioxide (**ThioMorPR-Me$_2$**) bearing derivatives of phospha-rhodamine with phenyl phosphine oxide bridging moiety and 2,6-dimethylphenyl at 9-position were synthesized (Table 6:E15-16). **ThioMorPR-Me$_2$** exhibited a 3.7-fold increase in quantum yield with a blue shift of 25 nm compared to **EtPR-Me$_2$**. The results revealed that the -I effect utilized by the electron-withdrawing groups develops the IP values of amino auxochromes and decreases the TICT phenomenon.

Another important study was reported by Sauer *et al.* containing 14 symmetrical and 5 unsymmetrical phospha-rhodamine derivatives.[82] The tetramethyl rhodamine analogs with phenyl phosphine oxide as bridging moiety bearing phenyl group (**MePOPhR**), 2-methylphenyl group (**MePOPhR-Me**), 2-methoxyphenyl group (**MePOPhR-OMe**), 2,6-dimethylphenyl group (**MePOPhR-OMe$_2$**) , 2-ethylphenyl group (**MePOPhR-Et**), 2,6-dimethyl group (**MePOPhR-Me$_2$**) , 2,4,6-trimethylphenyl group (**MePOPhR-Me$_3$**) and 2-trifluoromethylphenyl group (**MePOPhR-CF$_3$**) at the **9** position were synthesized (Table 6:E17-24, Scheme 10). The trimethyl rhodamine analogs with tert-butyl phosphine oxide as bridging moiety bearing 2-methylphenyl group (**MePO$^t$BuR-Me**) and 2,6-dimethoxyphenyl group (**MePO$^t$BuR-OMe$_2$**) at the 9-position (Table 6:E25-26, Scheme 10) were also studied. All tetramethyl rhodamine derivatives showed NIR characteristics, and most had relatively high fluorescence quantum yields. Additionally, diazetidine rhodamine analogs with phenyl phosphine oxide as bridging moiety bearing 2-methylphenyl group (**AzPOPhR-Me**), 2-trifluoromethylpheny group (**AzPOPhR-CF$_3$**); and dipyrrolidine rhodamine analogs with phenyl phosphine oxide as bridging moiety bearing 2-methylphenyl group (**PyPOPhR-Me**), 2-trifluoromethylpheny group (**PyPOPhR-CF$_3$**) at the 9-position (Table 6:E27-28, Scheme 10) were synthesized. These results revealed that azetidine substitution had no significant impact on absorption and emission maxima for the electron-poor derivatives but caused a drastic increase in fluorescence quantum yields. The pyrrolidine substitution (**PyPOPhR-Me** and **PyPOPhR-CF$_3$**, Table 6:E29-30, Scheme 10) caused an overall 8 nm red-shift in absorption and emission maxima but led to nearly a 0.5-fold decrease in the fluorescence quantum yields. It should be highlighted that other than the previously reported phospha-rhodamine derivatives, a different synthetic approach was reported for the first time with the following route; Li-halogen exchange of the corresponding brominated triphenylphosphine oxide derivative, the introduction of the corresponding ester, followed by treatment with HCl. Using a similar synthetic approach, the unsymmetrical dimethyl pyrrolidine rhodamines with phenyl phosphine oxide as bridging moiety bearing 2-methylphenyl group (**PyMePOPhR-Me**), 2-ethylphenyl group (**PyMePOPhR-Et**), 2-trifluoromethylphenyl group (**PyMePOPhR-CF$_3$**), 2-methoxyphenyl group (**PyMePOPhR-OMe**) and 2,6-dimethoxyphenyl group (**PyMePOPhR-OMe$_2$**) at the 9-position (Table 6:E31-35, Scheme 10) were synthesized. The mono pyrrolidine introduction into the rhodamine core caused a red shift in absorption and emission maxima compared to the tetramethyl rhodamine analogs. For the analogs **PyMePOPhR-Me** and **PyMePOPhR-Et**, a slight decrease in quantum yields were observed compared to those of their tetramethyl analogs. A significant increase was observed in the fluorescence quantum yields for the analogs **PyMePOPhR-CF$_3$** and **PyMePOPhR-OMe** compared to their tetramethyl rhodamine derivatives. However, the introduction of mono pyrrolidine into the rhodamine core decreased the fluorescence quantum yield of the corresponding analog **PyMePOPhR-OMe$_2$**. Overall, this study revealed a series of NIR phospha-rhodamines which resulted in understanding the chemistry behind their photophysical properties via a robust and flexible synthetic method. Later in 2020, Stains' group further functionalized the Nebraska Red dye series with HaloTag ligands and synthesized **HT-NR$_{666}$** (Table 6:E36, Scheme 10) and **mHT-spiroNR$_{666}$** (Table 6:E37, Scheme 10) for live-cell imaging experiments.[83] Utilizing amide coupling, the haloalkene derivative of the phosphinate-containing tetramethylrhodamine derivative **NR$_{666}$**, named **HT-NR$_{666}$**, was realized since **NR$_{666}$** itself does not cross the cell membrane. **HT-NR$_{666}$** showed the ability to label recombinant HaloTag protein *in vitro*. To test **HT-NR$_{666}$**'s ability to label the membrane population of hOX2R selectively, Chinese Hamster Ovary (CHO) cells were transfected with **HT-NR$_{666}$**, followed by washing. Clear labeling of the cytosolic protein was observed, and cells were transfected with a HaloTag-EGEP construct which showed no labeling of cytosolic protein, validating that **HT-NR$_{666}$** selectively label cell surface proteins. To create a fluorogenic labeling reagent for no-wash imaging, a modified version of **HT-NR$_{666}$** at the 2′-position of rhodamine was synthesized, yielding a spirolactonized derivative called **mHT-spiroNR$_{666}$** which the HaloTag ligand was placed at the 4′-position. Labeling the HaloTag with **mHT-spiroNR$_{666}$** increased fluorescence intensity in the presence of 10% fetal bovine serum (FBS), demonstrating the potential of **mHT-spiroNR$_{666}$** as a turn-on label for HaloTag containing proteins in cell culture media containing 10% FBS. However, the live-cell imaging experiments to evaluate **mHT-spiroNR$_{666}$** as a fluorogenic probe failed since no cytosolic labeling or nonspecific binding to the membrane was observed since **mHT-spiroNR$_{666}$** didn't cross the cell membrane. Further investigations were performed by labeling CHO cells transfected with **HaloTag-hOX2R** in the presence of 10% FBS. A membrane-localized fluorescent signal was observed without washing, showing that **mHT-NR$_{666}$** can be used as a non-wash protein labeling reagent. HaloTag mechanics is a well-studied area in the literature that permits the covalent linkage of a fluorophore with a protein of interest. This study provided promising NR-based HaloTag-ligands for live-cell imaging.

Another development in the phospha-rhodamine arena in 2020 was reported by Yamaguchi's group, which concentrated on investigating the effects of amino group substitution of P-Rhodamines on the photophysical properties.[84] For this purpose, seven P-rhodamines with phenyl phosphine oxide bridging group were synthesized. The first set of POR dyes synthesized contained a 2,6-dimethoxyphenyl group at the 9-position and dimethylamine (**MePR 705**), azetidine (**AzPR 710**), pyrrolidine (**PyPR 710**), and 7ABH (**BcPR 705**) as the amine moieties (Table 6:E38-41, Scheme 10). In addition to these five PORs, indoline-based analogs containing fused five-membered rings were synthesized with a 2-methylphenyl group at 9-position, **InPR 770,** and a 2,6-dimethoxyphenyl group at 9-position, **InPR 773** (Table 6:E42-43, Scheme 10). The photophysical properties revealed that the introduction of cyclic and acyclic amine donors had a subtle effect on the absorption maxima creating a redshift in the order of increasing electron-donating ability of the amine moieties with comparable fluorescence quantum yields, excluding **BcPR 705,** which showed a 1.6-fold increase in fluorescence quantum yield and a much-broadened absorption band. The fused analogs showed significant, red-shifted absorption, whereas a large drop in the fluorescence quantum yields compared to the non-fused derivatives. Next, the stability of the dyes was investigated. It has been demonstrated that the doubly ortho-substituted aromatics at the 9-position provided steric protection of the 9-position, which ensured chemical stability at a biologically relevant pH range of 4 to 10. However, hydrolytic deamination was observed at higher pH values.[80] To examine the relative stability of PORs, studies under strongly basic conditions were carried out. Results revealed that the resistance to the basic hydrolysis of the POR dyes was related to the steric hindrance of the amino groups. The bulkier pyrrolidine groups protected the adjacent carbon atom from nucleophilic attacks, yielding **PyPR 710** as the most stable under basic conditions, followed by **MePR 705**, **AzPR 710,** and **BcPR 705**. The fused analog bearing *ortho*-tolyl group (**InPR 770**) underwent a nucleophilic attack of a hydroxide ion due to the insufficient steric protection of the 9-position and reversibly converted into its colorless C9-carbinol form. In contrast, a similar conversion was prevented for the fused analog bearing 2,6-dimethoxyphenyl group (**InPR 775**). **InPR 775** was also shown to be stable to the nucleophilic attack of biological thiols. The absorbance signal of **InPR775** did not decrease in the presence of 10 mM glutathione (GSH), whereas a 90% decrease was observed for **InPR 770** under the same conditions. Lastly, the performance of the PORs in bioimaging in living cells. All the dyes stained lysosomes exclusively, but **PyPR 710** also stained mitochondria. The highest brightness and selectivity of lysosome staining were observed for **BcPR 705**, associated with its high fluorescence quantum yield. Overall, this dye series showed good photostability, NIR-shifted absorption, and selective lysosome staining, indicating their promising utility in the area.

In 2020, Deng *et al.* reported an o-tolyl group bearing tetramethyl phospha-rhodamine containing methyl phosphine oxide moiety as the bridging group, which was synthesized using the bis-nucleophile - electrophile match strategy yielding **TMPOR-methyl** (Table 6:E44, Scheme 10).[85] Compared to the photophysical properties of its tetraethyl analog PR synthesized by Wang's group in 2015, **TMPOR-methyl** revealed a 6 nm blue shift; however, the fluorescence quantum yield was increased by a factor of 2.5 in PBS buffer (pH = 7.4).

In 2020, Lavis' group expanded the Janelia Fluor (JF) palette.[68] Four azetidine phospha-rhodamine derivatives with different phosphine oxide bridging moieties were included in the study. The fluorinated derivatives of phospha-rhodamines with $PO_2H$- and P(O)Ph- groups as bridging moieties were obtained by replacing the phthalic anhydride moiety with tetrafluorophthalic anhydride. This modification universally increased $K_{L-Z}$ and $\boldsymbol{\varepsilon}$ values with a red shift in $\boldsymbol{\lambda}_{max}$ for oxygen- and sulfur-containing rhodamines; however, this modification was not generalizable for phospha rhodamines. The HaloTag ligands from the selected phospha-rhodamine dyes, **$JF_{690}$** (Table 6:E47, Scheme 10) and **$JF_{722}$** (Table 6:E48, Scheme 10), were synthesized. The compounds selectively labeled HaloTag fusions in cells. Among these dyes, introducing a fluorine substituent on each azetidine ring of **$JF_{722}$** yielded 3-fluoroazetidinyl **$JF_{711}$** (Table 6:E49, Scheme 10), which showed a further improvement in fluorescence quantum yield and a fivefold increase on binding HaloTag. The cell experiments revealed that **$JF_{711}$** derivatives were useful for the experiments where high brightness and low background were crucial; however, the $JF_{722}$ derivatives were better for live-cell applications.

In 2021 Gonzalez *et al.* utilized the phospha-rhodamine derivatives for voltage imaging.[86] Four tetraethyl phospha-rhodamine derivatives with methylphosphine oxide as the bridging moiety were synthesized (**EtPOMeR-*meta*$X_1SO_3^-$**, **EtPOMeR-*meta*$X_2SO_3^-$**, **EtPOMeR-*para*$X_1SO_3^-$** and **EtPOMeR-*para*$X_2SO_3^-$**) to serve as phosphine oxide rhodamine voltage reporters (**poRhoVRs**, Table 6:E50-55, Scheme 10). All derivatives displayed excitation and emission profiles above 700 nm, were localized to the membrane and displayed differing cellular brightness in HEK cells with good voltage sensitivity. **EtPOMeR-*meta*$X_2SO_3^-$** derivative showed spontaneous action potentials in rat hippocampal neurons and enabled all-optical electrophysiological manipulations with ChR2 (figure 8). This derivative further provided the first direct visualization of voltage dynamics alongside simultaneous $Ca^{2+}$ imaging in the retina.

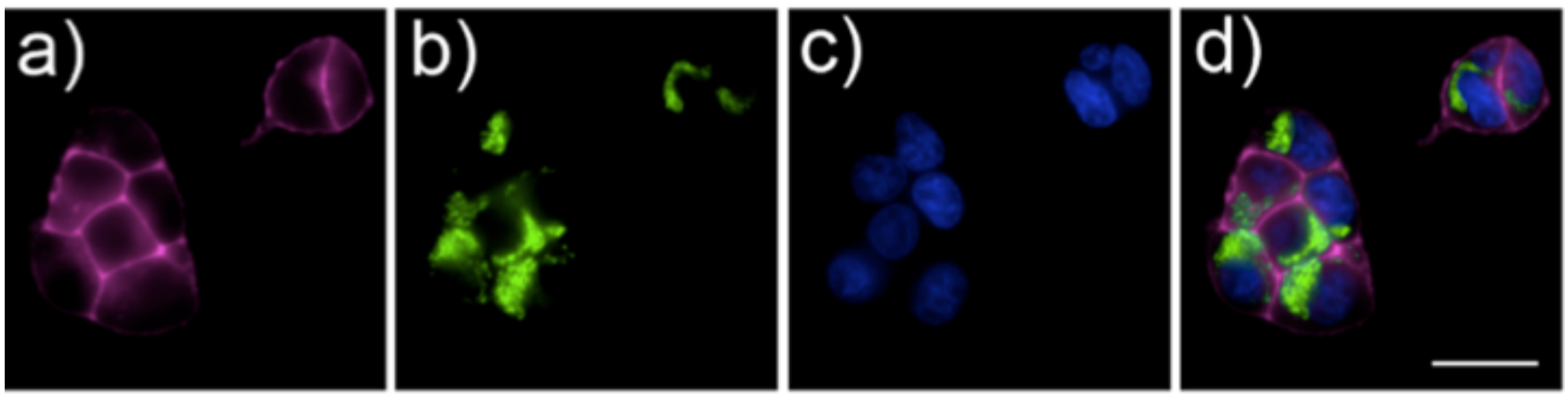

**Figure 8.** Cellular characterization of EtPOMeR-*para*$X_1SO_3^-$ indicators in HEK cells. Widefield, epifluorescence images of (a) EtPOMeR-*para*$X_1SO_3^-$ (1 μM) in HEK cells. Cells are counter-stained with (b) rhodamine 123 (1 μM) and (c) Hoechst 33342 (1 μM) to visualize mitochondria and nuclei, respectively. (d) An overlay of EtPOMeR-*para*$X_1SO_3^-$, rhodamine 123, and Hoechst 33342. Scale bar for parts a−d is 20 μm. Reproduced with permission [86] copyright 2021, American Chemical Society

The last work on phospha-rhodamines in 2021 was reported by Yamaguchi's group.[41] They synthesized a series of off-on-off type pH probes, including fluorescein, rhodamine, and rhodol cores functionalized with 2,6-dimethoxyphenyl at the 9-position bearing phenylphosphine oxide as the bridging moiety. For this series of NIR phospha-xanthene dyes, 4-ethynylphenyl group was installed on the phosphorous atom. This newly inserted terminal alkyne was available to be modified with various azide-containing molecules of interest via copper-catalyzed azide-alkyne cycloaddition (CuAAC). CuAAc reaction of the alkynylated phospha-rhodamine derivatives was POR-Hoechst (Hoechst = bis-benzimide) (Table 6:E56, Scheme 10) for nucleus targeting and POR-$SO_3H$ (Table 6:E57, Scheme 10). Fluorescence images of A431 cells showed that selective labeling of the target organelle with POR-Hoechst was achieved successfully.

## 3.3. Pyronine Derivatives

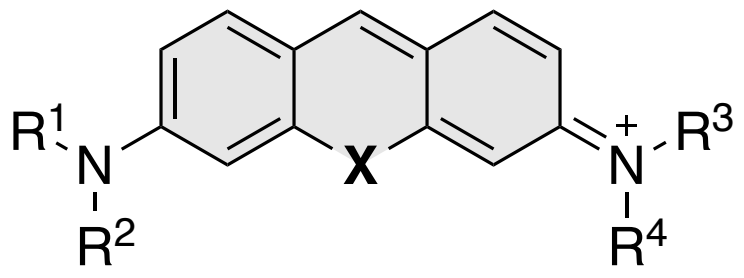

**X:** $SiR_2$, $GeR_2$, $SnR_2$, $SO_2$, $BR_2$

In classical xanthene dyes, oxygen occupies the 10th position, replacing oxygen with any other element, such as the pnictogens (group 4), the chalcogens (group 5), or boron have seen significant interest due to several orbital interactions leading to a decrease in the band-gap. This bathochromic shift is crucial for imaging and treating deep tumors. Up to the present, carbon, silicon, germanium, tin, sulfur, and boron substituted pyronine scaffolds (Scheme 11) have been proposed with red-shifted emission, and a wide range of biological applications was reported where classical pyronines were not effective.

Silicon substituted pyronine (**Si-Pyronine, SiP**, Table 7:E1, Scheme 11) is one of the classes of pyronine, and the Fu group reported the first silicon incorporation into the commercial xanthene core.[88] They showed that single silicon atom substitution resulted in a simultaneous decrease in LUMO energy level and increased HOMO energy level due to σ* (silicon atom) π* (adjacent carbon atom) interaction that resulted in a 90 nm bathochromic shift. Besides the red-shift, narrow absorption with a 2-fold increase in molar absorption coefficient was observed compared to classical pyronine, which bears oxygen atom at the 10th position. It is crucial to mention that polar protic solvents resulted in lower quantum fluorescence yield. This observation was attributed to the electron-deficient character of the carbon atom at position 9. Thus, increasing the bulkiness at this position became one of the endeavors in the community for reaching higher quantum yields.

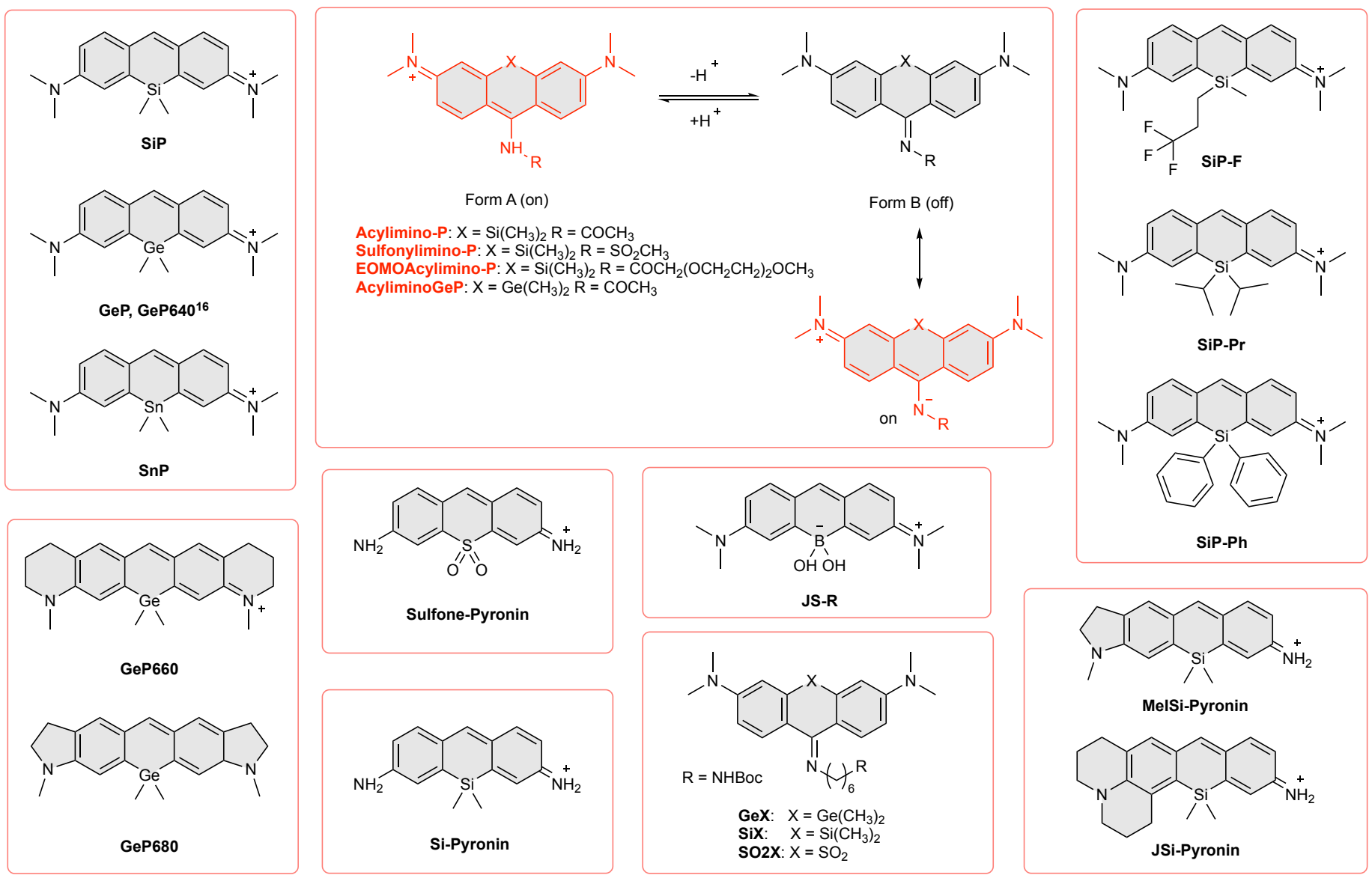

**Scheme 11.** Chemical structures of Pyronine derivatives.

Koide *et al.* synthesized silicon, germanium **(Ge-Pyronine, GeP,** Table 7:E2, Scheme 11), and tin (**Sn-Pyronine, SnP,** Table 7:E3, Scheme 11) substituted pyronines (Scheme 12).[43] Among them, germanium and tin derivatives are the first examples in the pyronine class and synthesized through metal-halogen exchange reaction of bis(2-bromo-4-dimethylaminophenyl) followed by the addition of corresponding dialkyl dichlorides of silicon, germanium, and tin. Then, oxidation of fluorophores was performed with chloranil to yield **SiP** (62% yield), **GeP** (44% yield), and **SnP** (21% yield), respectively. Their photophysical properties were evaluated in PBS buffer at pH 7.4, and the results are summarized in Table 7:E1-3, Scheme 11). It was noted that bathochromic shifts were almost identical. This observation could be ascribed to Yamaguchi and coworkers' theoretical work in which silicon, germanium, and tin in metalloles affect the LUMO energy levels of the π system to almost the same extent. Besides that, they reported the instability of SnP in air or DMSO and that it slowly decomposes to the corresponding xanthone. Tracy's group synthesized tetra phenyl substituted metallacyclopentadiene of silicon, germanium, and tin. It revealed that dimethyl tin derivatives were the least stable in the solution among the other derivatives and subject to decomposition in acetonitrile. Experimental evidence suggested an acid-catalyzed ring-opening reaction was in effect.[89]

Horváth *et al.* designed and synthesized **9-iminopyronin** derivatives (**Acylimino-P, Sulfonylimino-P, EOMOAcylimino-P,** Scheme 12) and performed theoretical and experimental analyses.[90] Photophysical properties were evaluated (Table 7:E4-6, Scheme 11) , revealing that synthesized 9-acylimino and 9-syl pyronines show significant Stokes shift up to 210 nm in physiological conditions. While, inulfon acidic frm **(A)**, fluorophore shows about 20 nm Stokes shift. Furthermore, they found that fluorophores did not or barely emit in aprotic solvents, whereas in protic solvents, they showed strong emission. This relationship is unusual because emission decreases dramatically or quenches in protic solvents due to the intermolecular solvent-solute H-bonding induced PET process.[6] Based on their theoretical work, they proposed that this H-bonding stabilized intramolecular charge-transfer (ICT) in one of the basic forms **(B)** of the pyronine, which was the fluorogenic state with enormous Stokes shift in the 9-(monoalkylimino) pyronine case. On the other hand, they found out that the locally excited (LE) character of **B** rather than ICT was observed in aprotic solvents, the non-emissive state. Furthermore, their calculations show that excitation of the **B**, followed by the elongation of the C9-N bond and solvent reorganization, is closely related to significant Stokes shifts.[91]

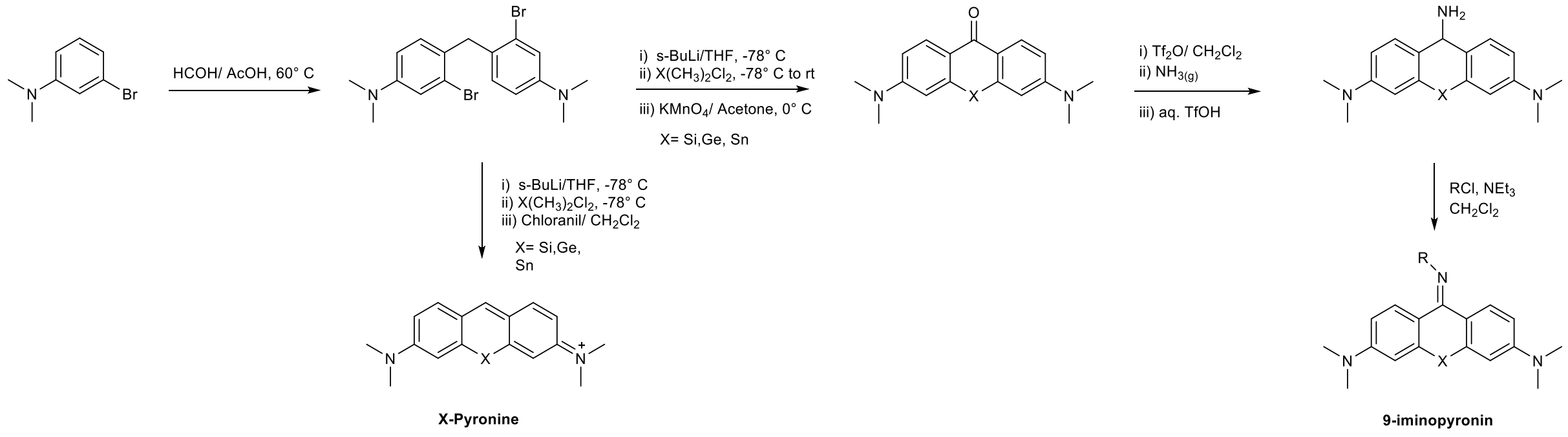

**Scheme 12.** General synthetic route for pyronine derivatives.

In 2019, Horváth *et al.*[92] synthesized another set of **9-iminopyronin** derivatives containing germanium "**AcyliminoGeP**" at position 10 as adapted from their previous work.[90] (Table 7:E7, Scheme 11). They proposed synthesized **9-iminopyronin** derivatives could be used as a fluorescent probe at alkaline pHs. The fluorophores existed in two acid-base forms, which are spectroscopically distinct, with $pK_a$s ranging between 6.8 and 8.0. This allowed dual emission ratiometric response with highly sensitive $H^+$ analysis (detection of pH changes on the order of 0.06 pH units). Selected sensors were also successfully utilized as sensors for haloalkane dehalogenase (HLD Lin B).

In 2016, Nie and coworkers[93] synthesized novel **GePyronine** dyes, **GeP640**, **GeP660**, and **GeP680** (Table 7:E2,8-9, Scheme 11) adapted from Koide and coworkers' work.[43] They found that **Ge-Pyronines** have a great affinity toward thiols (GSH, Cys, Hcy), ascribed to the partial positive charge density at position 9, which is significantly higher than parent O-pyronine (+0.318 vs. -0.008). Thus, $10^3 \sim 10^4$ greater affinity of **Ge-Pyronines** toward GSH was observed compared to O-Pyronine. This difference motivated them to utilize these reversible on-off probes in ROS monitoring. HeLa and HL-60 cell bioimaging experiments via **GeP640** showed good cell permeability, negligible cytotoxicity, and no photobleaching. Furthermore, their fast response (20-30 sec.) toward ROS makes them great candidates for real-time dynamic ROS imaging. It is also crucial to mention that **GeP**s are susceptible to the nucleophilic attack of hydroxy groups at position 9, with their pKa's ranging from 8.6 to 9.5 (much lower than O-Pyronine, pKa = 11.5). pH titration studies showed that the fluorescence intensity of **Ge640** was decreasing above pH at physiological conditions. The same group synthesized several **Si-Pyronine** derivatives **SiP**, **SiP-F**, **SiP-Pr,** and **SiP-Ph** (Table 7:E1,10-12, Scheme 11) designed to monitor the changes in GSH concentration in living cells.[94] Synthetically, the final oxidation step, commonly achieved with p-chloranil, was substituted with UV-irradiation in open air since reduction by-products of p-chloranil interfere during preparative HPLC purifications. **SiP-Pr** was chosen for conducting live cell imaging due to having the highest pKa among synthesized **SiPs** which minimizes the interfering effects of hydroxyl group nucleophilic attacks. In addition, 80% cell viability (5µM for 24h), good cell permeability, and water-solubility were illustrated with **SiP-Pr**.

*Table 7. Selected features of Pyronine derivatives*

| Entry (E) | Fluorophore* | Absorption | | Emission | $\Phi_f$ | Intracellular Localization | Properties | Applications | Ref. |
|---|---|---|---|---|---|---|---|---|---|
| | | $\lambda_{max}$ (nm) | ε ($M^{-1}\cdot cm^{-1}$) | $\lambda_{max}$ (nm) | | | | | |
| 1 | SiP[a] | 634 | $1x10^5$ | 648 | 0.42[b] | n.d. | pH-dependent, | - | [43] |
| | | | | | | | Thiol and base sensitive | | |
| 2 | **GeP[a], GeP640** | 621 | nd | 634 | 0.40[b] | n.d. | pH-dependent, Thiol sensitive | In vitro real-time dynamic ROS imaging | [43,93] |
| 3 | **SnP[a]** | 614 | nd | 628 | 0.43[b] | n.d. | Not stable in air, in DMSO | - | [43] |
| 4 | **Acylimino-P**<br>**Basic form[c]**<br>**Acidic form[e]** | 387<br>658 | 1.7x10$^{4}$ | 597<br>678 | 0.56[d] | n.d. | pH-dependent<br>Large Stokes shift | - | [90] |
| 5 | **Sulfonylimino-P[c]** | 444 | 1.5x10$^{4}$ | 617 | 0.23[d] | n.d. | Large Stokes shift | - | [90] |
| 6 | **EOMOAcylimino-P**<br>**Basic form[c]**<br>**Acidic form[e]** | 400<br>666 | 1.6x10$^{4}$ | 596 | 0.47[d]<br>681 | n.d. | pH-dependent<br>Large Stokes shift | - | [90] |
| 7 | **AcyliminoGeP**<br>**Basic form[f]**<br>**Acidic form[g]** | 397<br>648 | 1.9x10$^{4}$<br>1.2x10$^{5}$ | 597<br>667 | 0.23[d] | n.d. | Water soluble / pH-dependent/ Large Stokes shift | - | [92] |
| 8 | **GeP660[a]** | 646 | nd | 665 | 0.44[d] | n.d. | pH-dependent, Thiol sensitive | - | [93] |
| 9 | **GeP680[a]** | 662 | nd | 682 | 0.32[d] | n.d. | pH-dependent, Thiol sensitive | - | [93] |
| 10 | **SiP-F[a]** | 637 | nd | 651 | 0.20[b] | n.d. | pH-dependent, Thiol and base sensitive | - | [94] |
| 11 | **SiP-Pr[a]** | 632 | nd | 652 | 0.21[b] | n.d. | pH-dependent, Thiol and base sensitive | Monitoring GSH changes in a living cell | [94] |
| 12 | **SiP-Ph[a]** | 649 | nd | 665 | 0.23[b] | n.d. | pH-dependent, Thiol and base sensitive | - | [94] |
| 13 | **JS-R[h]** | 611 | 1.x10$^{5}$ | 631 | 0.59[b] | n.d. | pH-independent (pH 5.5-11) Polyol selective | Sugar and Polyol Sensor | [95] |
| 14 | **GeX[i]** | 454 | 2.4x10$^{4}$ | 618 | 0.17[d] | Upon target moiety | Large Stokes shift, Cell permeable | multichannel applications | [96] |
| 15 | **SiX[i]** | 458 | 1.7x10$^{4}$ | 623 | 0.28[d] | Upon target moiety | Large Stokes shift, Cell permeable | multichannel applications | [96] |
| 16 | **SO2X[j]** | 509 | 1.3x10$^{4}$ | 647 | 0.15[d] | n.d. | Works pH < 5 | - | [96] |
| 17 | **Sulfone-Pyronin[k]** | 632 | 2.0 x10$^{4}$ | 671 | 0.14[l] | n.d. | Poor aq. stability | - | [97] |
| 18 | **Si-Pyronin[m]** | 580 | 4.8 x10$^{4}$ | 597 | 0.89[n] | n.d. | Works pH < 5 | - | [98] |
| 19 | **MeISi-Pyronin[o]** | 625 | 4.7 x10$^{4}$ | 651 | 0.27[p] | n.d. | Enhanced aq. stability, Optimum pH <7 | Smart on-off probe theranostic drug delivery system candidate | [99] |
| 20 | **JSi-Pyronin[o]** | 620 | 4.1 x10$^{4}$ | 640 | 0.42[p] | n.d. | Enhanced aq. stability, Optimum pH<8 | Smart on-off probe theranostic drug delivery system candidate | [99] |

[a]Photophysical properties were measured in PBS at Ph 7.4. [b]For determination of the fluorescence quantum yield ($\Phi_f$), cresyl violet in MeOH ($\Phi_{fl}$ = 0.54) was used as a fluorescence standard. [c]Photophysical properties were measured in MeOH. [d]Absolute fluorescence quantum yield was given. [e]1% HCl in MeOH (v/v). [f]NaOH (*c* = 1 mmol dm$^{-3}$) in MeOH. [g]HCl (*c* = 1 mmol dm$^{-3}$) in MeOH. [h]In HEPES at pH 7.4. [i]10% Methanol in PBS at pH 7.4. [j]In 10% $CH_3CN$ in $H_2O$. [k]10% TFA in $CH_3CN$. [l]Sulfoindocyanine dye Cy 5.0 as standard ($\Phi_{fl}$=20% in PBS). [m]In NaOAc buffer at pH 4.8. [n]SR101 as a standard ($\Phi_{fl}$ = 95% in EtOH). [o]In PBS, 100 mM + 150 mM NaCl at pH 7.5. [p]SR101 was used as a fluorescence standard ($\Phi_{fl}$ = 0.95) in EtOH. n.d.: not detected.* all the fluorophores listed above are emphasized with bold font in the text.

Shimomura *et al.* synthesized sugar sensors based on **Boron-Pyronine (B-Pyronine, BP)** named JoSai-Red **(JS-R,** Table 7:E13, Scheme 11)**.**[95] This **B-Pyronine** dye revealed red-shifted emission (60nm) compared to O-pyronine. This phenomenon was ascribed absence of orbital interaction between the sp$^3$ boron unit and adjacent carbon, resulting in a lower LUMO level. **JS-**

R contains two hydroxy groups on the boron atom, making it possible to bind to the sugars or catechol. Upon sugar binding to **JS-R**, a red shift of about 25 nm was observed, where catechol binding activates the PET mechanism and quenches the fluorescence emission dramatically. They also stated that sugar sensors based on boronic acids work at a wide pH range, where **JS-R** gives the same response between pH 5.5-11, which ascribed to the $sp^3$ hybridization at the boron center as opposed to sensors with $sp^2$ boron, which allows hydroxy coordination and make them pH-sensitive.

Butkevich *et al.* synthesized **GeX, SiX, SO2X (9-iminopyronine)** probes (Table 7:E14-16, Scheme 11) with their pepstatin A, triphenylphosphonium (TPP), jasplakinolide ligand derivatives for targeting lysosome, mitochondria, and F-actin respectively.[96] These derivatives showed a large Stokes shift (>140 nm), allowing the use of multichannel applications as Horváth and coworkers proposed as the potential of 9-iminopyronine scaffolds.[90] **GeX** and **SiX** were preferred for investigating the targeting ability with the selected ligands under single-color live-cell STED nanoscopy and live-cell multicolor confocal and STED imaging. The probes were shown to be cell-permeable, compatible with STED nanoscopy, and non-toxic at recommended concentrations except **SO2X** derivative since it was shown to be fluorescent only at pHs lower than 5, which in turn is applicable in the intracellular acidic environment such as lysosome (figure 9).

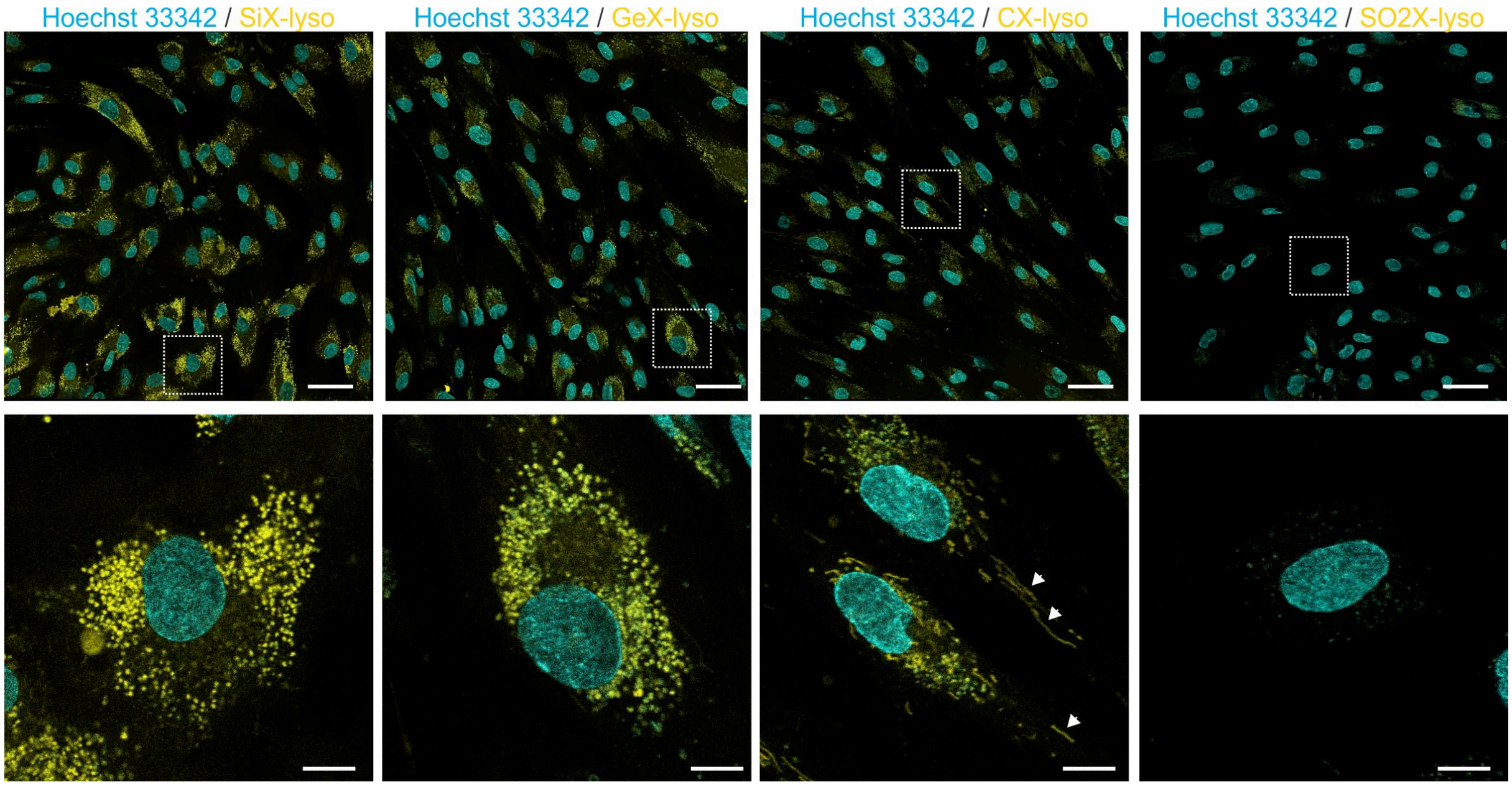

**Figure 9.** Confocal microscopy images of human fibroblasts stained with lysosomal probes. The upper row shows the large field of view of stained human fibroblasts. Dotted squares correspond to the zoomed-in areas shown in the lower row. White arrows indicate off-target mitochondrial staining of the CX-lyso probe. Cells were stained with 2 μM probes **SiX-lyso, GeX-lyso, CX-lyso, SO2X-lyso,** and Hoechst 33342 (0.1 μg/ml) for 1 h at 37°C in the growth medium, washed twice with HBSS and imaged immediately in the growth medium. Scale bars are 50 μm (upper row) and 10 μm (lower row). Reproduced with permission [96] copyright 2017, American Chemical Society

Dejouy and coworkers studied the stability and spectral behavior of **Sulfone-Pyronin** (Table 7:E17, Scheme 11), which contains primary amines instead of secondary or tertiary amines, that allowed optical tunability.[97] In the semi-preparative RP-HPLC purification, they observed that **Sulfone-Pyronin** turns to the hydroxyl substituted product. This was ascribed to the strong electron-withdrawing ability of sulfone. Electron-deficient carbon atom at position 9 is susceptible to hydroxyl attack, as mentioned in the case of **GePs.**[93] To minimize this hydroxyl attack and make the fully π-conjugated skeleton dominant, 10% TFA in $CH_3CN$ was used as a proper solvent for photophysical characterization (Table 7:E17, Scheme 11). The reactivity with hydroxyl groups at physiological conditions hindered the use of these probes in further *in-vitro* studies. The same group[98] adopted a new synthetic approach from Egawa and coworkers[25] to **Si-Pyronin** (Table 7:E18, Scheme 11) (primary amine instead of secondary/tertiary amine) due to the time-consuming and difficult purification of the final probes and low solubility of the intermediates (Scheme 13). They found that **Si-Pyronin** is unstable in psychological conditions, and pH titration showed that hydrolytic stability could be achieved at pH<5 due to electron deficiency at the carbon atom at position 9.

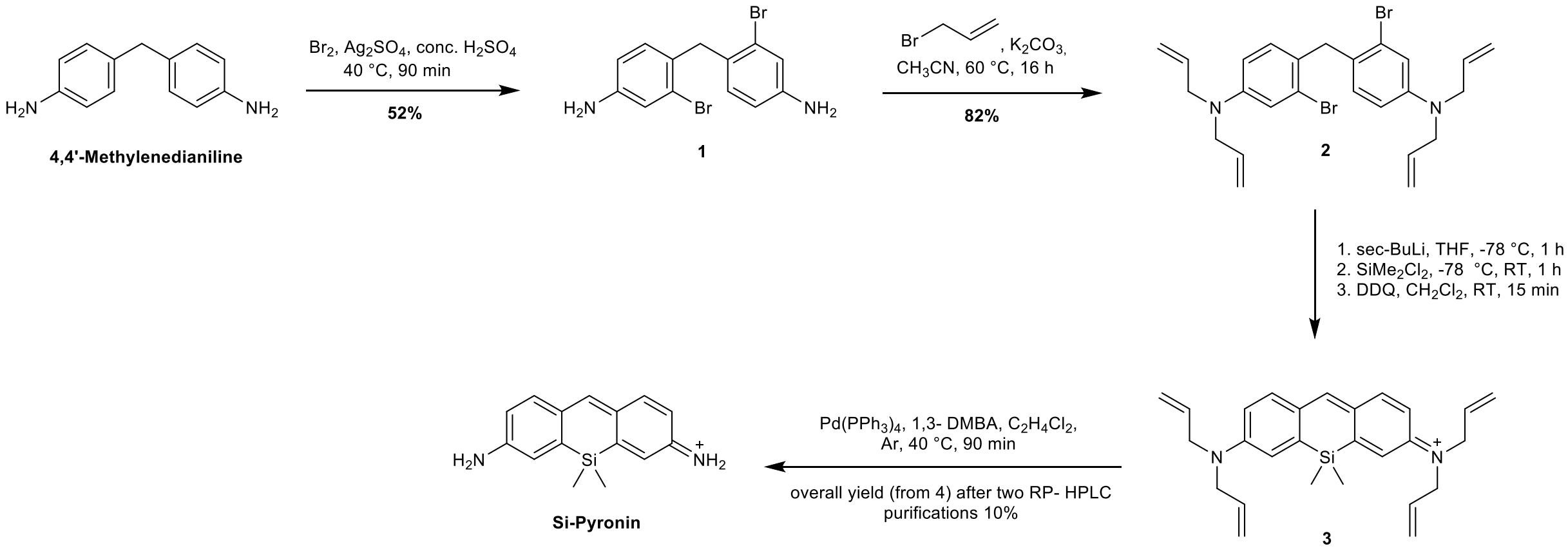

**Scheme 13.** Alternative synthetic route for Si-pyronin

In the light of poor aqueous stability of the primary amines at the **Si-** and **$SO_2$-** Pyronines at physiological pH, Dejouy and coworkers designed unsymmetrical **Si-pyronines** (**MeISi-Pyronin, JSi-Pyronin**, Table 7:E19-20, Scheme 11**),** which consist of tertiary amine on one side (*N*-methylindoline and julolidine) and primary amine on the other.[99] Consequently, electron donation of the tertiary amine led **MeISi-Pyronin** and **JSi-Pyronin** to have optimal stability in an aqueous medium up to pH=7 and pH=8, respectively (its primary amine analog does up to pH:5) as well as keeps the rigidity of the xanthene skeleton for higher fluorescence quantum efficiency. Secondly, since primary amine sides at **MeISi-Pyronin** and **JSi-Pyronin** can be modified with targeting units, these systems were called "smart" probes.

# 4. Rhodol Derivatives

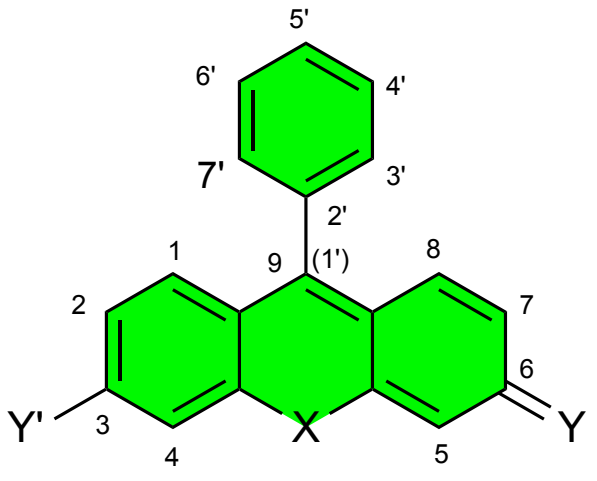

Y= —O , Y'= —$NR_2$
X= —O, —$SiMe_2$, —Se, —$CMe_2$, —POR

The "rhodol" term is used for molecules with oxygen in the Y position and an amine in Y' position in the xanthene family. Unlike rhodamines, having two different atoms in the Y and Y' positions allows the delocalization of the charge along these systems. With this property, rhodols can have two forms; neutral and charged due to resonance. Although rhodols have been around for a long time, they only have been used widely in the last decade as fluorescent imaging and sensing platforms.[100–102] Similar to the rhodamines, the rhodols can possess silicon, phosphorous, and selenium in the bridging moiety with enhanced π- conjugation around the skeleton instead of oxygen or carbon.

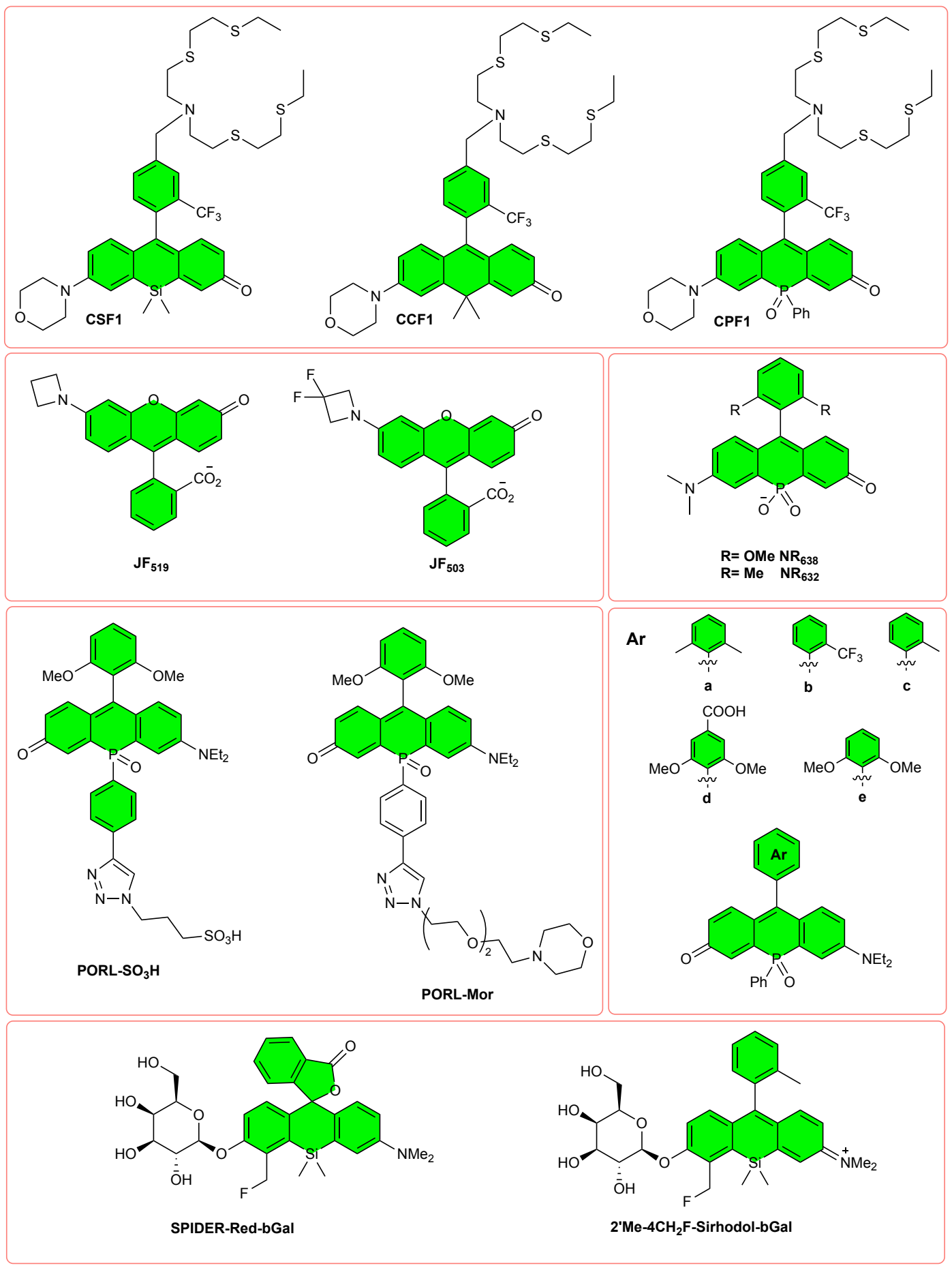

Scheme 14. Chemical structures of Rhodol derivatives.

In 2018, Chang and coworkers introduced copper sensors based on rhodol core.[103] The 10 position was modified with different groups ($SiMe_2$, P(O)Ph, $CMe_2$) for tuning the spectroscopic properties and yielded emission profiles ranging from orange to deep red. **CCF1** possessed $CMe_2$ group on X, oxygen in Y, morpholine on Y', $CF_3$ unit on 9 position. The trifluoromethyl group at position 2' was used to enhance the emission and dynamic range by diminishing the rotational motion of the bond between position 9 and $CF_3$ carbon. Similar to this design strategy, phosphorous (**CPF1**) and silicon (**CSF1**) derivatives were synthesized. The photophysical properties of these derivatives are summarized in Table 8 (E1-3, Scheme 14). All fluorophores were modified with the same thioether-based ligand. In the buffered physiological pH and $Cu^+$ medium, **CCF1** showed 5-fold, **CPF1** showed 7-fold, and **CSF1** showed a 17-fold increase in turn-on responses. This change could be explained by the variations in rotational freedom of the binding of $Cu^+$. Moreover, as the HUMO-LUMO gap decreases, the nonradiative relaxation pathways become more notable, and lower quantum yields are observed. The cell studies were performed in HEK 293T cells with copper depletion and/or supplementation (figure 10). This study revealed that changes in the 10 position of the rhodol derivatives could broaden the fluorescent emission spectra while preserving the recognition and reactivity towards the analyte.

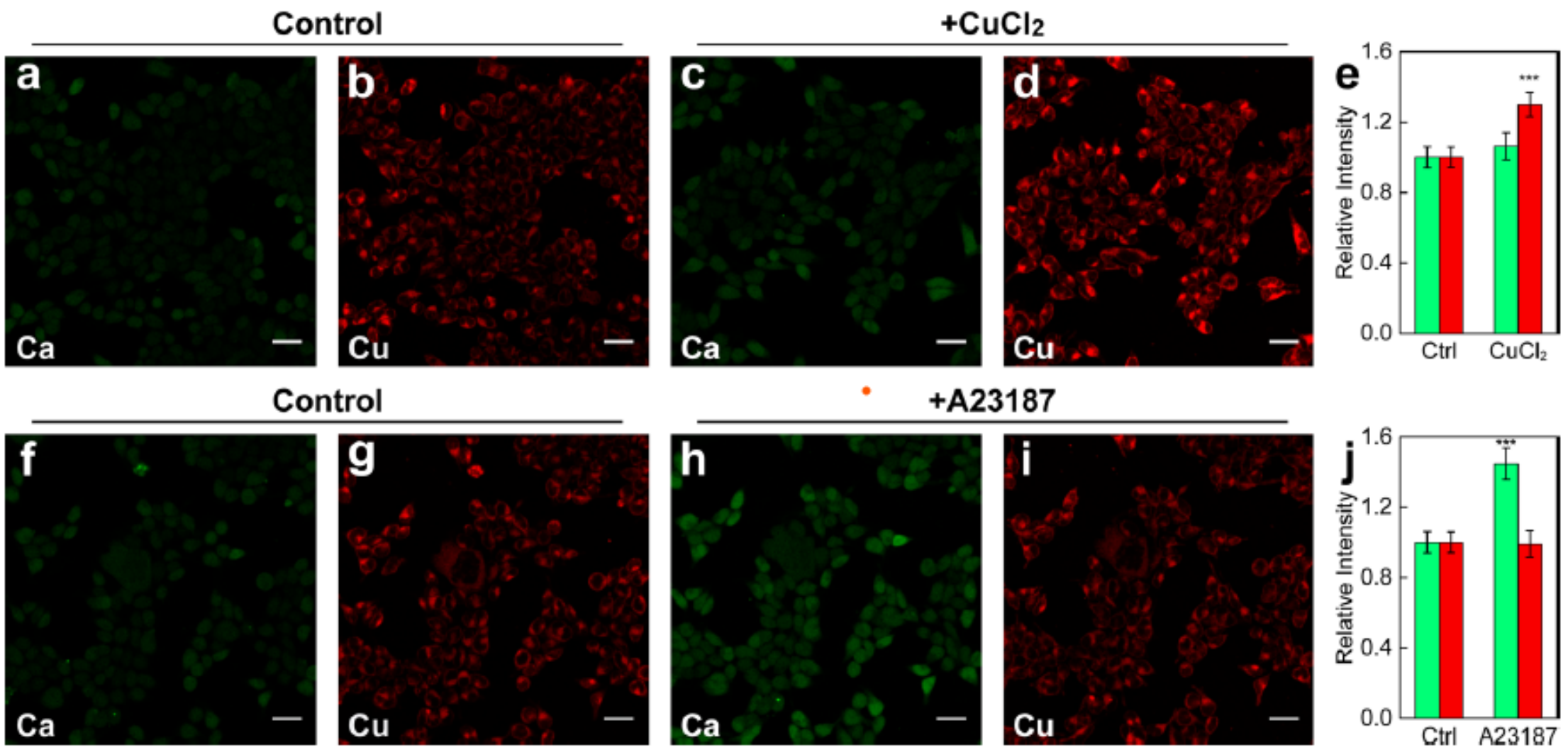

Figure 10. Dual-channel fluorescence imaging of calcium and copper pools in live HEK 293T cells with Calcium Green-1 in the green channel (a, c, f, h) and CPF1 in the red channel (b, d, g, i). Control cells (a, b) and cells treated with 100 μM CuCl2 for 12 h (c, d) were incubated with both dyes in HBSS and imaged; quantification is shown in e. Cells incubated with both dyes in HBSS prior to (f, g) and after (h, i) treatment of 1 μM calcium ionophore A23187 were imaged; quantification is shown in j. Green bars represent the Calcium Green-1 channel, and red bars are CPF1 channel. Scale bars: 40 μm. Data were normalized to control cells and shown as average ± s.d. (n = 4). ***P ≤ 0.001; two-tailed Student's t-test. Reproduced with permission [103] copyright 2018, American Chemical Society

In 2017, Yamaguchi's group introduced selective conversion of P=O-bridged rhodamines to related rhodols (Table 8:E4-7, Scheme 14).[37] The synthesis of the rhodols is a notorious challenge due to the unsymmetric nature of the molecule, yet their conversion from the relative rhodamines was shown to be straightforward by the hydrolysis in a basic aqueous solution. Additionally, they pointed out that steric protection at the C9-position is vital for transforming the rhodol from the rhodamine, where PORL-a, which bears the o-tolyl group on the C9-position, was not observed. It was shown that the absorption of the P=O-bridged rhodols shifted from 545 to 670 nm, and the emission shifted from 628 to 698 nm with the increasing solvent polarity (Toluene: 545 nm, PBS: 670) nm. The fluorescence quantum yields are also affected by solvent type drastically. For example, while **Φf** was 0.77 in MeCN, in PBS buffer, **Φf** was found to be 0.11 (Table 8:E1-4, Scheme 14). These solvent effects, which are quite specific to the P=O-bridged rhodols, were attributed to the decrease in the π- π* transitions in the polar medium. Theoretical calculations revealed that solvent effects, specifically the solvents' hydrogen bonding ability, cause a significant change in the LUMO level, and they attributed this effect to increased σ*- π * orbital interaction.

Quite recently, Yamaguchi and coworkers developed alkyne-modified phospha-rhodamine, phospha-rhodol, and phospha-fluorescein on-off dyes for late-stage functionalization and lysosomal imaging.[41] The rhodol and fluorescein were synthesized according to their previous work.[104] The rhodol core **PORL** (Table 8:E8, Scheme 14) exhibited absorption at 670 nm and emission at 698 nm with a **Φf** of 0.11. Functionalization through click chemistry was shown to be straightforward and demonstrated that the introduction of the triazole ring did not perturb the electronic structure of the core. When the terminal alkyne of PORL-CCH was modified with morpholine containing azide **PORL-Mor** (Table 8:E9, Scheme 14) was attained. With **PORL-Mor**, the cell viability was determined by MTT assay derivate, and no dark toxicity was observed up to 5μM. Furthermore, **PORL-Mor** showed selective accumulation and labeling of the target organelle lysosome.

In 2019 Stains' group developed phosphinate-containing rhodol and fluorescein derivatives for imaging probes based on the Nebraska Red scaffold[40], in which rhodol derivatizations of the corresponding rhodamines were modified with Yamaguchi's method.[37] Photophysical properties are summarized in (Table 8:E10-11, Scheme 14). Compared to the relative rhodamines (Nebraska Red derivatives), the rhodols displayed around 30 nm blue-shift for both absorption and emission maxima. These new rhodamines, rhodol, and fluorescein derivatives of Nebraska Red were robust and highly resistant to photobleaching.

*Table 8. Selected features of Rhodol derivatives*

| Entry (E) | Fluorophore* | Absorption | | Emission | $\Phi_f$ | Intracellular Localization | Properties | Applications | Ref. |
|---|---|---|---|---|---|---|---|---|---|
| | | $\lambda_{max}$ (nm) | ε ($M^{-1}\cdot cm^{-1}$) | $\lambda_{max}$ (nm) | | | | | |
| 1 | CCF1 | 516 | 14100 | 608 | 0.048[e] | n.d. | - | Copper sensor | [103] |
| 2 | CSF1 | 568 | 30000 | 638 | 0.004[e] | n.d. | - | Copper sensor | [103] |
| 3 | CPF1 | 569 | 17600 | 679 | 0.00018[e] | n.d. | - | Copper sensor | [103] |
| 4 | PORL-b | 670[c] | 66100 | 698 | 0.11 | n.d. | - | - | [37] |
| 5 | PORL-c | 666[c] | 49900 | 692 | 0.11 | n.d. | - | - | [37] |
| 6 | PORL-d | 671[c] | 60600 | 701 | 0.11 | n.d. | - | - | [37] |
| 7 | PORL-e | 673[c] | 50200 | 699 | 0.10 | n.d. | - | - | [37] |
| 8 | PORL | 670[g] | 66100 | 698 | 0.11 | n.d. | Off-on-off type probe | imaging | [41] |
| 9 | PORL-Mor | n.d. | n.d. | n.d. | n.d. | Lysosome | Off-on-off type probe | imaging | [41] |
| 10 | $NR_{638}$ | 638[c] | 55630 | 662 | 0.21 | n.d. | - | Bioprobe | [40] |
| 11 | $NR_{632}$ | 632[c] | 63970 | 655 | 0.26 | n.d. | - | Bioprobe | [40] |
| 12 | SPiDER-Red-βGal | 610[d] | - | 638 | - | Cytosol | selective toward LacZ(+) cells | Activable probe | [105] |
| 13 | 2'Me-4$CH_2$F-Sirhodol-βGal | 612, 528 | - | 630 | 0.28 | n.d. | - | - | [105] |

[a] Photophysical measurements were done in 10 mM HEPES, pH 7.3 solution. [b] Maximal extinction coefficients measured in 0.01% $Et_3N$ solution. [c] Measured in pH=7.4 PBS buffer and 1%(v/v) DMSO used as co-solvent. [d] pH=7.4 PBS buffer with 0.1% DMSO. [e] Cresyl violet in methanol ($\Phi_f$=0.54) was used as standard. [f] Measured in 200 mM pH=2.0 sodium phosphate buffer containing 0.1% DMSO as cosolvent. [g] measurements done in pH=7.4 HEPES buffer containing 0.1% DMSO as a cosolvent. n.d.: not detected.* all the fluorophores listed above are emphasized with bold font in the text.

Urano group synthesized rhodol derivative (**SPiDER-βGal**) to detect *lac*Z-positive cells in 2016.[106] As a further study, the group synthesized and examined a new silicon-substituted rhodol derivative for the same purpose in 2017.[105] Among these derivatives, **2'Me-4$CH_2$F-Sirhodol-βGal** exhibited absorption maxima at 612, and 528 nm and emission at 630 nm with a **$\Phi$f** of 0.28; (Table 8:E13, Scheme 14) and 2'COOH-4CH2F-Sirhodol-**β**Gal (**SPiDER-Red-βGal,** (Table 8:E12, Scheme 14)) showed absorption maxima around 610 nm and emission around 638 nm. **SPiDER-Red- βGal** was designed as an on-off probe, utilizing a common spiro-lactone system. Upon interaction with **β**-galactosidase (**βGal**), which is over-expressed in LacZ-positive cells, a reactive quinone methide derivative forms, and upon reaction with the nucleophiles in the media (proteins, $H_2S$, water, etc.), the highly fluorescent probe was attained within the immobilized in the cell. This approach and emission in the red region allowed the use of **SPiDER-Red- βGal** to visualize different **β**Gal expressed cell types in combination with GFP markers (figure 11).

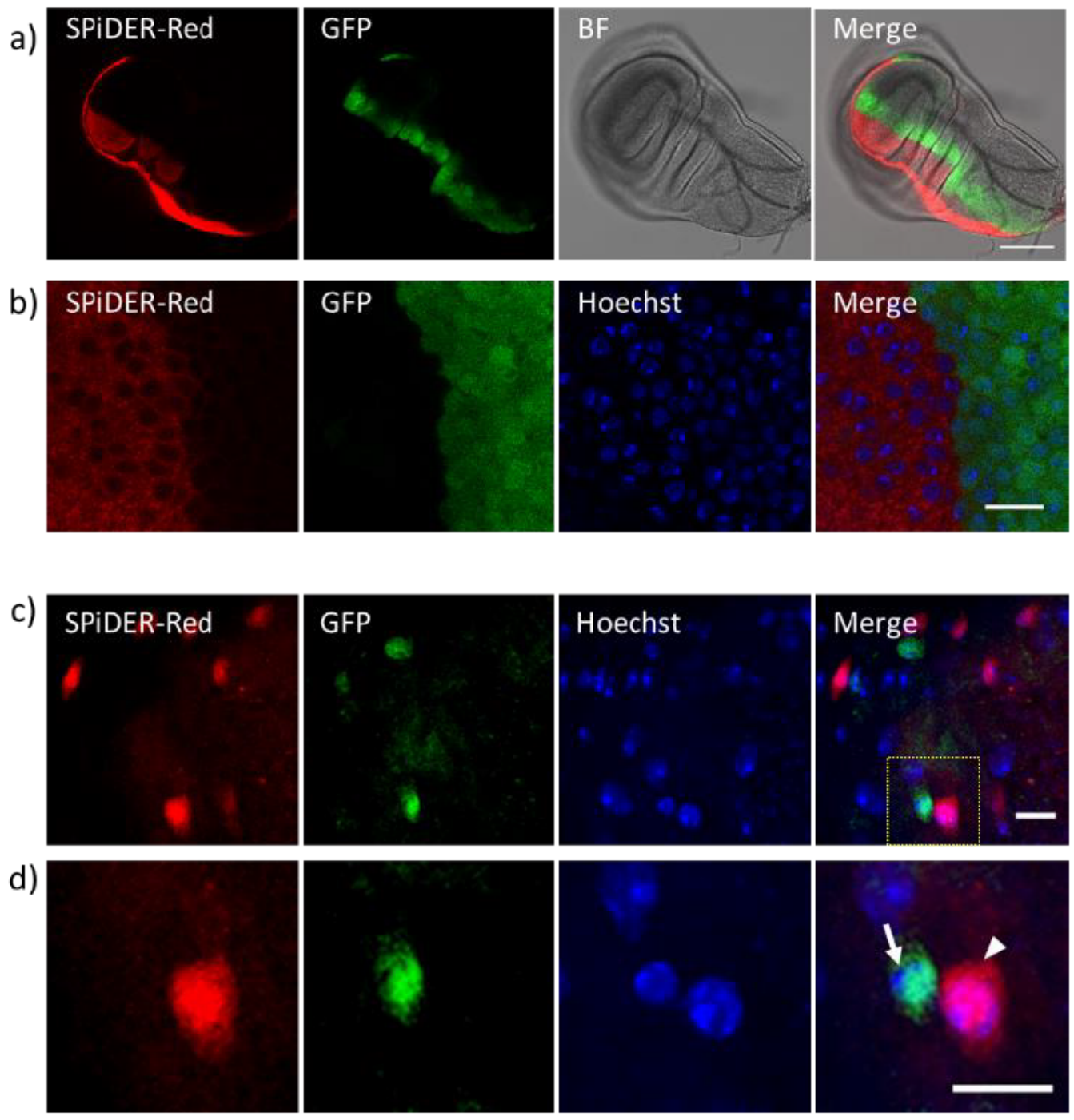

**Figure 11.** Fluorescence imaging of live en-lacZ/dpp-GFP Drosophila third larval wing discs after incubation with 10 μM SPiDER-Red-βGal and 4 μM Hoechst 33342 for 1 hr at room temperature. β-Galactosidase was expressed in the posterior part of wing discs under the engrailed promoter, whereas GFP is expressed in the anterior border cells under the control of the dpp promoter. Ex/Em = 488 nm/500- 550 nm for GFP, 552 nm/600-702 nm for SPiDER-Red-βGal, and 405 nm/430- 481 nm for Hoechst 33342. Scale bars, 100 μm for (a) and 10 μm for (b). (c,d) Fluorescent imaging of esgts-GFP and Su(H)-lacZ intestinal progenitor cells in Drosophila digestive tracts by incubation with 10 μM SPiDER-Red-βGal for 1 hr at 29 °C. β-Galactosidase is expressed in enteroblasts (EB), whereas GFP is expressed in intestinal stem cells (ISC). Nuclei were stained with 4 μM Hoechst 33342. Images in (d) show magnification of the highlighted area in (c). Arrow indicates ISC, and the arrowhead indicates EB. Ex/Em = 488 nm/500-570 nm for GFP, 594 nm/605-681 nm for SPiDER-Red-βGal, and 405 nm/420-490 nm for Hoechst 33342. Scale bars, 10 μm. Reproduced with permission [106] copyright 2016, John Wiley & Sons, Inc.

## 5. Miscellaneous

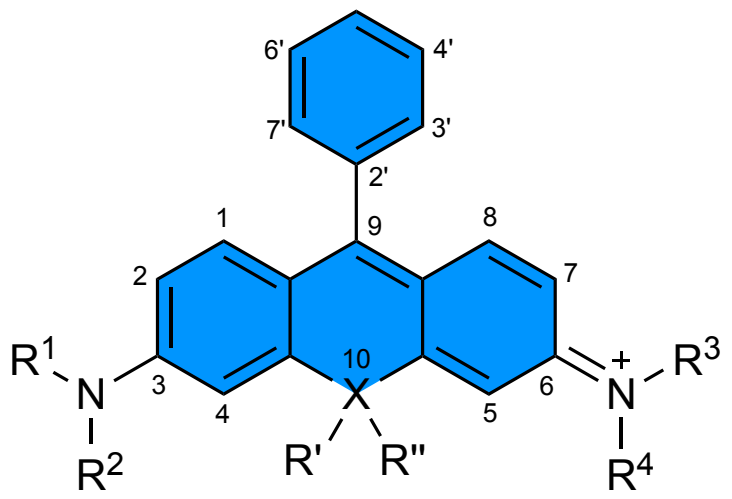

X: $GeR_2$, $SnR_2$, BiR, S, Se, Te

In 2010, 14 metalloles, silicon, germanium, or tin-containing metallocyclopentadiens, received renewed attention due to their unusual electronic structures and photophysical properties. They were utilized in a variety of optoelectronic devices. The electronic and optical behavior of group 14 metalloles was lying under its low-lying LUMO levels, resulting in a bathochromic shift of the excitation and emission wavelengths.[107] In 2011 Koide *et al.*[43] reported a series of pyronine and rhodamine derivatives by utilizing Si, Ge, and Sn. Two tetramethyl rhodamine derivatives bearing 2-methyphenyl group at the 9-position with Ge and Sn bridging moieties were synthesized and named **GeR** (Table 9:E1, Scheme 15) and **SnR** (Table 9:E2, Scheme 15), respectively. **SnR** was an unstable scaffold that resulted in decomposition before the isolation. **GeR** showed tolerance to photobleaching and emitted strong fluorescence intercellularly in HeLa cells. The cyclic voltammetry and Stern-Volmer quenching experiments demonstrated that far-red to NIR fluorescence of **GeR** could be controlled through the PeT mechanism, and the fluorescence quantum yield of **GeR** could be nearly equal to zero. A **GeR** derivative containing *m*-dimethylaminotoluene was also synthesized to examine the utility of group 14 rhodamines as a fluorophore for PeT-based probes. The photophysical data revealed that its quantum yield was regulated by a PeT process. Additionally, fluorescence images of HeLa cells showed that **GeR** localized in mitochondria.

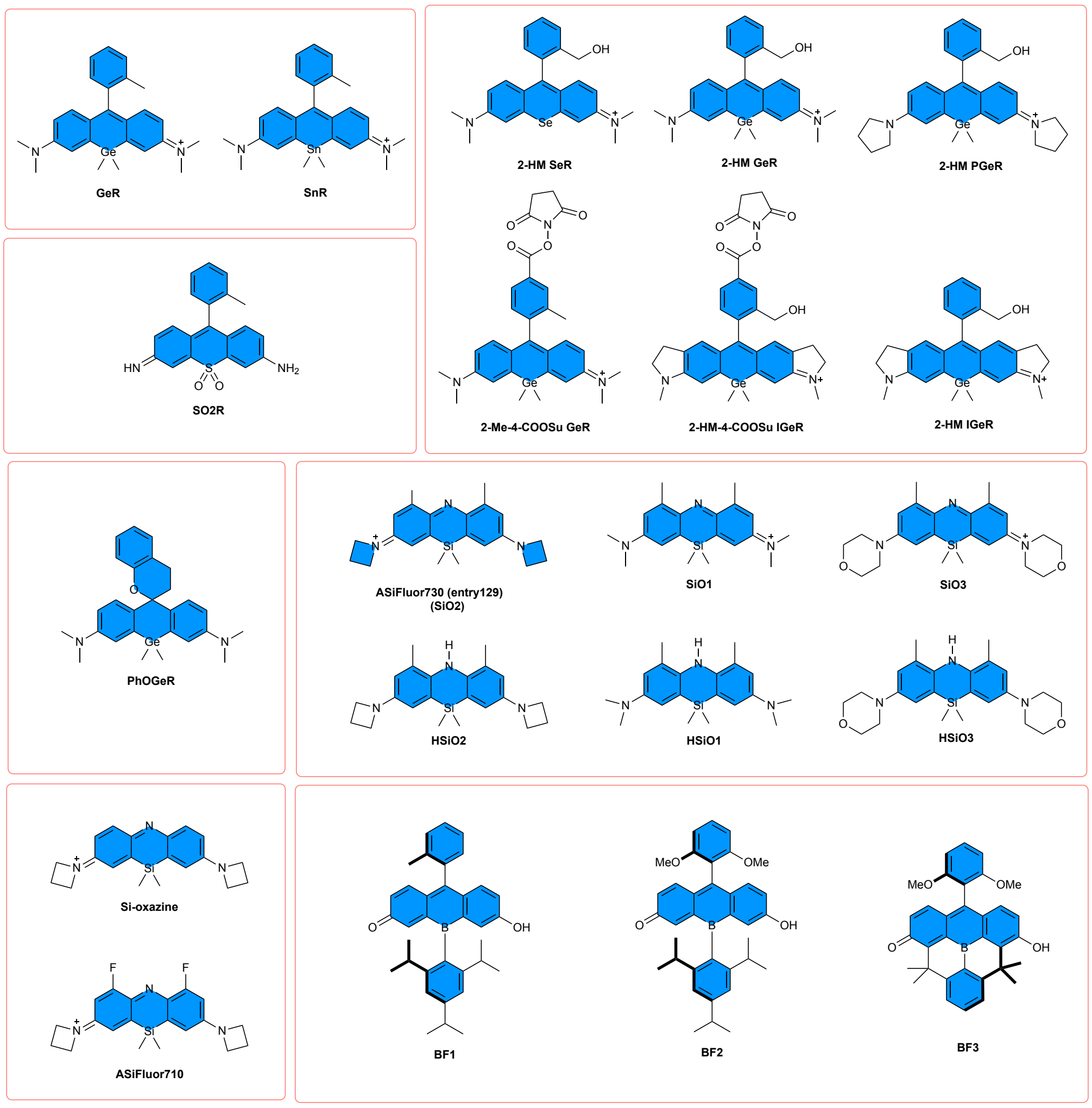

**Scheme 15.** Chemical structures of Miscellaneous fluorophore derivatives.

Later in 2018 Dejouy *et al.*[97] reported a rhodamine derivative, **SO2R** (Table 9:E3, Scheme 15), with a sulfonyl group as the bridging moiety. The sulfone-rhodamine bearing 2-methylphenyl group at the 9-position was synthesized. The analysis of its photophysical properties with various solvents revealed its poor aqueous stability, especially at physiological pH. Pka of the sulfone rhodamine was too low compared to other heteroatom fused rhodamine core, so hydroxyl attack on *meso*-position is inevitable due to high electrophilic character. This work highlighted the stability problem of heteroatom fused xanthene fluorophores encountered in aqueous media.

In 2019, Koide *et al.*[108] reported tetramethyl selenorhodamine analog of the hydroxymethyl selenorhodamine (**2-HM SeR,** Table 9:E4, Scheme 15) was synthesized, and its photophysical properties were studied and compared with its Si (**2-HM SiR**, Table 4:E15, Scheme 8), Ge (**2-HM GeR,** Table 9:E5, Scheme 15) and O bearing analogs. The studies revealed that **2-HM SiR** and **2-HM GeR** showed different behavior from **2-HM OR** and **2-HM-SeR,** including the formation of a spirolactone form under strongly acidic conditions (pH<4) due to the high electrophilicity of the corresponding xanthene skeleton, resulted in quaternary ammonium salts. To compensate electrophilicity of the structure, indoline and pyrrolidine-bearing amino groups were designed (**2-HM PGeR** and **2-HM IGeR**, Table 9:E6-7, Scheme 15), which led to robust structures and the increased pKa. **2-HM PGeR** and **2-HM IGeR** showed pH-dependency and bathochromic shift to realize activable NIR fluorescence probes with desirable quantum yields. Additionally, introducing a succinimidyl ester to **2-HM IGeR** and **GeR** yielded **2-HM-4-COOSu IGeR** and **2-Me-4-COOSu GeR** (Scheme10), respectively. Their Herceptin **Her-2-HM IGeR** (Table 9:E8, Scheme 15), and Avidin **Avi-2-HM IGeR** (Table 9:E9, Scheme 15) conjugates were prepared to investigate pH-dependency of the fluorescence of these labeling reagents when bound to biomolecules. As a result, 2-methylphenyl derivatives did not respond to varying pH, whereas 2-methylhydroxyphenyl derivatives did as expected. Then, **Her-2-HM IGeR** and **Avi-HMIGeR were** used in the fluorescence imaging of the process of endocytosis and *in vivo* cancer imaging, respectively. Regarding photobleaching, **Her-2-HM IGeR** showed superior results compared to commercially available cyanine-based conjugate. Lastly, a peritoneal dissemination model of SHIN3 tumors in BALB/c nude mice was prepared, and **Avi-HMIGeR** was injected. The imaging results showed that the Ge-probe-biomolecule conjugate was successful for cancer imaging and discriminated between intestine autofluorescence and probe fluorescence (figure 12). This work highlighted that the amino group on the xanthene skeleton could be designed so that the pKa of the fluorophore could be manipulated to activate at the intracellular region of interest.

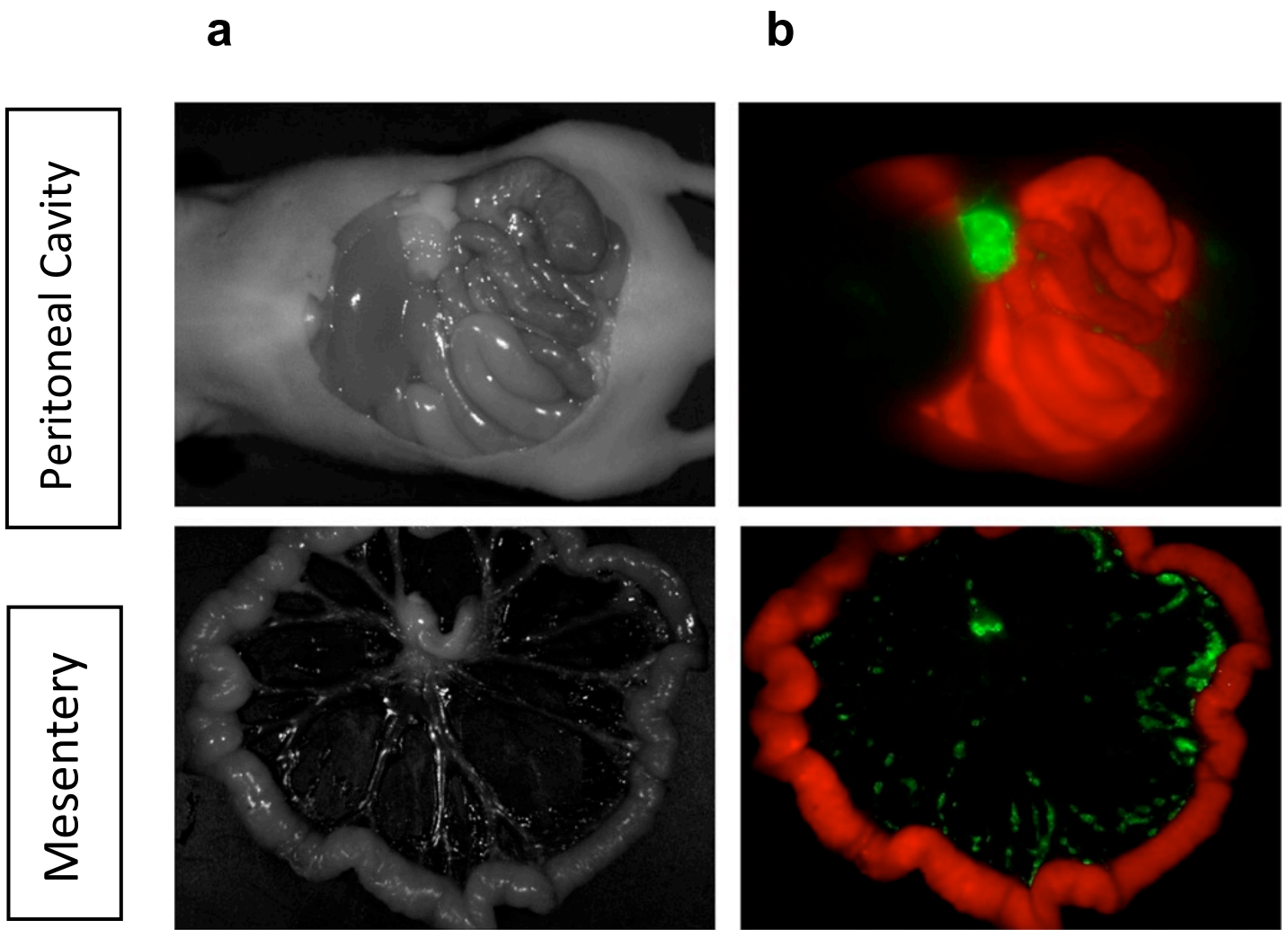

**Figure 12.** Fluorescence imaging of peritoneal metastases of SHIN3 cells in a mouse model. The mice were sacrificed at 4 h after peritoneal injection of 100 µg Avi-HMIGeR in 300 µL PBS, and the images were captured after dissection. a. White light image. b Unmixed image. In the unmixed image, fluorescence from the probe and autofluorescence was assigned green and red, respectively. Ex/Em= 616−661 nm/675 nm LP filter. Scale bars represent 1 cm Reproduced with permission [108] copyright 2019, Nature

In 2021, Alexey N. Butkevich[78] reported a synthetic approach to Si-Rhodamine derivatives. The approach included a regioselective double nucleophilic addition of aryllanthanum reagents to esters, anhydrides, and lactones. A non-fluorescent tetramethyl Ge-rhodamine analog (Scheme 1) was also synthesized utilizing this approach. This method provided an expanded substrate scope compared to the synthetic method from bis-aryllithium reagents, and the synthesis of the Ge-rhodamine analog proved that the scope was not only limited to Si derivatives.

For the development of NIR-emissive xanthene dyes, there were several structural modifications. The most typical approach was by extending the π-conjugation of the xanthene core. Another approach was replacing the endocyclic oxygen atom at the 10-position with other heteroatoms such as Si, P for fluorescein cores. However, boron-containing fluorescein derivatives did not receive much attention. In 2019 Ando *et al*. reported bora-fluorescein derivatives, **BFs**, containing a tricoordinate boron atom at the 10-position of the fluorescein skeleton. The values obtained from the photophysical studies revealed that the deprotonated derivatives of **BF2** (Table 9:E11, Scheme 15) and **BF3** (Table 9:E12, Scheme 15) showed the most red-shifted results of all hitherto reported heteroatom-substituted fluorescein dyes whereas, Lewis base coordination with boron contributed to the significant hypsochromic shift of the absorption and emission. Lewis acid-base equilibrium on the boron center and Brønsted acid-base equilibrium on the fluorescein skeleton created multistage responses, resulting in fluctuations in absorption spectra. Although **BF1** (Table 9:E10, Scheme 15) was synthesized and isolated under ambient conditions, it decomposes in polar solvents immediately, highlighting the importance of steric hindrance on the C9 position to design a robust xanthene skeleton. Additionally, the utilization of tricoordinate boron atoms revealed the significant impact of the boron atom on the electronic structure of fluorescein dye and provided insight into the design principle of fluorescein dyes.[109]

*Table 9 Selected features of Miscellaneous derivatives*

| Entry E | Fluorophore* | Absorption | | Emission | $\Phi_f$ | Intracellular Localization | Properties | Applications | Ref. |
|---|---|---|---|---|---|---|---|---|---|
| | | λmax (nm) | ε ($M^{-1}\cdot cm^{-1}$) | λmax (nm) | | | | | |
| 1 | **GeR**[a] | 635 | n.d. | 649 | 0.34[b] | MT | Photostable, Not phototoxic, Cell permeable | - | [43] [108] |
| 2 | **SnR**[a] | nd | n.d. | nd | nd | n.d. | Not stable | - | [43] |
| 3 | **SO2R**[c] | 600 | n.d. | 685 | 0.06[d] | n.d. | Stable in acidic medium. | - | [97] |
| 4 | **2-HM SeR**[a] | 586 | n.d. | 604 | Nd | n.d. | pKa =7.86 | - | [108] |
| 5 | **2-HM GeR**[a] | 639 | n.d. | 653 | 0.34[b,e] | n.d. | $pK_a$=5.70 | - | [108] |
| 6 | **2-HM PGER**[a] | 646 | n.d. | 658 | 0.34[b,e] | n.d. | $pK_a$=5.44, pH-dependent | - | [108] |
| 7 | **2-HM IGER**[a] | 679 | n.d. | 693 | 0.25[f,e] | Intracellular acidic environment | $pK_a$=6.24, pH-dependent | - | [108] |
| 8 | **Her-HMIGeR**[a] | 686 | n.d. | 700 | n.d. | Intracellular acidic environment/ Her2 receptor | $pK_a$=6.61, pH-dependent | fluorescence imaging of endocytosis | [108] |
| 9 | **Avi-HMIGeR**[a] | 643 | n.d. | 700 | n.d. | Intracellular acidic environment/ Lectin expressed cancer cells | $pK_a$=5.46, pH-dependent | *In vivo* Cancer imaging | [108] |
| 10 | **BF1** | nd | n.d. | n.d. | n.d. | n.d. | Not stable under polar medium | - | [109] |
| 11 | **BF2**[g]<br>**$[BF2]^{-}$**[i]<br>**$[BF2\cdot F_2]^{2-}$**[j] | 538<br>851<br>585 | $8.4X10^3$<br>$1.9x10^4$<br>$1.5X10^5$ | 704<br>907<br>595 | 0.011[h]<br>0.003[h]<br>0.89[h] | n.d. | Multistage responses to Brønsted and Lewis bases. | - | [109] |
| 12 | **BF3**[g]<br>**$[BF3]^{-}$**[i]<br>**$[BF3\cdot F_2]^{2-}$**[j] | 524<br>801<br>608 | $1.1x10^4$<br>$3.4x10^4$<br>$8.56x10^4$ | 660<br>843<br>621 | 0.004[l]<br>0.029[l]<br>0.87[l] | n.d. | Multistage responses to Brønsted and Lewis bases. | - | [109] |
| 13 | **ASiFluor710**[k] | 709 | $6.0x10^4$ | 719 | 0.11[l] | n.d. | Not stable in aq. medium | - | [110] |
| 14 | **ASiFluor730**[m] **or (SiO2) HSiO2** | 725<br>727 | $5.3x10^4$<br>$8.2x10^4$ | 739<br>747 | 0.01[n]<br>0.06[f] | MT | $HClO/ONOO^-$ selective, Cell membrane permeable | $HClO/ONOO^-$ imaging (inflammation-related diseases) | [110,111] |
| 15 | **SiO1**[m]<br>**HSiO1** | 724 | $7.0x10^4$ | 746 | 0.01[f] | MT | $HClO/ONOO^-$ selective, | $HClO/ONOO^-$ imaging | [111] |
| | | | | | | | Cell membrane permeable | (inflammation-related diseases) | |
| **16** | **SiO3**[m] **HSiO3** | 732 | 8.3x10$^4$ | 760 | 0.01[f] | Lysosome | HClO/$ONOO^-$ selective, Cell membrane permeable | HClO/$ONOO^-$ imaging (inflammation-related diseases) | [111] |

[a]Measured in PBS at pH 7.4. [b]Cresyl violet in methanol ($\Phi_f$=0.54) was used as standard. [c]For the photophysical measurements, 0.1% FA in water solution was used. [d]Sulfoindocyanine dye Cy 5.0 as a fluorescence standard ($\Phi_f$ = 0.20 in PBS). [e]Fluorescence quantum efficiency ($\Phi_f$) of the 2-Me analog of each probe in PBS (pH 7.4) was given. [f]Cy5.5 in PBS ($\Phi_f$ = 0.23) was used as a fluorescence standard. [g]Photophysical properties were measured in $CH_3CN$. [h]The absolute fluorescence quantum yield. [i]Measured in $CH_3CN$ with 1,8-Diazabicyclo[5.4.0]undec-7-ene (DBU). [j]Measured in $CH_3CN$ with excess TBAF. [k]Photophysical properties were measured in EtOH. [l]For determination of the fluorescence quantum yield ($\Phi_f$), Cy5.5 in EtOH ($\Phi_f$ = 0.20) was used as a fluorescence standard. [m]Photophysical properties were measured in PBS. [n]HITC in EtOH ($\Phi_f$ = 0.30) was used as a fluorescence standard. n.d.: not detected. MT: mitochondria.[*] all the fluorophores listed above are emphasized with bold font in the text.

In 2018 Miller's group[110] foreseen that in addition to the modification of classical xanthene core by replacing the bridging oxygen moiety with heteroatoms, and replacing the carbon atom at the 9-position with nitrogen could yield a further NIR shift in the absorption maximum. Hence, they reported a series of novel azasilane fluorophores. Initially, an azetidine-bearing azasilane **(Si-oxazine,** Scheme 13) dye was synthesized starting from diphenylamine. The synthetic pathway included the tetrabromination of diphenylamine, protection of the amine with a p-methoxybenzyl group (PMB), formation of the silicon bridge, the introduction of the azetidine groups with Buchwald chemistry,[112] followed by deprotection with iodine which also readily oxidized the compound. However, after the deprotection step, the dye was decomposed upon solvent removal. To realize a stable core, the introduction of electronically different substituents was utilized in the azasilane core: inductively donating but π-donating fluorine groups and electron-donating bulkier methyl groups. Difluoro and dimethyl derivatives of the azetidine-bearing azasilane dyes were synthesized. The difluoro derivative **ASiFluor710** (Table 9:E13, Scheme 15) and dimethyl derivative **AsiFluor730** (Table 9:E14, Scheme 15) showed red-shifted absorption maxima beyond 700 nm. The fluorescence quantum yield of **ASiFluor730** was low in aqueous medium however showed good photostability, where irradiation at its peak absorption wavelength for 1 h did not result in any decomposition.

In 2020, Wang *et al.*[111] expanded the scope of the azasilane dye series with a new synthetic methodology. Four dimethyl azasilane dye derivatives bearing dimethylamine, azetidine, and morpholine were synthesized from 3,5-dibromotoluene and named **SiO1** (Table 9:E15, Scheme 15), **SiO2** (**AsiFluor730,** Table 9:E14, Scheme 15) , and **SiO3** (Table 9:E16, Scheme 15), respectively. The synthetic pathway included the Buchwald-Hartwig coupling of 3,5-dibromotoluene with the corresponding secondary amine, formation of the silicon bridge, followed by the nitrosation-cyclization using $NaNO_2$/HCl to produce the corresponding azasilane derivative. The Si-oxazine dyes showed absorption maxima centered above 700 nm but very low fluorescence quantum yields. The reduced products of the derivatives, **HSiO1-3** (Table 9, E14-16, Scheme 15), showed selective and rapid fluorescence off-on response for hypochlorous acid (HClO) and peroxynitrite ($ONOO^{-1}$) over other ROS/RNS sources. **HSiO3** showed low cytotoxicity, and bioimaging was tested on RAW264.7 murine macrophages. A bright intracellular fluorescence was observed only for the $ClO^-$ and SIN-1 (commercial $ONOO^-$ donor) treated **HSiO3** loaded RAW264.7 cells. Furthermore, the ability of HSiO3 for selective imaging was also demonstrated *in vivo* (figure 13).

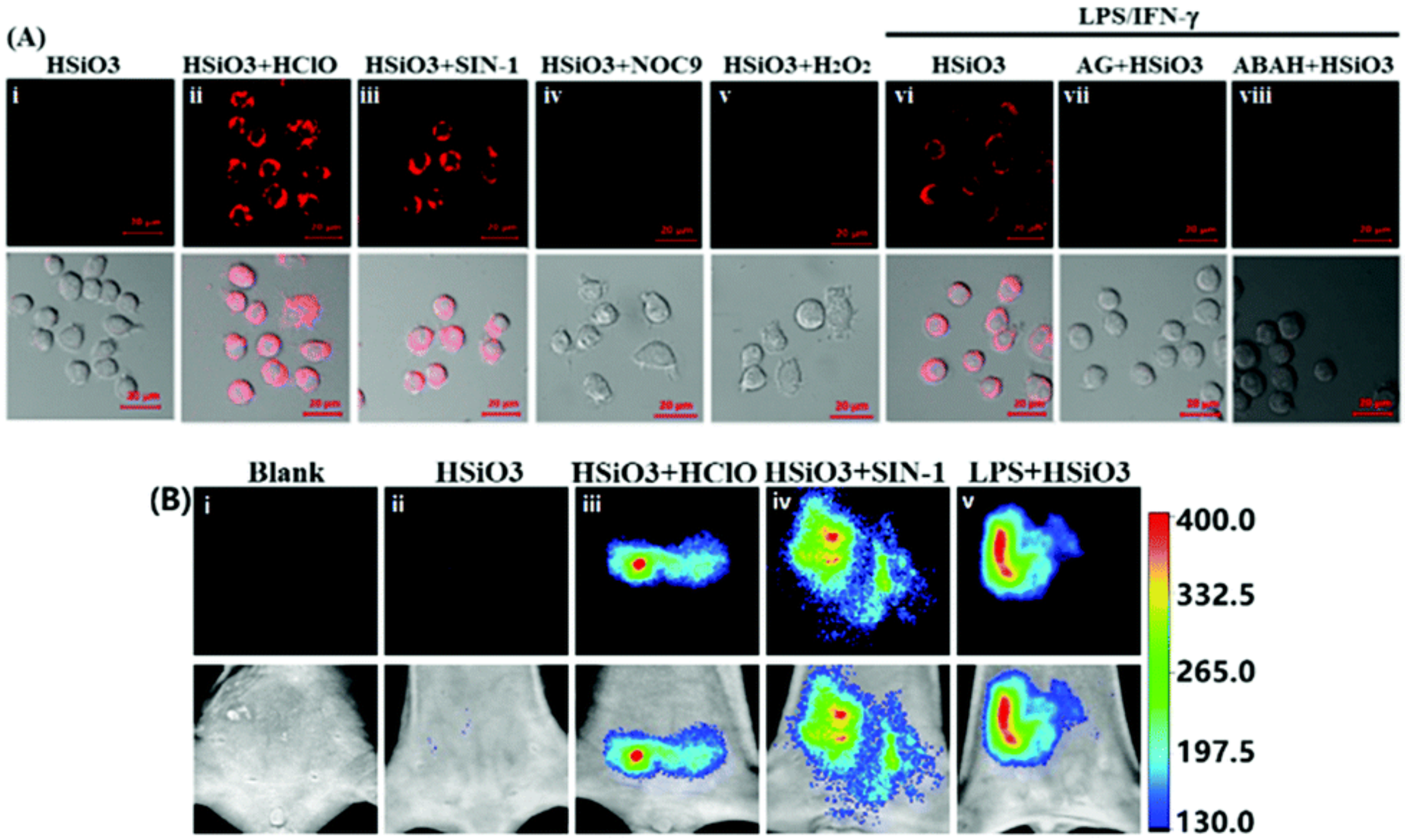

Figure 13. (A) Fluorescence images of RAW264.7 cells under different conditions. Scale bar: 20 µm. (B) Fluorescence images of BALB/c nude mice under different conditions. Reproduced with permission [111] copyright 2020, The Royal Society of Chemistry

# 6. PDT candidates and agents

In this review, it was clearly demonstrated that the photophysical characteristics of the xanthene-based dyes could be tuned to longer wavelengths by replacing the oxygen atom at the 10-position with third-row elements such as Silicon and Phosphorus. Compared to this modification, incorporating fourth- or fifth-row elements such as Selenium and Tellurium in the xanthene structure yielded a relatively small red shift together with the generation of an excited triplet state and subsequent singlet oxygen ($^1O_2$) generation. The excited triplet state was generated due to the presence of heavy atoms. Chromophores with the singlet oxygen generation ability upon irradiation could be used as photosensitizers in PDT. The properties of an ideal photosensitizer include absorption maximum centered between 650 nm and 900 nm (since only red to deep-red absorption enables tissue penetration efficiently), water solubility, a high singlet oxygen quantum yield, and no dark toxicity.[113]

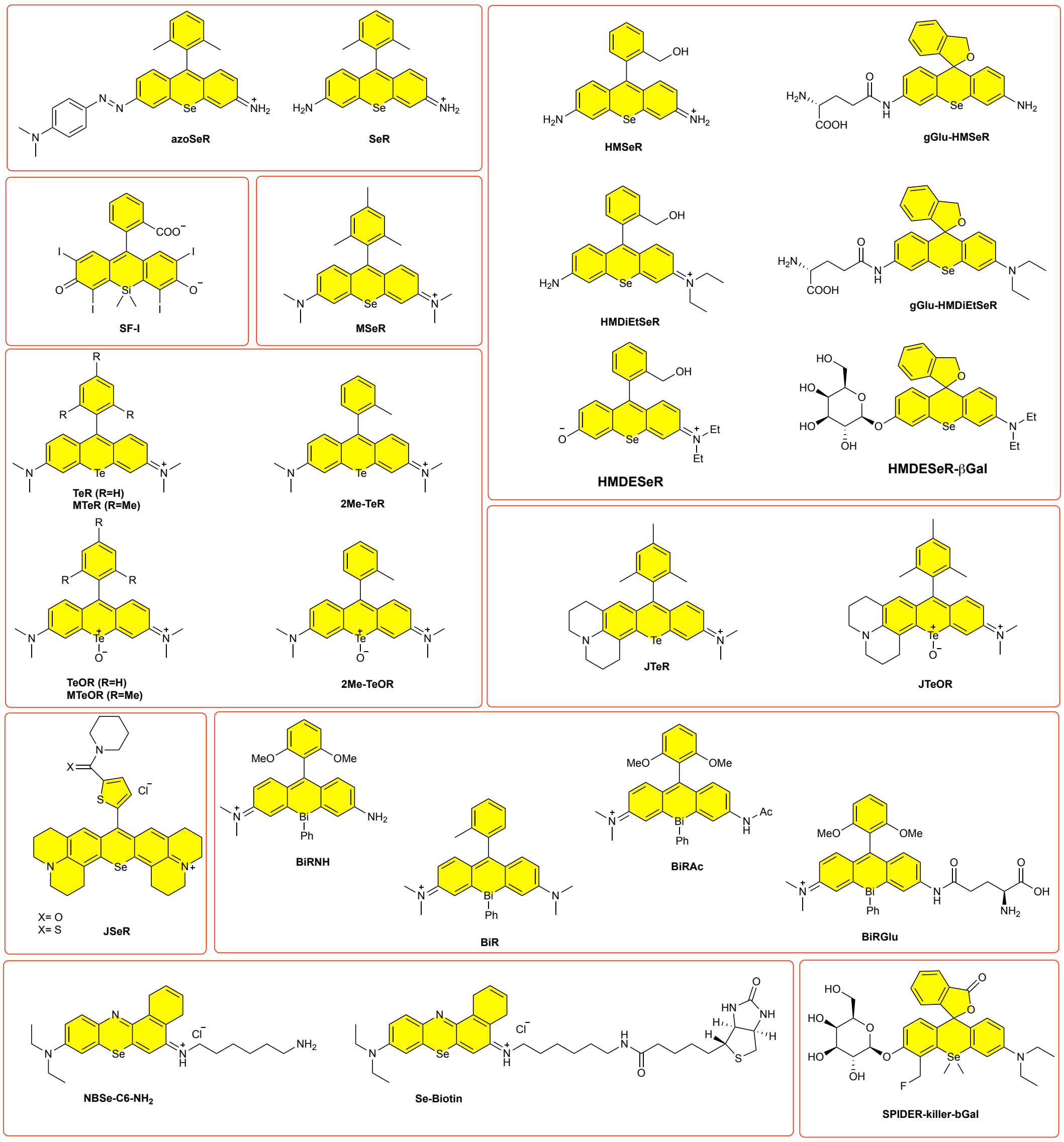

**Scheme 16.** Chemical structures of PDT candidates and agents.

In 2014, Hill *et al*.[114] reported a series of selenorhodamine derivatives either bearing amide (**SeRAa**, **SeRAb**, Table 10:E27,29, Scheme 16) or thioamide (**SeRTAa**, **SeRTAb**, Table 10:E28,30, Scheme 16) functional group at 5th position of a 9-(2-thienyl) moiety on the selenorhodamine core. Photophysical properties of synthesized photosensitizers were investigated, and all photosensitizers have absorption maxima above 600 nm. Furthermore, $\Phi_\Delta$ was found in the range of 0.44-0.54 in the MeOH medium. These selenorhodamine derivatives were tested as photosensitizers in the Colo-26 cell line expressing P-gp. Thioamide derivatives (**SeRTAa** and **SeRTAb**) exhibit better inhibition of P-gp than amide analogs (**SeRAa** and **SeRAb**). Moreover, thioamide analogs showed synergistic therapeutic effects on Colo-26 cell treatment in vitro after exposure to PDT and treatment with the cancer drug doxorubicin.

Kryman *et al*. [115] reported a series of selenorhodamine photosensitizers (**JSeR**, Table 10:E1-2, Scheme 16) in 2015. The selenorhodamines bearing amide and thioamide units at the 9-position with two julolidyl fragments containing nitrogen atoms were synthesized and utilized as effective photosensitizers for PDT towards Colo-26 cells in culture.

Chiba *et al*.[116] reported an activatable selenorhodamine targeted to γ-glutamyltranspeptidase (GGT). Hydroxymethyl selenorhodamine green (**HMSeR** (Table 10:E3, Scheme 14)), hydroxymethyl *N*,*N*-diethyl selenorhodamine **HMDiEtSeR** (Table 10:E4, Scheme 16), and their γ-glutamyl derivatives were synthesized. The photochemical studies of **HMSeR** and **HMDiEtSeR** confirmed the pH-dependent change in the absorption spectrum. γ-glutamyl derivatives showed little $^1O_2$ production upon irradiation with a 532 nm laser. Also, they showed that γ-glutamyl derivatives could be converted into **HMSeR** and **HMDiEtSeR** upon reaction with γ-glutamiyltranspeptitase (GGT) in vitro, which resulted in the recovery of absorption and ability of $^1O_2$ production. **gGlu-HMDiEtSeR** (Table 10:E5, Scheme 16) showed phototoxicity to two different cell lines with varying activities of GGT (SHIN3 cells with high GGT activity and SKOV3 cells with low GGT activity), indicating that the cell death was independent of GGT activity. **gGlu-HMSeR** (Table 10:E6, Scheme 16) was further tested with tumor spheroid, proving that high-GGT-expressing cells were specifically eliminated by **gGlu-HMSeR**.

*Table 10 Selected features of PDT agents and Candidates*

| Entry (E) | Fluorophore* | Absorption | | Emission | $\Phi_f$/ $\Phi_\Delta$ ($^1O_2$) | Intracellular Localization | Properties | Applications | Ref. |
|---|---|---|---|---|---|---|---|---|---|
| | | λmax (nm) | ε ($M^{-1}.cm^{-1}$) | λmax (nm) | | | | | |
| 1 | **JSeR[a], x=O** | 626 | $1.35x10^5$ | 660 | 0.012[b]/ 0.64[c] | MT | Interact with P-glycoprotein | NIR photosensitizer | [115] |
| 2 | **JSeR[a], x=S** | 626 | $9.9x10^4$ | 660 | 0.010[b]/ 0.51[c] | MT | Interact with P-glycoprotein | NIR photosensitizer | [115] |
| 3 | **HMSeR[d]** | 434 | n.d. | 562 | 0.002[e]/ 0.40[f] | Lysosome | Non-specific phototoxicity | - | [116] |
| 4 | **HMDiEtSeR[d]** | 567 | n.d. | 602 | 0.007[e]/ 0.74[f] | n.d. | Non-specific phototoxicity | - | [116] |
| 5 | **gGlu-HMDiEtSeR[d]** | 453, 551, 588 | n.d. | n.d. | n.d./ n.d. | n.d. | Non-specific phototoxicity | - | [116] |
| 6 | **gGlu-HMSeR[d]** | 450, 521, 551 | n.d. | n.d. | n.d./ n.d. | Lysosome | selective phototoxicity toward GGT expressing tumors | Dual targeted activatable photosensitizer | [116] |
| 7 | **SeR[g]** | 533 | $5.9x10^4$ | 577 | 0.006[e]/ 0.56[f] | MT | nd | Active photosensitizer | [117] |
| 8 | **azoSeR[g]** | 630 | $3.6x10^4$ | n.d. | 0.001[e]/ 0.03[f] | MT | Hypoxic cell selectivity | Dual targeted activatable photosensitizer | [117] |
| 9 | **HMDESeR-βGal** | 543, 580 | $4.2x10^2$ | n.d. | n.d. | Cytosol | selective phototoxicity toward LacZ(+) cells | Activable photosensitizer | [118] |
| | Activated form (HMDESeR) | 558 | $9.2x10^4$ | 590 | 0.006 | | | | |
| 10 | **SPiDER-killer-βGal** | 493, 525 | n.d. | 560 | <0.01 | Cytosol | selective phototoxicity toward LacZ(+) cells | Activable photosensitizer | [119] |
| | Activated form (4-CH2OH-HMDESeR) | 557 | n.d. | 594 | 0.01/ 0.36 | | | | |
| 11 | **MSeR[a]** | 568 | $1.03x10^5$ | n.d. | n.d./ 0.85[c] | n.d. | Self-sensitized | Photocatalytic aerobic thiol oxidation | [120] |
| 12 | **TeR[g]** | 600 | $1.0x10^5$ | n.d. | 0.005[h,i]/ 0.43[c] | n.d. | Self-sensitizer | - | [121] |
| 13 | **2Me-TeR[g]** | 600 | $9.3x10^4$ | n.d. | 0.005[h,i]/ 0.70[c] | n.d. | Self-sensitizer | - | [122] |
| 14 | **MTeR[g]** | 600 | $8.6x10^4$ | n.d. | 0.005[h,i]/ 0.75[c] | n.d. | Self-sensitizer | Photocatalytic aerobic thiol oxidation | [122] |
| 15 | **JTER[g]** | 608 | $8.9x10^4$ | n.d. | 0.005[h,i]/ 0.90[c] | n.d. | Self-sensitizer | Photocatalytic aerobic thiol oxidation | [122] |
| **16** | **TeOR**[g] | 664, 680 | ≥5x$10^4$ | n.d. | n.d. | n.d. | Not stable | - | [122] |
| **17** | **2-Me-TeOR**[g] | 670 | 8.6x$10^4$ | 685[b] | 0.21[h,i]/ n.d. | n.d. | Not stable | - | [122] |
| **18** | **MTeOR**[g] | 664 | 9.5x$10^4$ | 685[b] | 0.31[h,i]/ n.d. | n.d. | Stable in nucleophilic solvent/ Able to regenerate its reduced form | Photocatalytic aerobic thiol oxidation | [122] |
| **19** | **JTEOR**[g] | 680 | 7.4x$10^4$ | 702[b] | 0.20[h,i]/ n.d. | n.d. | Stable in nucleophilic solvent, Able to regenerate its reduced form | Photocatalytic aerobic thiol oxidation | [122] |
| **20** | **BiR**[j] | 635 | 7.76x$10^4$ | 658[k] | 0.039[l]/ 0.66[m] | Endoplasmic reticulum | Cell permeable, No dark-toxicity | NIR theranostic photosensitizer | [113] |
| **21** | **BiRNH**[j] | 615 | 9.0x$10^4$ | 634[k] | 0.033[l]/ 0.74[m] | Endoplasmic reticulum | Water soluble/ Inert against varying pH | Active Photosensitizer | [123] |
| **22** | **BiRAc**[j] | 526 | 3.2x$10^4$ | n.d. | n.d./ n.d. | n.d. | Water soluble/ Inert against varying pH | Inactive Photosensitizer | [123] |
| **23** | **BiRGlu**[j] | 525 | 4.8x$10^4$ | n.d. | n.d./ n.d. | Endoplasmic reticulum | Specific phototoxicity toward GGT expressing tumors | Dual targeted activatable theranostic photosensitizer | [123] |
| **24** | **NBSe-C6-NH$_2$**[n] | 670 | 5.3x$10^4$ | 710 | 0.021[o]/ (0.68)[p] | Lysosome | Apoptotic cell death | NIR photosensitizer | [124] |
| **25** | **Se-Biotin**[n] | 670 | 5.2x$10^4$ | 710 | 0.026[o]/ (0.69)[p] | Lysosome | Tumor-selective/ Apoptotic cell death | Dual targeted NIR photosensitizer | [124] |
| **26** | **SF-1**[q] | 614 | 7.7x$10^4$ | 630 | 0.11[o]/ 45[r,s],30[r,t] | Cytosol | Photostable/ No dark-toxicity/ Necrotic cell/ death | NIR Theranostic photosensitizer | [125] |
| **27** | **SeRAa**[a] | 605 | 7.18x$10^4$ | 636 | 0.009 / 0.50 | MT | Phototoxic toward P-gp | PDT agent | [114] |
| **28** | **SeRTAb**[a] | 608 | 9.78x$10^4$ | 635 | 0.008 / 0.54 | MT | Phototoxic toward P-gp | PDT agent and synergistic therapeutic effect with doxorubicin | [114] |
| **29** | **SeRAa**[a] | 609 | 8.73x$10^4$ | 634 | 0.009 / 0.48 | MT | Phototoxic toward P-gp | PDT agent | [114] |
| **30** | **SeRTAb**[a] | 608 | 8.11x$10^4$ | 634 | 0.008 / 0.44 | MT | Phototoxic toward P-gp | PDT agent and synergistic therapeutic effect with doxorubicin | [114] |

[a]Measured in MeOH. [b]TMR-Se used as a reference standard for fluorescence quantum yield [c]Relative singlet oxygen ($^1O_2$) quantum yield in MeOH, TMR-Se was employed as a reference, $\Phi_\Delta(^1O_2)$= 0.87., ($\Phi_f$) = 0.009 [d]Measurements were done in pH=2.0 sodium phosphate buffer. [e]Rhodamine B ($\Phi_f$=0.65) was used as a fluorescence standard. [f]Rose bengal ($\Phi_\Delta(^1O_2)$= 0.75 in $H_2O$/MeOH = 1:1) was used as standard. [g]Photophysical measurements were done at pH 7.4 PBS buffer. [h]In pH 6.3, phosphate buffer. [i]Alexa Fluor 700 ($\Phi_f$ = 0.25), was used as a reference standard., [j]Measurements were done in HEPES at pH 7.4 buffer with 0.5% DMSO as a cosolvent. [k]Measurements were done in HEPES at pH 7.4 buffer with 0.2% DMSO as a cosolvent. [l]Absolute fluorescence quantum yield was given. [m]Methylene Blue ($\Phi_{\Delta\ MB}$= 0.52 in acetonitrile) as reference. [n]Photophysical properties were measured in PBS (1%DMSO).[o]The absolute fluorescence quantum yield.[p]Relative singlet oxygen ($^1O_2$) quantum yield in PBS/ethanol (1:1 v/v), Methylene Blue ($\Phi_{\Delta\ MB}$ = 0.52 in PBS/ethanol (1:1 v/v) as the reference. [q]Measured in PBS buffer (pH 7.4, 0.5% DMSO). [r]Methylene Blue as a reference standard ($\Phi_{\Delta\ MB}$ = 0.52 in PBS buffer) [s]Upon irradiation with a 595 nm LED. [t]Upon irradiation with a 630 nm LED. MT: mitochondria. P-gp: P-glycoprotein n.d.: not detected.[*] all the fluorophores listed above are emphasized with bold font in the text.

In 2017, Piao *et al.* demonstrated another utilization of selenorhodamines as a photosensitizer scaffold. [117] They synthesized a selenorhodamine derivative bearing 2,6-dimethylpheny group at the 9-position, **SeR** (Table 10:E7, Scheme 16), and the desired photosensitizer **azoSeR** (Table 10:E8, Scheme 16) by azo coupling reaction. **azoSeR** showed a red-shifted and broader absorption compared to **SeR,** and no emission was observed as expected. The introduction of azo moiety into **SeR** effectively blocked the intersystem crossing process. In the cells under hypoxia, the azo group was cleaved, and the production of $^1O_2$ was achieved. **AzoSeR** also showed high selectivity for cells under hypoxia.

Urano and coworkers proposed β-galactosidase targeted selenium-based activable **HMDESeR-βGal** [118] (Table 10:E9, Scheme 16) and its novel derivative, **SPiDER-killer-βGal**, which has the capability of being immobilized within the cell to overcome the issue related to the escaping of activated photosensitizer from the target cell/tissue in long incubation time. [119] (Table 10:E10, Scheme 16) According to the cell studies done on the cocultured HEK293 cells and *lac*Z(+) cells, **HMDESeR-βGal** and **SPiDER-killer-βGal** showed selective phototoxicity on *lac*Z(+) cells with the indicators of apoptosis, and no dark toxicity. Additionally, *in vivo* studies were demonstrated on *Drosophila* and concluded that **HMDESeR-βGal** and **SPiDER-killer-βGal** not only showed selective phototoxicity in cultured cells but also at the tissue level (figure 14).

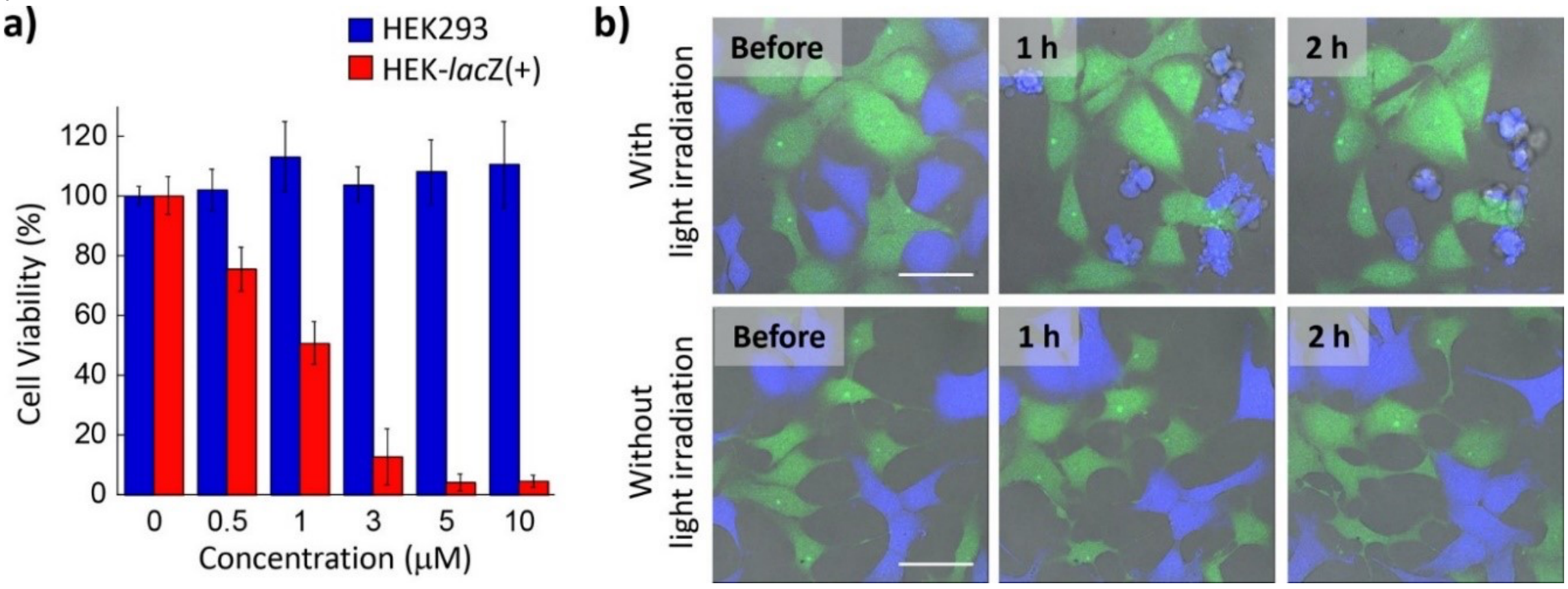

**Figure 14.** (a) Viability assay of cultured HEK293 and HEK/lacZ cells loaded with SPiDER-killer-**β**Gal. Cells were preincubated with SPiDER-killer**β**Gal for 4 h and then light-irradiated (550 nm, 8.0 mW/cm$^2$, 3 min). The cell counting kit-8 (CCK-8) assay was performed 24 h post irradiation to measure cell viability. Error bars stand for SD (n = 4). (b) Time-lapse fluorescence imaging of a coculture of HEK293 and HEK-lacZ(+) cells. HEK-lacZ(+) cells and HEK293 cells were pre-stained with CellTracker Blue and CellTracker Green, respectively. The coculture was incubated with 1 μM SPiDER-killer-**β**Gal for 1 h at 37 °C, followed by light irradiation (WLL laser (561 nm), 5 min). Immediately thereafter, time-lapse fluorescence imaging was started using a confocal microscope. HEK293 cells (green) were attached to the imaging dish, while HEK-lacZ(+) cells (blue) started to form blebs, followed by cell shrinkage and membrane rupture. Scale bars, 50 μm. Reproduced with permission [118] copyright 2014, John Wiley & Sons, Inc.

In 2017, Lutkus *et al.*[120] reported a tetramethyl selenorhodamine bearing 2,4,6-trimethylphenyl group at the 9-position, **MSeR** (Table 10:E11, Scheme 16). Its photochemical reactivity was tested, and the selenorhodamine derivative was used as a photocatalyst for the selective oxidation of thiophenol and 2-naphthalenethiol. The derivative showed extremely low disulfide conversion yields compared to the tellurorhodamine analogue, **MTeR** (Table 10:E13, Scheme 16). Compared to the thio- and selenoxanthylium rhodamine analogs, telluroxanthylium bearing derivatives and their tellurium oxide analogs with Te(IV) oxidation state was studied by Kryman *et al.*[126] reported a series of 9-aryl-3,6-diaminotelluroxanthylium dyes and their telluroxides in 2013 including two known and two novel 9-mesityl derivatives of tellurororhodamines, **TeR**[121] **2Me-TeR,**[122] **MTeR, JTeR**, (Table 10:E12-15, Scheme 16) and their corresponding telluroxides, **TeOR**, **2Me-TeOR**, **MTeOR, JTeOR**, (Table 10:E16-19, Scheme 1). The singlet oxygen quantum yield experiments showed that rotational freedom of the 9-phenyl substituent decreases the efficiency as expected, and the derivative bearing 2,4,6-trimethyphenyl at the 9-position with julolidyl ring system, **JTeR**, showed the highest efficiency. Photooxidation experiments were performed for the tellurorhodamine derivatives, and oxidation with hydrogen peroxide was also demonstrated. Lastly, 9-(4-bromophenyl) tellurorhodamine was synthesized, and its oxidation characteristics were examined.

In 2017, Hirayama *et al.*[113] reported a bismuth-rhodamine derivative. The bismuth atom was fused into the tetramethyl rhodamine core bearing 2-methylphenyl at the 9-position yielding **BiR**(Table 10:E20, Scheme 16). In the **BiR** structure, the bridging moiety was phenlybismuthane. The derivative showed both red-shifted absorption and high $^1O_2$ generation yields. The photosensitization of **BiR** was demonstrated in live cells, including HepG2, A549, HEK293, and TIG-3, with $EC_{50}$ values in the nano-molar range. The application of **BiR** in the photodynamic treatment of a mouse xenograft model revealed the degradation of **BiR** upon irradiation, although the fluorescence signal was stable in the xenograft tumor during *in vitro* imaging.

Later in 2021 Mukaimine *et al.*[123] reported a series of asymmetric bismuth rhodamines. The two derivatives synthesized were trimethyl rhodamine bearing 2,4-dimethoxyphenyl group at the 9-position with phenylbismuthane as the bridging moiety, **BiRNH** (Table 10:E21, Scheme 16) and its acetylated derivative **BiRAc** (Table 10:E22, Scheme 16). **BiRNH** showed absorption in the red region and efficient photoinduced generation of singlet oxygen, whereas **BiRAc** showed blue-shifted absorption and did not show a photosensitizing effect. By taking advantage of these properties, a new fluorogenic sensitizer, **BiRGlu** (Table 10:E23, Scheme 16), was generated. The fluorescence and photosensitization by **BiRGlu** were both activated in the presence of enzymatic activity of GGT. Cell studies revealed that **BiRGlu** was converted into **BiRNH** in the GGT-active A549 cell line, which increased fluorescence and phototoxicity upon red light irradiation. The results obtained from the SKOV3ip1 cell line with low GGT activity showed no activation of fluorescence or phototoxicity, revealing that in addition to being a fluorogenic indicator, **BiRGlu** was also a selectively activatable photosensitizer targeting GGT activity.

Gebremedhin *et al.*[124] reported benzo[a]phenoselenazine-based dye with 1,6-diaminohexane linker, **NBSe-C6-NH$_2$**, (Table 10:E24, Scheme 16) and functionalized with biotin ligand yielding **Se-Biotin** (Table 10, E25, Scheme 16). The photophysical studies revealed that be-Biotin and its precursor, NBSe-C6-NH$_2$, showed the same absorption and emission maxima with identical singlet oxygen generation efficiency and $IC_{50}$ values. Se-Biotin demonstrated good tumor targeting ability by showing complete discrimination between MCF-7 tumor cells and COS-7 health cells under different co-culture conditions. Both benzo[a]phenoselenazine dyes showed lysosome localization due to their hydrophobic and weak basic characters inducing the lysosome-disruption pathway, which resulted in cell apoptosis and necrosis. Compared to its non-targeted precursor, NBSe-C6-NH$_2$, the dark toxicity of the Se-Biotin was slightly lower against MCF-7 cells (figure 15). The results revealed that Se-Biotin could be a promising PDT agent as a novel tumor and lysosome dual-targeting PDT agent.

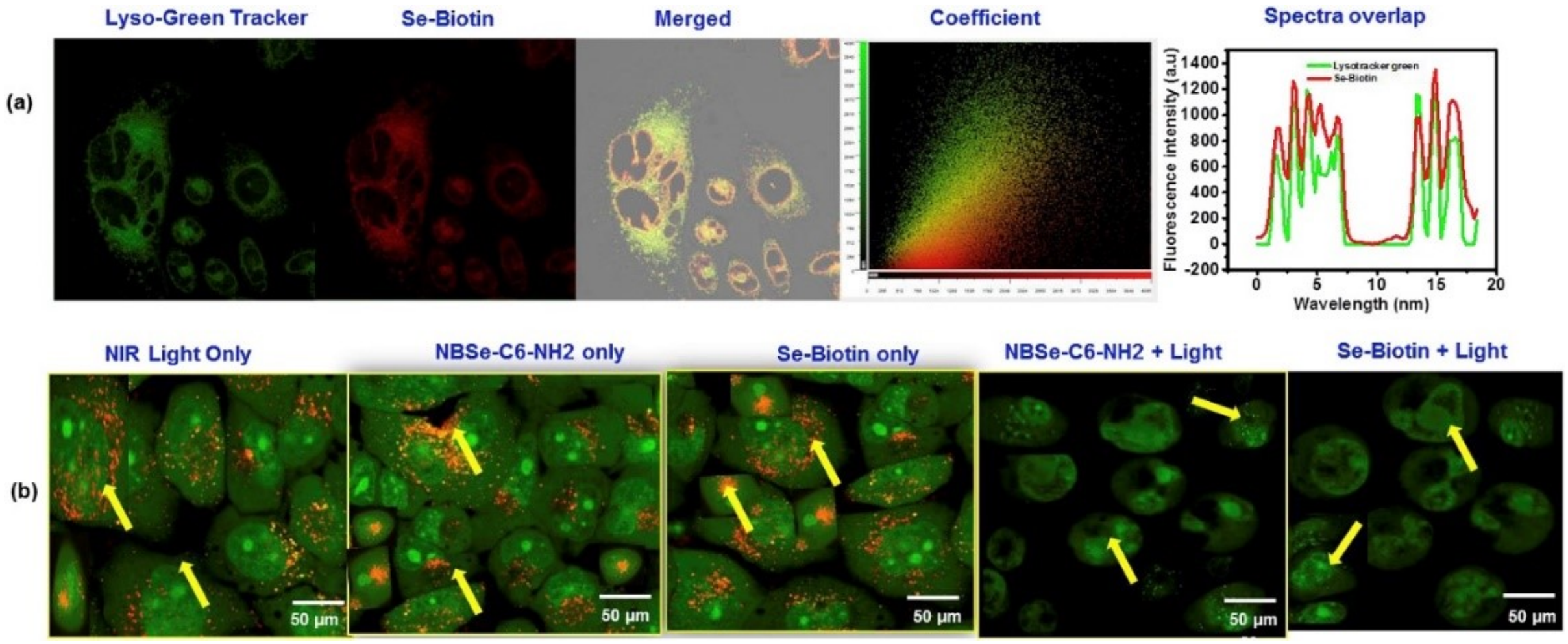

**Figure 15.** (a) Co-localization of Se-Biotin (2.5 µM) with Lyso-Green tracker (1.0 µM). Green Channel ($\lambda_{ex}$ = 488 nm, λem = 510–550 nm), and Red Channel ($\lambda_{ex}$ = 640 nm, $\lambda_{em}$ = 650–700 nm). (b) Acridine orange (AO) staining and imaging of MCF-7 cells treated with Se-Biotin or NBSe–C6–NH2 with or without PDT light at 660 nm light (6 J/cm2). Cells were stained with AO (2.5 µM). All images were acquired from CLSMth, the excitation wavelength was fixed at 488 nm, and the emission region was scanned between 515 and 545 nm (green channel) and 610–640 nm (red channel). Reproduced with permission [124] copyright 2019, Elsevier

Despite the efforts to employ Si-fluoresceins in many different bioimaging studies, the dye structure has not been utilized in photodynamic therapy (PDT) applications. Gunbas and Kolemen groups reported a highly cytotoxic photosensitizer (PS) from the Si-fluorescein core for the first time in literature.[125] The classical spirolactonized Si-fluorescein core with dimethylsilane group as the bridging moiety was tetra iodinated from positions 2, 4, 5, and 7 to facilitate effective heavy atom mediated intersystem crossing (ISC) upon excitation. The Si-fluorescein PS, **SF-I** (Table 10:E26, Scheme 14) showed absorption and emission maxima positioned at 614 and 630 nm ($\Phi_f$ = 0.11 in PBS buffer containing 0.5% DMSO). Compared to its non-iodinated form, **SF**, a red shift of 34 and 32 nm was observed for the absorption and emission maxima, respectively, for **SF-I**. The singlet oxygen generation efficiency was calculated as 45% in PBS buffer upon irradiation with a 595 nm LED using methylene blue as reference. Cytotoxicity was tested in two cancer cells, in triple-negative breast (TNBS, MDA MB-231) and colorectal (HCT-116) cancers with limited chemotherapy options. *In vitro* studies showed that the **SF-I** induces cell death with PDT action and acts as a fluorophore simultaneously (figure 16). Apart from being the first example of a PS with a Si-fluorescein core, **SF-I** displays high water solubility, red-shifted absorption/emission signals, high photostability, and negligible dark toxicity, fulfilling the desirable properties of a theranostic agent.

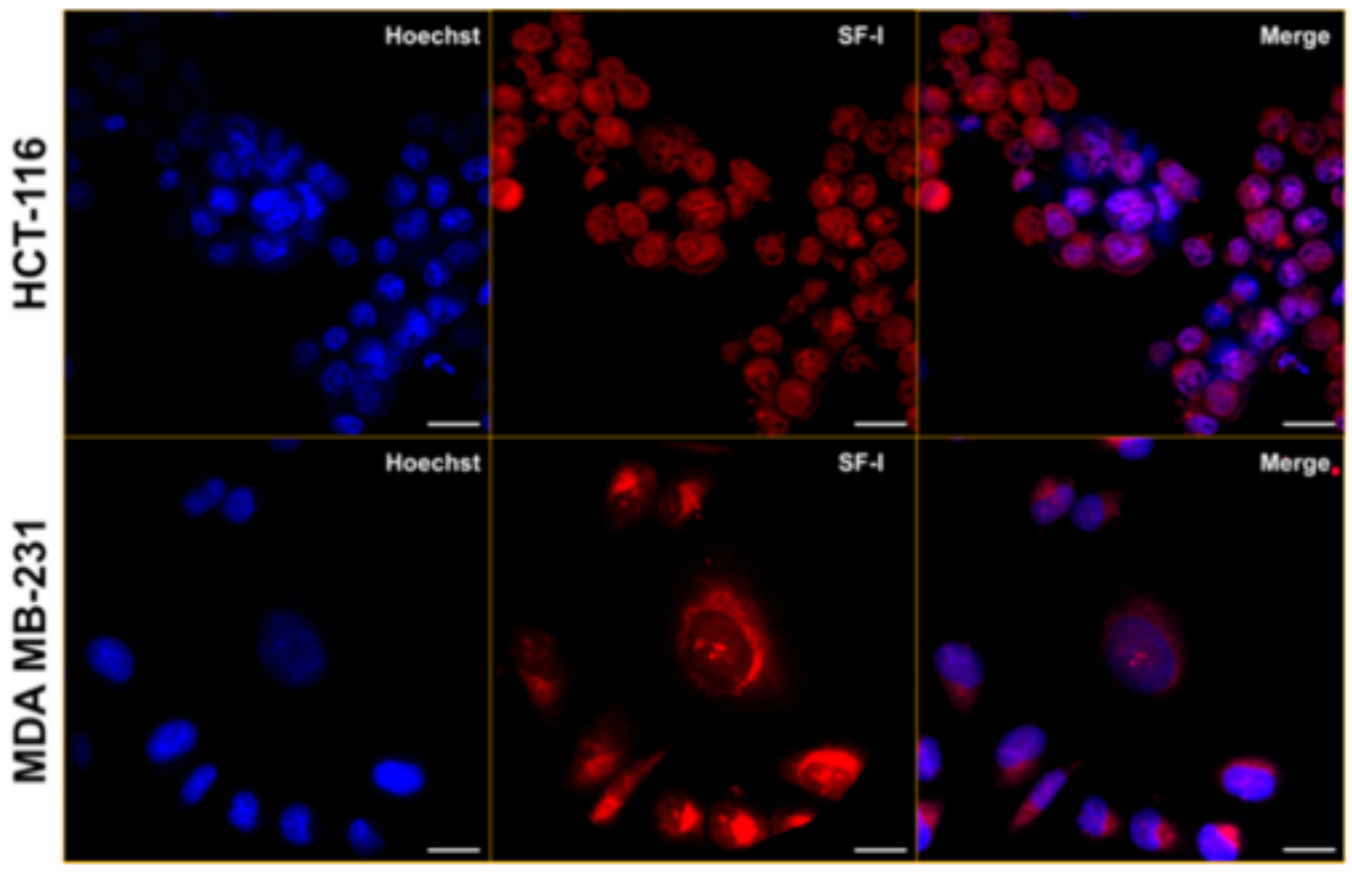

**Figure 16.** Confocal images of SF-I (5 µM) in HCT-116 (a) and MDA MB-231 (b) cancer cells after 2 h incubation. Blue, Hoechst 33342; red, SF-I. Scale bar: 20 µm. Reproduced with permission [125] Copyright 2021, American Chemical Society

# 7. Outlook

This review presents the design strategies, photophysical properties, fluorescence imaging applications, and utilization as PDT agents of selected hybrid xanthene dyes that emerged in the last decade. Hybrid xanthene dyes, particularly ones bearing silicon or phosphine oxide on the 10th position of xanthene moiety, became increasingly popular due to their excellent photochemical properties in aqueous media, high fluorescence quantum yield, photostability, and suitable absorption and emission maxima for deep tissue imaging and treatment. Based on these advantages, several xanthene-based dyes have been synthesized, and their bio-imaging capability for various applications were extensively and successfully demonstrated. However, the collected data in the tables presented here clearly show that the in-vivo demonstrations, a key data required for successful implementation of fluorescence imaging as an alternative tool to the current state of the art, are rare. Further studies should strive to implement in-vivo analyses toward more impactful growth in the field of fluorescence imaging. Additionally, deep tumor visualization required significantly red-shifted emission, which is rarely demonstrated with xanthene-based dyes. Additional design principles and molecular structures need to be unveiled for the realization of dyes with over 800 nm emission, as well as molecular systems that can demonstrate NIR-II imaging utilizing xanthene-based cores for implementing their aforementioned advantages. Although a plethora of successful fluoresce imaging studies have been published, the outlook for utilization of xanthene-based systems in treatment, specifically for PDT application, is quite grim. The body of work published related to PDT studies based on xanthene dyes, although quite promising, is in its infancy, and there is much ground to cover towards utilizing these materials in clinical work in the coming future. The immense potential of PDT and treatment modalities where PDT and other methods such as PTT and chemotherapy are utilized together are far from being fulfilled. One of the leading reasons behind this is the significant number of characteristic properties that a candidate PDT agent needs to satisfy, including but not limited to water solubility, photostability, effective targeting, long-wavelength absorption, low cytotoxicity, appropriate pharmacokinetics combined with cost-effective synthesis and implementation. The list can be daunting, but we believe xanthene-based molecular frameworks are poised to greatly impact the field based on their extremely successful implementation in fluoresce imaging and initial performance as PDT agents. Additionally, with the guiding light of immense work on PDT action combined with recent drug delivery systems advances, success should be within reach of the community in the coming years.

## Acknowledgements

This work received funding from the European Research Council (ERC) under the European Union's Horizon 2020 research and innovation program (grant agreement no. [852614]).

## References

[1] S.L. Anwar, G. Adistyawan, W. Wulaningsih, C. Gutenbrunner, B. Nugraha, Rehabilitation for cancer survivors: How we can reduce the healthcare service inequality in low- and middle-income countries, Am. J. Phys. Med. Rehabil. 97 (2018) 764–771.

[2] A. Jemal, F. Bray, M.M. Center, J. Ferlay, E. Ward, D. Forman, Global cancer statistics, CA Cancer J. Clin. 61 (2011) 69–90.

[3] P.R. Srinivas, B.S. Kramer, S. Srivastava, Trends in biomarker research for cancer detection, Lancet Oncol. 2 (2001) 698–704.

[4] X. Chen, J. Gole, A. Gore, Q. He, M. Lu, J. Min, Z. Yuan, X. Yang, Y. Jiang, T. Zhang, C. Suo, X. Li, L. Cheng, Z. Zhang, H. Niu, Z. Li, Z. Xie, H. Shi, X. Zhang, M. Fan, X. Wang, Y. Yang, J. Dang, C. McConnell, J. Zhang, J. Wang, S. Yu, W. Ye, Y. Gao, K. Zhang, R. Liu, L. Jin, Non-invasive early detection of cancer four years before conventional diagnosis using a blood test, Nat. Commun. 11 (2020).

[5] T. Terai, T. Nagano, Fluorescent probes for bioimaging applications, Curr. Opin. Chem. Biol. 12 (2008) 515–521.

[6] D.E.J.G.J. Dolmans, D. Fukurmura, R.K. Jain, Photodynamic therapy for cancer, Nat. Rev. Cancer. 3 (2003) 380–387.

[7] T.J. Dougherty, C.J. Gomer, B.W. Henderson, G. Jori, D. Kessel, M. Korbelik, J. Moan, Q. Peng, REVIEW Photodynamic Therapy, 1998. http://jnci.oxfordjournals.org/.

[8] T. Hasan, B. Ortel, A.C.E. Moor, B.W. Pogue, Photodynamic Therapy of Cancer, in: 2003: pp. 605–622.

[9] Z. Guo, S. Park, J. Yoon, I. Shin, Recent progress in the development of near-infrared fluorescent probes for bioimaging applications, Chem. Soc. Rev. 43 (2014) 16–29.

[10] T.C. Pham, V.N. Nguyen, Y. Choi, S. Lee, J. Yoon, Recent Strategies to Develop Innovative Photosensitizers for Enhanced Photodynamic Therapy, Chem. Rev. 121 (2021) 13454–13619.

[11] J. Zheng, Y. Liu, F. Song, L. Jiao, Y. Wu, X. Peng, A nitroreductase-activatable near-infrared theranostic photosensitizer for photodynamic therapy under mild hypoxia, Chem. Commun. 56 (2020) 5819–5822.

[12] S. Kolemen, E.U. Akkaya, Reaction-based BODIPY probes for selective bio-imaging, Coord. Chem. Rev. 354 (2018) 121–134.

[13] Y. Ni, J. Wu, Far-red and near infrared BODIPY dyes: Synthesis and applications for fluorescent pH probes and bio-imaging, Org. Biomol. Chem. 12 (2014) 3774–3791.

[14] Z. Liu, Z. Jiang, M. Yan, X. Wang, Recent Progress of BODIPY Dyes With Aggregation-Induced Emission, Fron.t Chem. 7 (2019).

[15] M. Liu, S. Ma, M. She, J. Chen, Z. Wang, P. Liu, S. Zhang, J. Li, Structural modification of BODIPY: Improve its applicability, Chin. Chem. Lett. 30 (2019) 1815–1824.
[16] K. Chansaenpak, S. Tanjindaprateep, N. Chaicharoenaudomrung, O. Weeranantanapan, P. Noisa, A. Kamkaew, Aza-BODIPY based polymeric nanoparticles for cancer cell imaging, RSC Adv. 8 (2018) 39248–39255.
[17] X. Chen, T. Pradhan, F. Wang, J.S. Kim, J. Yoon, Fluorescent chemosensors based on spiroring-opening of xanthenes and related derivatives, Chem. Rev. 112 (2012) 1910–1956.
[18] B. Chen, C. Lv, X. Tang, Chemoselective reduction-based fluorescence probe for detection of hydrogen sulfide in living cells, Anal. Bioanal. Chem. 404 (2012) 1919–1923.
[19] T. Almammadov, G. Atakan, O. Leylek, G. Ozcan, G. Gunbas, S. Kolemen, Resorufin Enters the Photodynamic Therapy Arena: A Monoamine Oxidase Activatable Agent for Selective Cytotoxicity, ACS Med. Chem. Lett. 11 (2020) 2491–2496.
[20] H.J. Chen, X. bin Zhou, A.L. Wang, B.Y. Zheng, C.K. Yeh, J.D. Huang, Synthesis and biological characterization of novel rose bengal derivatives with improved amphiphilicity for sono-photodynamic therapy, Eur. J. Med. Chem. 145 (2018) 86–95.
[21] P. Pal, H. Zeng, G. Durocher, D. Girard, T. Liz, A.K. Gupta, R. Giasson, L. Blanchard, L. Gaboury, A. Balassy, C. Turme, A. Laperriere, L. Villeneuve, Phototoxicity of Some Bromine-Substituted Rhodamine Dyes: Synthesis, Photophysical Properties and Application as Photosensitizers, 1996.
[22] Y. Xiao, X. Qian, Substitution of oxygen with silicon: A big step forward for fluorescent dyes in life science, Coord. Chem. Rev. 423 (2020).
[23] V. Balzani, A. Credi, M. Venturi, Photochemistry and photophysics of coordination compounds: An extended view, Coord. Chem. Rev. 171 (1998) 3–16.
[24] C.-X. Yin, K.-M. Xiong, F.-J. Huo, J.C. Salamanca, R.M. Strongin, Fluoreszenzsonden mit mehreren Bindungsstellen unterscheiden zwischen Cys, Hcy und GSH, Angew. Chem. 129 (2017) 13368–13379.
[25] T. Egawa, Y. Koide, K. Hanaoka, T. Komatsu, T. CooTeraiper, T. Nagano, Development of a fluorescein analogue, TokyoMagenta, as a novel scaffold for fluorescence probes in red region, Chem. Commun. 47 (2011) 4162–4164.
[26] L. Crovetto, A. Orte, J.M. Paredes, S. Resa, J. Valverde, F. Castello, D. Miguel, J.M. Cuerva, E.M. Talavera, J.M. Alvarez-Pez, Photophysics of a Live-Cell-Marker, Red Silicon-Substituted Xanthene Dye, J. Phys. Chem. A. 119 (2015) 10854–10862.
[27] K. Hirabayashi, K. Hanaoka, T. Takayanagi, Y. Toki, T. Egawa, M. Kamiya, T. Komatsu, T. Ueno, T. Terai, K. Yoshida, M. Uchiyama, T. Nagano, Y. Urano, Analysis of Chemical Equilibrium of Silicon-Substituted Fluorescein and Its Application to Develop a Scaffold for Red Fluorescent Probes, Anal. Chem. 87 (2015) 9061–9069.
[28] J.B. Grimm, T.A. Brown, A.N. Tkachuk, L.D. Lavis, General Synthetic Method for Si-Fluoresceins and Si-Rhodamines, ACS Cent. Sci. 3 (2017) 975–985.
[29] L.P. Smaga, N.W. Pino, G.E. Ibarra, V. Krishnamurthy, J. Chan, A Photoactivatable Formaldehyde Donor with Fluorescence Monitoring Reveals Threshold to Arrest Cell Migration, J. Am. Chem. Soc. 142 (2020) 680–684.
[30] L. Espinar-Barranco, P. Luque-Navarro, M.A. Strnad, P. Herrero-Foncubierta, L. Crovetto, D. Miguel, M.D. Giron, A. Orte, J.M. Cuerva, R. Salto, J.M. Alvarez-Pez, J.M. Paredes, A solvatofluorochromic silicon-substituted xanthene dye useful in bioimaging, Dyes Pigm. 168 (2019) 264–272.
[31] T.E. Bearrood, G. Aguirre-Figueroa, J. Chan, Rational Design of a Red Fluorescent Sensor for ALDH1A1 Displaying Enhanced Cellular Uptake and Reactivity, Bioconjug. Chem. 31 (2020) 224–228..
[32] J. Catalán, Toward a generalized treatment of the solvent effect based on four empirical scales: Dipolarity (SdP, a new scale), polarizability (SP), acidity (SA), and basicity (SB) of the medium, J. Phys. Chem. B. 113 (2009) 5951–5960.
[33] A. Fukazawa, S. Suda, M. Taki, E. Yamaguchi, M. Grzybowski, Y. Sato, T. Higashiyama, S. Yamaguchi, Phospha-fluorescein: A red-emissive fluorescein analogue with high photobleaching resistance, Chem. Commun. 52 (2016) 1120–1123.

[34] Y. Urano, M. Kamiya, K. Kanda, T. Ueno, K. Hirose, T. Nagano, Evolution of fluorescein as a platform for finely tunable fluorescence probes, J. Am. Chem. Soc. 127 (2005) 4888–4894.
[35] X. Chai, X. Cui, B. Wang, F. Yang, Y. Cai, Q. Wu, T. Wang, Near-Infrared Phosphorus-Substituted Rhodamine with Emission Wavelength above 700 nm for Bioimaging, Chem.--Eur. J. 21 (2015) 16754–16758.
[36] X. Zhou, R. Lai, J.R. Beck, H. Li, C.I. Stains, Nebraska Red: A phosphinate-based near-infrared fluorophore scaffold for chemical biology applications, Chem. Commun. 52 (2016) 12290–12293.
[37] A. Fukazawa, J. Usuba, R.A. Adler, S. Yamaguchi, Synthesis of seminaphtho-phospha-fluorescein dyes based on the consecutive arylation of aryldichlorophosphines, Chem. Commun. 53 (2017) 8565–8568.
[38] M. Grzybowski, M. Taki, S. Yamaguchi, Selective Conversion of P=O-Bridged Rhodamines into P=O-Rhodols: Solvatochromic Near-Infrared Fluorophores, Chem.--Eur. J. 23 (2017) 13028–13032.
[39] H. Ogasawara, M. Grzybowski, R. Hosokawa, Y. Sato, M. Taki, S. Yamaguchi, A far-red fluorescent probe based on a phospha-fluorescein scaffold for cytosolic calcium imaging, Chem. Commun. 54 (2018) 299–302.
[40] Y. Fang, G.N. Good, X. Zhou, C.I. Stains, Phosphinate-containing rhodol and fluorescein scaffolds for the development of bioprobes, Chem. Commun. 55 (2019) 5962–5965.
[41] H. Ogasawara, Y. Tanaka, M. Taki, S. Yamaguchi, Late-stage functionalisation of alkyne-modified phospha-xanthene dyes: lysosomal imaging using an off–on–off type of pH probe, Chem. Sci. 12 (2021) 7902–7907.
[42] Yang Youjun, Lowry Mark, Xu Xiangyang, Escobedo Jorge O., Sibrian-Vazquez Martha, Wong Lisa, Schowalter Corin M., Jensen Timothy J., Fronczek Frank R., Warner Isiah M, Strongin Robert M., Seminaphthofluorones are a family of water-soluble, low molecular weight, NIR-emitting fluorophores, PNAS. 105 (2008) 8829–8834.
[43] Y. Koide, Y. Urano, K. Hanaoka, T. Terai, T. Nagano, Evolution of group 14 rhodamines as platforms for near-infrared fluorescence probes utilizing photoinduced electron transfer, in: ACS Chem. Biol, 2011: pp. 600–608.
[44] T. Myochin, K. Hanaoka, T. Komatsu, T. Terai, T. Nagano, Design strategy for a near-infrared fluorescence probe for matrix metalloproteinase utilizing highly cell permeable boron dipyrromethene, J. Am. Chem. Soc. 134 (2012) 13730–13737.
[45] T. Myochin, K. Hanaoka, S. Iwaki, T. Ueno, T. Komatsu, T. Terai, T. Nagano, Y. Urano, Development of a Series of Near-Infrared Dark Quenchers Based on Si-rhodamines and Their Application to Fluorescent Probes, J. Am. Chem. Soc. 137 (2015) 4759–4765.
[46] Y. Koide, Y. Urano, K. Hanaoka, W. Piao, M. Kusakabe, N. Saito, T. Terai, T. Okabe, T. Nagano, Development of NIR fluorescent dyes based on Si-rhodamine for in vivo imaging, J. Am. Chem. Soc. 134 (2012) 5029–5031.
[47] G. Lukinavičius, K. Umezawa, N. Olivier, A. Honigmann, G. Yang, T. Plass, V. Mueller, L. Reymond, I.R. Corrêa, Z.G. Luo, C. Schultz, E.A. Lemke, P. Heppenstall, C. Eggeling, S. Manley, K. Johnsson, A near-infrared fluorophore for live-cell super-resolution microscopy of cellular proteins, Nat. Chem. 5 (2013) 132–139.
[48] S.N. Uno, M. Kamiya, T. Yoshihara, K. Sugawara, K. Okabe, M.C. Tarhan, H. Fujita, T. Funatsu, Y. Okada, S. Tobita, Y. Urano, A spontaneously blinking fluorophore based on intramolecular spirocyclization for live-cell super-resolution imaging, Nat. Chem. 6 (2014) 681–689.
[49] J.B. Grimm, B.P. English, J. Chen, J.P. Slaughter, Z. Zhang, A. Revyakin, R. Patel, J.J. Macklin, D. Normanno, R.H. Singer, T. Lionnet, L.D. Lavis, A general method to improve fluorophores for live-cell and single-molecule microscopy, Nat. Methods. 12 (2015) 244–250.
[50] S. Kim, T. Tachikawa, M. Fujitsuka, T. Majima, Far-red fluorescence probe for monitoring singlet oxygen during photodynamic therapy, J. Am. Chem. Soc. 136 (2014) 11707–11715.

[51] J.B. Grimm, T. Klein, B.G. Kopek, G. Shtengel, H.F. Hess, M. Sauer, L.D. Lavis, Synthesis of a Far-Red Photoactivatable Silicon-Containing Rhodamine for Super-Resolution Microscopy, Angew. Chem. 128 (2016) 1755–1759.
[52] B. Wang, X. Chai, W. Zhu, T. Wang, Q. Wu, A general approach to spirolactonized Si-rhodamines, Chem. Commun. 50 (2014) 14374–14377.
[53] B. Wang, S. Yu, X. Chai, T. Li, Q. Wu, T. Wang, A Lysosome-Compatible Near-Infrared Fluorescent Probe for Targeted Monitoring of Nitric Oxide, Chem.--Eur. J. 22 (2016) 5649–5656.
[54] B. Wang, X. Cui, Z. Zhang, X. Chai, H. Ding, Q. Wu, Z. Guo, T. Wang, A six-membered-ring incorporated Si-rhodamine for imaging of copper(II) in lysosomes, Org. Biomol. Chem. 14 (2016) 6720–6728.
[55] L.G. Meimetis, R.J. Giedt, H. Mikula, J.C. Carlson, R.H. Kohler, D.B. Pirovich, R. Weissleder, Fluorescent vinblastine probes for live cell imaging, Chem. Commun. 52 (2016) 9953–9956.
[56] T.F. Brewer, C.J. Chang, An Aza-Cope Reactivity-Based Fluorescent Probe for Imaging Formaldehyde in Living Cells, J. Am. Chem. Soc. 137 (2015) 10886–10889.
[57] X. Zhou, L. Lesiak, R. Lai, J.R. Beck, J. Zhao, C.G. Elowsky, H. Li, C.I. Stains, Chemoselective Alteration of Fluorophore Scaffolds as a Strategy for the Development of Ratiometric Chemodosimeters, Angew. Chem. 129 (2017) 4261–4264.
[58] J.B. Grimm, A.K. Muthusamy, Y. Liang, T.A. Brown, W.C. Lemon, R. Patel, R. Lu, J.J. Macklin, P.J. Keller, N. Ji, L.D. Lavis, A general method to fine-tune fluorophores for live-cell and in vivo imaging, Nat. Methods. 14 (2017) 987–994.
[59] J. Tang, Z. Guo, Y. Zhang, B. Bai, W.H. Zhu, Rational design of a fast and selective near-infrared fluorescent probe for targeted monitoring of endogenous nitric oxide, Chem. Commun. 53 (2017) 10520–10523.
[60] H. Zhang, J. Liu, C. Liu, P. Yu, M. Sun, X. Yan, J.P. Guo, W. Guo, Imaging lysosomal highly reactive oxygen species and lighting up cancer cells and tumors enabled by a Si-rhodamine-based near-infrared fluorescent probe, Biomaterials. 133 (2017) 60–69.
[61] K. Umezawa, M. Yoshida, M. Kamiya, T. Yamasoba, Y. Urano, Rational design of reversible fluorescent probes for live-cell imaging and quantification of fast glutathione dynamics, Nat. Chem. 9 (2017) 279–286.
[62] J. Tang, Q. Li, Z. Guo, W. Zhu, A fast-response and highly specific Si-Rhodamine probe for endogenous peroxynitrite detection in living cells, Org. Biomol. Chem. 17 (2019) 1875–1880.
[63] A. Ogasawara, M. Kamiya, K. Sakamoto, Y. Kuriki, K. Fujita, T. Komatsu, T. Ueno, K. Hanaoka, H. Onoyama, H. Abe, Y. Tsuji, M. Fujishiro, K. Koike, M. Fukayama, Y. Seto, Y. Urano, Red Fluorescence Probe Targeted to Dipeptidylpeptidase-IV for Highly Sensitive Detection of Esophageal Cancer, Bioconjug. Chem. 30 (2019) 1055–1060.
[64] J. Sung, J.G. Rho, G.G. Jeon, Y. Chu, J.S. Min, S. Lee, J.H. Kim, W. Kim, E. Kim, A New Infrared Probe Targeting Mitochondria via Regulation of Molecular Hydrophobicity, Bioconjug. Chem. 30 (2019) 210–217.
[65] Y. Chu, M.C. Shin, J. Sung, J. Park, E. Kim, S. Lee, Development of Theragnostic Tool Using NIR Fluorescence Probe Targeting Mitochondria in Glioma Cells, Bioconjug Chem. 30 (2019) 1642–1648.
[66] H. Zhang, K. Li, L.L. Li, K.K. Yu, X.Y. Liu, M.Y. Li, N. Wang, Y.H. Liu, X.Q. Yu, Pyridine-Si-xanthene: A novel near-infrared fluorescent platform for biological imaging, Chin. Chem. Lett. 30 (2019) 1063–1066.
[67] K. Numasawa, K. Hanaoka, N. Saito, Y. Yamaguchi, T. Ikeno, H. Echizen, M. Yasunaga, T. Komatsu, T. Ueno, M. Miura, T. Nagano, Y. Urano, A Fluorescent Probe for Rapid, High-Contrast Visualization of Folate-Receptor-Expressing Tumors In Vivo, Angew. Chem.132 (2020) 6071–6076.
[68] J.B. Grimm, A.N. Tkachuk, L. Xie, H. Choi, B. Mohar, N. Falco, K. Schaefer, R. Patel, Q. Zheng, Z. Liu, J. Lippincott-Schwartz, T.A. Brown, L.D. Lavis, A general method to optimize and functionalize red-shifted rhodamine dyes, Nat. Methods. 17 (2020) 815–821.

[69] K. Numasawa, K. Hanaoka, T. Ikeno, H. Echizen, T. Ishikawa, M. Morimoto, T. Komatsu, T. Ueno, Y. Ikegaya, T. Nagano, Y. Urano, A cytosolically localized far-red to near-infrared rhodamine-based fluorescent probe for calcium ions, Analyst. 145 (2020) 7736–7740.
[70] O. Murata, Y. Shindo, Y. Ikeda, N. Iwasawa, D. Citterio, K. Oka, Y. Hiruta, Near-infrared fluorescent probes for imaging of intracellular mg2+ and application to multi-color imaging of mg2+, atp, and mitochondrial membrane potential, Anal. Chem. 92 (2020) 966–974.
[71] Z. Zhou, X. Yuan, D. Long, M. Liu, K. Li, Y. Xie, A pyridine-Si-rhodamine-based near-infrared fluorescent probe for visualizing reactive oxygen species in living cells, Spectrochim. Acta. A. Mol. Biomol. Spectrosc. 246 (2021).
[72] X. Liu, X. Gong, J. Yuan, X. Fan, X. Zhang, T. Ren, S. Yang, R. Yang, L. Yuan, X.B. Zhang, Dual-Stimulus Responsive Near-Infrared Reversible Ratiometric Fluorescent and Photoacoustic Probe for in Vivo Tumor Imaging, Anal. Chem. 93 (2021) 5420–5429.
[73] T. Wang, Q.J. Zhao, H.G. Hu, S.C. Yu, X. Liu, L. Liu, Q.Y. Wu, Spirolactonized Si-rhodamine: A novel NIR fluorophore utilized as a platform to construct Si-rhodamine-based probes, Chem. Commun. 48 (2012) 8781–8783.
[74] Y. Kushida, K. Hanaoka, T. Komatsu, T. Terai, T. Ueno, K. Yoshida, M. Uchiyama, T. Nagano, Red fluorescent scaffold for highly sensitive protease activity probes, Bioorg. Med. Chem. Lett. 22 (2012) 3908–3911.
[75] E. Kim, K.S. Yang, R.J. Giedt, R. Weissleder, Red Si-rhodamine drug conjugates enable imaging in GFP cells, Chem. Commun.50 (2014) 4504–4507.
[76] E. Kim, K.S. Yang, R.H. Kohler, J.M. Dubach, H. Mikula, R. Weissleder, Optimized Near-IR Fluorescent Agents for in Vivo Imaging of Btk Expression, Bioconjug. Chem. 26 (2015) 1513–1518.
[77] A. Turetsky, E. Kim, R.H. Kohler, M.A. Miller, R. Weissleder, Single cell imaging of Bruton's Tyrosine Kinase using an irreversible inhibitor, Sci. Rep. 4 (2014).
[78] A.N. Butkevich, Modular Synthetic Approach to Silicon-Rhodamine Homologues and Analogues via Bis-aryllanthanum Reagents, Org Lett. 23 (2021) 2604–2609.
[79] A.N. Butkevich, V.N. Belov, K. Kolmakov, V. v. Sokolov, H. Shojaei, S.C. Sidenstein, D. Kamin, J. Matthias, R. Vlijm, J. Engelhardt, S.W. Hell, Hydroxylated Fluorescent Dyes for Live-Cell Labeling: Synthesis, Spectra and Super-Resolution STED, Chem.--Eur. J. 23 (2017) 12114–12119.
[80] M. Grzybowski, M. Taki, K. Senda, Y. Sato, T. Ariyoshi, Y. Okada, R. Kawakami, T. Imamura, S. Yamaguchi, A Highly Photostable Near-Infrared Labeling Agent Based on a Phospha-rhodamine for Long-Term and Deep Imaging, Angew. Chem., Int. Ed. 57 (2018) 10137–10141.
[81] X. Lv, C. Gao, T. Han, H. Shi, W. Guo, Improving the quantum yields of fluorophores by inhibiting twisted intramolecular charge transfer using electron-withdrawing group-functionalized piperidine auxochromes, Chem. Commun. 56 (2020) 715–718.
[82] M. Sauer, V. Nasufovic, H.D. Arndt, I. Vilotijevic, Robust synthesis of NIR-emissive P-rhodamine fluorophores, Org. Biomol. Chem. 18 (2020) 1567–1571.
[83] L. Lesiak, X. Zhou, Y. Fang, J. Zhao, J.R. Beck, C.I. Stains, Imaging GPCR internalization using near-infrared Nebraska red-based reagents, Org. Biomol. Chem. 18 (2020) 2459–2467.
[84] M. Grzybowski, M. Taki, K. Kajiwara, S. Yamaguchi, Effects of Amino Group Substitution on the Photophysical Properties and Stability of Near-Infrared Fluorescent P-Rhodamines, Chem.--Eur. J. 26 (2020) 7912–7917.
[85] F. Deng, L. Liu, W. Huang, C. Huang, Q. Qiao, Z. Xu, Systematic study of synthesizing various heteroatom-substituted rhodamines from diaryl ether analogues, Spectrochim. Acta. A. Mol. Biomol. Spectrosc. 240 (2020).
[86] M.A. Gonzalez, A.S. Walker, K.J. Cao, J.R. Lazzari-Dean, N.S. Settineri, E.J. Kong, R.H. Kramer, E.W. Miller, Voltage Imaging with a NIR-Absorbing Phosphine Oxide Rhodamine Voltage Reporter, J. Am. Chem. Soc. 143 (2021) 2304–2314.
[87] X. Song, A. Johnson, J. Foley, 7-azabicyclo[2.2.1]heptane as a unique and effective dialkylamino auxochrome moiety: Demonstration in a fluorescent rhodamine dye, J. Am. Chem. Soc. 130 (2008) 17652–17653.

[88] M. Fu, Y. Xiao, X. Qian, D. Zhao, Y. Xu, A design concept of long-wavelength fluorescent analogs of rhodamine dyes: Replacement of oxygen with silicon atom, Chem. Commun. (2008) 1780–1782.

[89] H.J. Tracy, J.L. Mullin, W.T. Klooster, J.A. Martin, J. Haug, S. Wallace, I. Rudloe, K. Watts, Enhanced photoluminescence from group 14 metalloles in aggregated and solid solutions, Inorg. Chem. 44 (2005) 2003–2011.

[90] P. Horváth, P. Šebej, T. Šolomek, P. Klán, Small-molecule fluorophores with large stokes shifts: 9-iminopyronin analogues as clickable tags, J. Org. Chem. 80 (2015) 1299–1311.

[91] E.L. Mertz, V.A. Tikhomirov, L.I. Krishtalik, A.N. Frumkin, Karpov Institute of Physical Chemistry, VorontsoVo pole, 1996.

[92] P. Horváth, P. Šebej, D. Kovář, J. Damborský, Z. Prokop, P. Klán, Fluorescent pH Indicators for Neutral to Near-Alkaline Conditions Based on 9-Iminopyronin Derivatives, ACS Omega. 4 (2019) 5479–5485.

[93] H. Nie, J. Jing, Y. Tian, W. Yang, R. Zhang, X. Zhang, Reversible and Dynamic Fluorescence Imaging of Cellular Redox Self-Regulation Using Fast-Responsive Near-Infrared Ge-Pyronines, ACS Appl. Mater. Interfaces. 8 (2016) 8991–8997.

[94] H. Nie, L. Qiao, W. Yang, B. Guo, F. Xin, J. Jing, X. Zhang, UV-assisted synthesis of long-wavelength Si-pyronine fluorescent dyes for real-time and dynamic imaging of glutathione fluctuation in living cells, J. Mater. Chem. B. 4 (2016) 4826–4831.

[95] N. Shimomura, Y. Egawa, R. Miki, T. Fujihara, Y. Ishimaru, T. Seki, A red fluorophore comprising a borinate-containing xanthene analogue as a polyol sensor, Org. Biomol. Chem. 14 (2016) 10031–10036.

[96] A.N. Butkevich, G. Lukinavičius, E. D'Este, S.W. Hell, Cell-Permeant Large Stokes Shift Dyes for Transfection-Free Multicolor Nanoscopy, J. Am. Chem. Soc. 139 (2017) 12378–12381.

[97] G. Dejouy, M. Laly, I.E. Valverde, A. Romieu, Synthesis, stability and spectral behavior of fluorogenic sulfone-pyronin and sulfone-rosamine dyes, Dyes Pigm. 159 (2018) 262–274.

[98] A. Romieu, G. Dejouy, I.E. Valverde, Quest for novel fluorogenic xanthene dyes: Synthesis, spectral properties and stability of 3-imino-3H-xanthen-6-amine (pyronin) and its silicon analog, Tetrahedron Lett. 59 (2018) 4574–4581.

[99] G. Dejouy, K. Renault, I.E. Valverde, A. Romieu, Synthetic routes to novel fluorogenic pyronins and silicon analogs with far-red spectral properties and enhanced aqueous stability, Dyes Pigm. 180 (2020).

[100] E.J. Kim, A. Podder, M. Maiti, J.M. Lee, B.G. Chung, S. Bhuniya, Selective monitoring of vascular cell senescence via Β-Galactosidase detection with a fluorescent chemosensor, Sens. Actuators B. Chem. 274 (2018) 194–200.

[101] S. Bhuniya, S. Maiti, E.-J. Kim, H. Lee, J.L. Sessler, K.S. Hong, J.S. Kim, An Activatable Theranostic for Targeted Cancer Therapy and Imaging, Angew. Chem. 126 (2014) 4558–4563.

[102] Y. Yuan, C.J. Zhang, S. Xu, B. Liu, A self-reporting AIE probe with a built-in singlet oxygen sensor for targeted photodynamic ablation of cancer cells, Chem. Sci. 7 (2016) 1862–1866.

[103] S. Jia, K.M. Ramos-Torres, S. Kolemen, C.M. Ackerman, C.J. Chang, Tuning the Color Palette of Fluorescent Copper Sensors through Systematic Heteroatom Substitution at Rhodol Cores, ACS Chem. Biol. 13 (2018) 1844–1852.

[104] H. Okuno, N. Ieda, Y. Hotta, M. Kawaguchi, K. Kimura, H. Nakagawa, A yellowish-green-light-controllable nitric oxide donor based on: N -nitrosoaminophenol applicable for photocontrolled vasodilation, Org. Biomol. Chem. 15 (2017) 2791–2796.

[105] H. Ito, Y. Kawamata, M. Kamiya, K. Tsuda-Sakurai, S. Tanaka, T. Ueno, T. Komatsu, K. Hanaoka, S. Okabe, M. Miura, Y. Urano, Red-Shifted Fluorogenic Substrate for Detection of lacZ-Positive Cells in Living Tissue with Single-Cell Resolution, Angew. Chem., Int. Ed. 57 (2018) 15702–15706.

[106] T. Doura, M. Kamiya, F. Obata, Y. Yamaguchi, T.Y. Hiyama, T. Matsuda, A. Fukamizu, M. Noda, M. Miura, Y. Urano, Detection of LacZ -Positive Cells in Living Tissue with Single-Cell Resolution , Angew. Chem. 128 (2016) 9772–9776.

[107] S. Yamaguchi, Y. Itami, K. Tamao, Group 14 metalloles with thienyl groups on 2,5-positions: Effects of group 14 elements on their π-electronic structures, Organometallics. 17 (1998) 4910–4916.
[108] Y. Koide, R. Kojima, K. Hanaoka, K. Numasawa, T. Komatsu, T. Nagano, H. Kobayashi, Y. Urano, Design strategy for germanium-rhodamine based pH-activatable near-infrared fluorescence probes suitable for biological applications, Commun. Chem. 2 (2019).
[109] N. Ando, H. Soutome, S. Yamaguchi, Near-infrared fluorescein dyes containing a tricoordinate boron atom, Chem. Sci. 10 (2019) 7816–7821.
[110] A. Choi, S.C. Miller, Silicon Substitution in Oxazine Dyes Yields Near-Infrared Azasiline Fluorophores That Absorb and Emit beyond 700 nm, Org Lett. 20 (2018) 4482–4485.
[111] L. Wang, J. Liu, S. Zhao, H. Zhang, Y. Sun, A. Wei, W. Guo, Fluorescence imaging of hypochlorous acid and peroxynitrite: In vitro and in vivo with emission wavelength beyond 750 nm, Chem. Commun. 56 (2020) 7718–7721.
[112] X. Zhang, H. Yu, Y. Xiao, Replacing phenyl ring with thiophene: An approach to longer wavelength aza-dipyrromethene boron difluoride (Aza-BODIPY) dyes, J. Org. Chem. 77 (2012) 669–673.
[113] T. Hirayama, A. Mukaimine, K. Nishigaki, H. Tsuboi, S. Hirosawa, K. Okuda, M. Ebihara, H. Nagasawa, Bismuth-rhodamine: A new red light-excitable photosensitizer, Dalton Trans. 46 (2017) 15991–15995.
[114] J.E. Hill, M.K. Linder, K.S. Davies, G.A. Sawada, J. Morgan, T.Y. Ohulchanskyy, M.R. Detty, Selenorhodamine photosensitizers for photodynamic therapy of P-glycoprotein-expressing cancer cells, J. Med. Chem. 57 (2014) 8622–8634.
[115] M.W. Kryman, K.S. Davies, M.K. Linder, T.Y. Ohulchanskyy, M.R. Detty, Selenorhodamine photosensitizers with the Texas-red core for photodynamic therapy of cancer cells, Bioorg. Med. Chem. 23 (2015) 4501–4507.
[116] M. Chiba, Y. Ichikawa, M. Kamiya, T. Komatsu, T. Ueno, K. Hanaoka, T. Nagano, N. Lange, Y. Urano, An Activatable Photosensitizer Targeted to γ-Glutamyltranspeptidase, Angew. Chem. 129 (2017) 10554–10558.
[117] W. Piao, K. Hanaoka, T. Fujisawa, S. Takeuchi, T. Komatsu, T. Ueno, T. Terai, T. Tahara, T. Nagano, Y. Urano, Development of an azo-based photosensitizer activated under mild hypoxia for photodynamic therapy, J. Am. Chem. Soc. 139 (2017) 13713–13719.
[118] Y. Ichikawa, M. Kamiya, F. Obata, M. Miura, T. Terai, T. Komatsu, T. Ueno, K. Hanaoka, T. Nagano, Y. Urano, Selective ablation of β-galactosidase-expressing cells with a rationally designed activatable photosensitizer, Angew. Chem. Int. Ed. 53 (2014) 6772–6775.
[119] M. Chiba, M. Kamiya, K. Tsuda-Sakurai, Y. Fujisawa, H. Kosakamoto, R. Kojima, M. Miura, Y. Urano, Activatable Photosensitizer for Targeted Ablation of lacZ-Positive Cells with Single-Cell Resolution, ACS Cent. Sci. 5 (2019) 1676–1681.
[120] L. v. Lutkus, H.E. Irving, K.S. Davies, J.E. Hill, J.E. Lohman, M.W. Eskew, M.R. Detty, T.M. McCormick, Photocatalytic Aerobic Thiol Oxidation with a Self-Sensitized Tellurorhodamine Chromophore, Organometallics. 36 (2017) 2588–2596.
[121] B. Calitree, D.J. Donnelly, J.J. Holt, M.K. Gannon, C.L. Nygren, D.K. Sukumaran, J. Autschbach, M.R. Detty, Tellurium analogues of rosamine and rhodamine dyes: Synthesis, structure, 125Te NMR, and heteroatom contributions to excitation energies, Organometallics. 26 (2007) 6248–6257.
[122] Y. Koide, M. Kawaguchi, Y. Urano, K. Hanaoka, T. Komatsu, M. Abo, T. Terai, T. Nagano, A reversible near-infrared fluorescence probe for reactive oxygen species based on Te-rhodamine, Chem. Commun. 48 (2012) 3091–3093.
[123] A. Mukaimine, T. Hirayama, H. Nagasawa, Asymmetric bismuth-rhodamines as an activatable fluorogenic photosensitizer, Org. Biomol. Chem. 19 (2021) 3611–3619.
[124] K.H. Gebremedhin, M. Li, F. Gao, B. Gurram, J. Fan, J. Wang, Y. Li, X. Peng, Benzo[a]phenoselenazine-based NIR photosensitizer for tumor-targeting photodynamic therapy via lysosomal-disruption pathway, Dyes Pigm. 170 (2019).

[125] S. Cetin, Z. Elmazoglu, O. Karaman, H. Gunduz, G. Gunbas, S. Kolemen, Balanced Intersystem Crossing in Iodinated Silicon-Fluoresceins Allows New Class of Red Shifted Theranostic Agents, ACS Med. Chem. Lett. 12 (2021) 752–757.

[126] M.W. Kryman, G.A. Schamerhorn, K. Yung, B. Sathyamoorthy, D.K. Sukumaran, T.Y. Ohulchanskyy, J.B. Benedict, M.R. Detty, Organotellurium fluorescence probes for redox reactions: 9-aryl-3,6-diaminotelluroxanthylium dyes and their telluroxides, Organometallics. 32 (2013) 4321–4333.